\documentclass[a4paper]{article}\usepackage[]{graphicx}\usepackage[dvipsnames]{xcolor}
\makeatletter
\def\maxwidth{ %
  \ifdim\Gin@nat@width>\linewidth
    \linewidth
  \else
    \Gin@nat@width
  \fi
}
\makeatother

\definecolor{fgcolor}{rgb}{0.345, 0.345, 0.345}

\usepackage{framed}
\makeatletter
 {\par\unskip\endMakeFramed%
 \at@end@of@kframe}
\makeatother

\definecolor{shadecolor}{rgb}{.97, .97, .97}
\definecolor{messagecolor}{rgb}{0, 0, 0}
\definecolor{warningcolor}{rgb}{1, 0, 1}
\definecolor{errorcolor}{rgb}{1, 0, 0}
\newenvironment{knitrout}{}{} % an empty environment to be redefined in TeX

\usepackage{alltt}
\usepackage[a4paper, lmargin=3.5cm, rmargin=3cm]{geometry}
\usepackage{array}
\usepackage{times}
\usepackage[authoryear, round]{natbib}
\usepackage{booktabs}
\usepackage{multirow}

\usepackage{caption}
\usepackage{siunitx}
\usepackage[normalem]{ulem}
\usepackage{colortbl}
\usepackage{hhline}
\usepackage{calc}
\usepackage{tabularx}
\usepackage{threeparttable}
\usepackage{wrapfig}
\usepackage{adjustbox}

\usepackage{graphicx}
\chardef\bslash=`\\ % p. 424, TeXbook

\usepackage{bm}
\usepackage{amsthm,amsmath,amssymb,amsfonts,amssymb}
\usepackage{listings}
\usepackage{enumerate}
\usepackage{enumitem}
\setlist[enumerate,1]{label={(\roman*)}}
\usepackage[figuresright]{rotating}
\usepackage{mathtools}
\usepackage[dvipsnames]{xcolor}
\usepackage{xurl}
\usepackage{hyperref}
\usepackage[capitalize]{cleveref}
\hypersetup{
colorlinks,
citecolor=black,
filecolor=black,
linkcolor=black,
urlcolor=black
}
\definecolor{backgroundColour}{rgb}{0.97,0.96,0.97}
\IfFileExists{upquote.sty}{\usepackage{upquote}}{}
\begin{document}
%%%%%%%%%%%%%%%%%%%%%%%%%%%%%%%%%%%%%%%%%%%%%%%%%%%%%%%%%%%%%%%%%%%%%%%

% Import R files here with proper labels
% include = FALSE removes any output from this chunk

%\title{\vspace{-3.0cm}Regime classification uncertainty and optimizer reliability on hidden Markov models parameter estimation using Template Model Builder}

\title{\vspace{-3.0cm} Computational issues in parameter estimation for hidden Markov models with Template Model Builder}

\author{
  \noindent Timoth\'ee Bacri \thanks{Corresponding author: e-mail: \sf{timot@bacri.eu}} $^{1}$,
  Geir D. Berentsen $^{2}$,
  Jan Bulla $^{1,3}$\\[1ex],
  B{\aa}rd St{\o}ve $^{1}$\\[1ex]
  $^{1}$ \small{Department of Mathematics, University of Bergen, Postbox 7803, 5007 Bergen, Norway}\\
  $^{2}$ \small{Department of Business and Management Science, Norwegian School of Economics, Helleveien 30, 5045 Bergen, Norway}\\
  $^{3}$ \small{Department of Psychiatry and Psychotherapy, University of Regensburg, 93053 Regensburg, Germany}
}

\maketitle

% \vspace{-0.5cm}
\begin{abstract}
% \fontsize{9pt}{11pt} % 2nd parameter = 1.2 times the font size, https://tex.stackexchange.com/a/48277
% \selectfont
\noindent
A popular way to estimate the parameters of a hidden Markov model (HMM) is direct numerical maximization (DNM) of the (log-)likelihood function. The advantages of employing the {\tt{TMB}} \citep{kristensen} framework in {\tt{R}} for this purpose  were illustrated recently \citet{bacri}. In this paper, we present extensions of these results in two directions.\\
First, we present a practical way to obtain uncertainty estimates in form of confidence intervals (CIs) for the so-called smoothing probabilities at moderate computational and programming effort via {\tt{TMB}}. Our approach thus permits to avoid computer-intensive bootstrap methods. By means of several examples, we illustrate patterns present for the derived CIs.\\
Secondly, we investigate the performance of popular optimizers available in {\tt{R}} when estimating HMMs via DNM. Hereby, our focus lies on the potential benefits of employing {\tt{TMB}}. Investigated criteria via a number of simulation studies are convergence speed, accuracy, and the impact of (poor) initial values. Our findings suggest that all optimizers considered benefit in terms of speed from using the gradient supplied by {\tt{TMB}}. When supplying both gradient and Hessian from {\tt{TMB}}, the number of iterations reduces, suggesting a more efficient convergence to the maximum of the log-likelihood. Last, we briefly point out potential advantages of a hybrid approach.

\end{abstract}

% \begin{abbreviations}
% Hidden Markov Model (HMM), Direct Numerical Maximization (DNM), negative log-likelihood (nll), Maximum Likelihood (ML), Baum-Welch (BW), Expectation-Maximization (EM), Confidence intervals (CIs), ``Track Your Tinnitus'' (TYT), Template Model Builder (TMB)
% \end{abbreviations}

\vspace*{1pc}
\\[2pt]
\noindent Supporting Information for this article is available from the authors or on the WWW under\break
\url{https://timothee-bacri.github.io/HMM_with_TMB}

%%%%%%%%%%%%%%%%%%%%%%%%%%%%%%%%%%%%%%%%%%%%%%%%%%%%%%%%%%%%%%%%%
\section{Introduction}
\label{sec:intro}
%%%%%%%%%%%%%%%%%%%%%%%%%%%%%%%%%%%%%%%%%%%%%%%%%%%%%%%%%%%%%%%%%

Hidden Markov models (HMMs) are a well-studied and popular class of models in many areas. While they have been used for speech recognition historically \citep[see, e.g.][]{juang, baum, rabiner, rabinera, fredkin}, these models also became important in many other fields due to their flexibility. These are, to name only a few, biology and bioinformatics \citep{schadt, durbin, eddy}, finance \citep{hamilton}, ecology \citep{mcclintock}, stochastic weather modeling \citep{lystig, ailliot}, and engineering \citep{mor}. In short, HMMs are a class of models where the given data is assumed to follow varying distributions according to an underlying unknown Markov chain.\\
Parameter estimation for HMMs is typically achieved either by the Baum-Welch (BW) algorithm \citep{bauma, dempster, rabiner, liporace, wu} - an Expectation-Maximization(EM)-type algorithm - or in a quite straightforward fashion by direct numerical maximization (DNM) of the (log-)likelihood function \citep[see, e.g.,][]{turner,macdonald}. A discussion of both approaches can be found in \cite[p.~358]{cappe}. Furthermore, \citep{lange} points out that it is challenging to fit complex models with the BW algorithm, \cite[pp. ~77-78]{zucchini} advises using DNM with HMMs due to the ease of fitting complex models, and \citep{altmana, turner} report a greater speed of DNM compared to the BW algorithm. Nevertheless, the BW is highly accepted and finds widespread application. However, we will only use DNM in the following due to the possibility of accelerating the estimation via the R-package {\tt{TMB}} \citep{kristensen, bacri}.

In this paper, we present a straightforward procedure to calculate uncertainty estimates of the \textit{smoothing probabilities} through appropriate use of {\tt{TMB}}. Hence, we quantify the uncertainty of the state classifications of the observations at a low computational cost. To the best of our knowledge, such results on confidence intervals for the aforementioned probabilities are not available in the literature.\\
Furthermore, the chosen optimization routine for DNM plays an important role in parameter estimation. While e.g.~\citet{zucchini} focuses on the unconstrained optimizer \texttt{nlm}, alternatives include the popular \texttt{nlminb} \citep{gay} and many others, such as those provided by the \texttt{optim} function \citep{R-optimr}. In this context, \citet{bulla} compare different estimation routines for two-state HMMs: a Newton-Type algorithm \citep{dennis, schnabel}, the Nelder-Mead algorithm \citep{nelder}, BW, and a hybrid algorithm successively executing BW and DNM.
To complement these studies, we investigate the speed of several optimization algorithms, many of which allow the gradient and Hessian of the objective function as input. Particular focus lies on the ability of the {\tt{TMB}} framework, which allows for easy retrieval of the exact gradient and Hessian (up to machine precision). This enriches the existing literature on parameter estimation for HMMs.\\
%\citet{oudelha} note that "the BW algorithm uses an initial random guess of the parameters, therefore after convergence the output tends to be close to this initial value of the algorithm, which is not necessarily the global optimum of the model parameters", and 
Among others, \citet{zucchini} report that ``Depending on the starting values, it can easily happen that the algorithm (i.e.~DNM) identifies a local, but not the global, maximum''. The literature on how to tackle this problem is rich, and includes among others: the use of artificial neural networks to guess the initial values \citep{hassan}, model induction \citep{stolcke}, genetic algorithms in combination with HMMs \citep{oudelha}, hybridizing both the BW and the DNM algorithms \citep{redner, lange, bulla}, specific assumptions on the parameters \citep{kato}, or educated guesses \citep[p. ~53]{zucchini}. We investigate how stable the various considered optimization routines are towards poor initial values. Again, this part takes under consideration the special features of {\tt{TMB}}, which have not been investigated yet.

The paper is organized as follows. In \autoref{sec:basics}, we provide a brief overview of parameter estimation for HMMs, including inference for CIs. In \autoref{sec:smoothing-uncertainty} we show how uncertainty estimates of the smoothing probabilities can be computed via {\tt{TMB}}. Then, we apply our results to a couple of data sets with different characteristics. We perform simulation studies in \cref{sec:optimization,sec:robustness}. Therein, we compare measures of performance and accuracy of various optimizers and check how well these optimizers perform in the presence of poor initial values. \autoref{sec:conclusion} provides some concluding remarks. All code necessary to reproduce our results is available in the supporting information.

%%%%%%%%%%%%%%%%%%%%%%%%%%%%%%%%%%%%%%%%%%%%%%%%%%%%%%%%%%%%%%%%%
\section{Basics on hidden Markov models}
\label{sec:basics}
%%%%%%%%%%%%%%%%%%%%%%%%%%%%%%%%%%%%%%%%%%%%%%%%%%%%%%%%%%%%%%%%%

The HMMs considered here are fit on observed time series $\{X_t: t = 1, \ldots, T\}$ where $t$ denotes the (time) index ranging from one to $T$. In this setting, a mixture of conditional distributions is assumed to be driven by an unobserved (hidden) homogeneous Markov chain, whose states will be denoted as $\{C_t : t = 1, \ldots, T\}$. We will use different conditional distributions in the paper. First, we specify an $m$-state Poisson HMM, i.e.~with the conditional distribution
\begin{equation*}
p_i(x) = \text{P}(X_t = x \vert C_t = i) = \frac{e^{-\lambda_i} \lambda_i^x}{x!}
\end{equation*}
with parameters $\lambda_i, i = 1,...,m$. Secondly, we consider Gaussian HMMs with conditional distribution specified by
\begin{equation*}
p_i(x) = \text{P}(X_t = x \vert C_t = i) = \frac{1}{\sigma \sqrt{2 \pi}} e^{-\frac{1}{2} \left( \frac{x - \mu_i}{ \sigma_i} \right)^2 },
\end{equation*}
with parameters $(\mu_i, \sigma_i), i = 1, \ldots, m$. In addition,  $\bm{\Gamma} = \{\gamma_{ij}\}, i,j=1,...,m$ denotes the transition probability matrix (TPM) of the HMM's underlying Markov chain, and $\bm{\delta}$ is a vector of length $m$ collecting the corresponding stationary distribution. We assume that the Markov chains underlying our HMMs are irreducible and aperiodic. This ensures the existence and uniqueness of a stationary distribution as the limiting distribution \citep{Feller}. However, these results are of limited relevance for most estimation algorithms, because the elements of $\bm{\Gamma}$ are generally strictly positive. Nevertheless, one should be careful when manually fixing selected elements of $\bm{\Gamma}$ to zero.

The basis for our estimation procedure is the (log-)likelihood function. We denote the ``history'' of observations $x_t$ and the observed process $X_t$ up to time $t$ by $\bm{x^{(t)}} = \{x_1, \ldots, x_t \}$ and $\bm{X^{(t)}} = \{X_1, \ldots, X_t \}$, respectively. In addition, let $\bm{\theta}$ be the vector of model parameters. As explained, e.g., by \cite[p.~37]{zucchini}, the likelihood function can then be represented as a product of matrices:  
\begin{equation}
\label{eq:hmm_likelihood}
L(\bm{\theta}) = \text{P}(\bm{X^{(T)}} = \bm{x^{(T)}}) = \bm{\delta} \bm{P}(x_1) \bm{\Gamma} \bm{P}(x_2) \bm{\Gamma} \bm{P}(x_3) \ldots \bm{\Gamma} \bm{P}(x_T) \bm{1}',
\end{equation}
where the $m$ conditional probability density functions evaluated at $x$ (we use this term for discrete support as well) are collected in the diagonal matrix
\begin{equation*}
\bm{P}(x) = \begin{pmatrix}
p_1(x)    &         &         & 0\\
          & p_2(x)  &         &\\
          &         & \ddots  &\\
0         &         &         & p_m(x)
\end{pmatrix},
\end{equation*}
and $\bm{1}'$ and $\bm{\delta}$ denote a transposed vector of ones and the stationary distribution, respectively. That is, we assume that the initial distribution corresponds to $\bm{\delta}$. %Funny! Moreover, alt g{\aa}r i grisen. 
The likelihood function given in \autoref{eq:hmm_likelihood} can be efficiently evaluated by a so-called forward pass through the observations, as illustrated in \autoref{sec:appendix-hmm_fwbw}. Consequently, it is possible to obtain $\hat{\bm{\theta}}$, the ML estimates of $\bm{\theta}$, by - in our case - unconstrained DNM. Since several of our parameters are constrained, we rely on known re-parametrization procedures (see \autoref{sec:appendix-reparameterization} for details).

Confidence intervals (CIs) for the estimated parameters of HMMs can be derived via various approaches. The most common ones are Wald-type, profile likelihood, and bootstrap-based CIs. \citet{bacri} shows that {\tt{TMB}} can yield valid Wald-type CIs in a fraction of the time required by classical bootstrap-based methods as investigated e.g.~by \citet{bulla, zucchini}. Furthermore, the likelihood profile method may fail to provide CIs \citep{bacri}. Therefore, we rely on Wald-type confidence intervals derived via {\tt{TMB}}. For details, see \autoref{sec:appendix-cis}.

%%%%%%%%%%%%%%%%%%%%%%%%%%%%%%%%%%%%%%%%%%%%%%%%%%%%%%%%%%%%%%%%%
\section{Uncertainty of smoothing probabilities}
\label{sec:smoothing-uncertainty}
%%%%%%%%%%%%%%%%%%%%%%%%%%%%%%%%%%%%%%%%%%%%%%%%%%%%%%%%%%%%%%%%%

When interpreting the estimation results of an HMM, the so-called smoothing probabilities often play an important role.
These quantities, denoted by $p_{it}(\bm{\theta})$ in the following, correspond to the probability of being in state $i$ at time $t$ given the observations, i.e., 
\begin{equation*}
\label{eq:sm_prob}
  p_{it}(\bm{\theta}) = \text{P}_{\bm{\theta}}(C_t = i \vert X^{(T)} = x^{(T)}) = \frac{\alpha_t(i) \beta_t(i)}{L(\bm{\theta})},
\end{equation*}
and can be calculated for $i=1, \ldots, m$ and $t=1, \ldots, T$ \citep[see, e.g.,][p.~87]{zucchini}. The involved quantities $\alpha_t(i)$ and $\beta_t(i)$, which also depend on $\bm{\theta}$, result directly from the forward- and backward algorithm, respectively, as illustrated in \autoref{sec:appendix-hmm_fwbw}.
Estimates of $p_{it}(\bm{\theta})$ are obtained by $p_{it}(\hat{\bm{\theta}})$ where $\hat{\bm{\theta}}$ is the ML estimate of $\bm{\theta}$.

An important feature of {\tt{TMB}} is that it not only permits obtaining standard errors for $\hat{\bm{\theta}}$ (and CIs for $\bm{\theta}$), but in principle also for any other quantity depending on the parameters. This is achieved by combining the delta method with automatic differentiation (see \citep{kristensen} for details).
A minor requirement for deriving standard deviations for $p_{it}(\hat{\bm{\theta}})$ via the \texttt{ADREPORT} function provided by {\tt{TMB}} consists in implementing the function $p_{it}(\bm{\theta})$ in \texttt{C++}. Then, this part has to be integrated into the scripts necessary for the parameter estimation procedure. It is noteworthy that, once implemented, the procedures related to inference of the smoothing probabilities do not need to change when the HMM considered changes, because the input vector for $p_{it}(\bm{\theta})$ remains the same. The \texttt{C++} code in the supporting information illustrates how to complete these tasks. Note that population value we are interested in constructing a CI for is $p_{it}(\bm{\theta}) $, which should not be confused with  $E_{X^{(T)}}\left(P_{\hat{\bm{\theta}}(X^{(T)})}(C_t = i\mid X^{(T)})\right)$. That is, we treat $p_{it}(\bm{\theta})$ as the probability of the event $C_t=i$ conditional on $X^{(T)}$ being equal to the particular sequence of observations $x^{(T)}$. After all, it is this sequence and the corresponding smoothing probabilities which are of interest to the researcher. 
% In the end, smoothing probabilities and corresponding confidence intervals are available as by-products of the parameter estimation procedure with {\tt{TMB}}.
 %, because only the parameters involved in the estimation need to be updated.
% Furthermore, the 95\% CIs can be formed via $\hat{\psi} \pm q_{0.975} \cdot \text{SE}(\hat{\psi})$ where $q_{0.975} \approx 1.96$ is the 97.5th percentile from the standard normal distribution, and $\text{SE}(\hat{\psi})$ denotes the standard error of $\hat{\psi}$, calculated from {\tt{TMB}}. 

Quantifying the uncertainty in the estimated smoothing probabilities can be of interest in multiple fields because the underlying states are often linked to certain events of interest (e.g., the state of an economy or the behavior of a customer). In the following, we illustrate the results of our approach through a couple of examples.

%%%%%%%%%%%%%%%%%%%%%%%%%%%%%%%%%%%%%%%%%%%%%%%%%%%%%%
\subsection{Track Your Tinnitus}
\label{sec:sm-tyt}
%%%%%%%%%%%%%%%%%%%%%%%%%%%%%%%%%%%%%%%%%%%%%%%%%%%%%%

The ``Track Your Tinnitus'' (TYT) mobile application gathered a large data set, a description of which is detailed in \citet{pryss} and \citet{pryssa}.
The data plotted in \autoref{fig:data-plot-tinn} shows the so-called ``arousal'' variable reported for an individual over 87 consecutive days.
This variable takes high values when a high level of excitement is achieved and low values when the individual is in a calm emotional state.
The values are measured on a discrete scale; we refer to \citet{probst, probsta} for details.

\begin{knitrout}
\definecolor{shadecolor}{rgb}{0.969, 0.969, 0.969}\color{fgcolor}\begin{figure}[htb]

{\centering \includegraphics[width=\maxwidth]{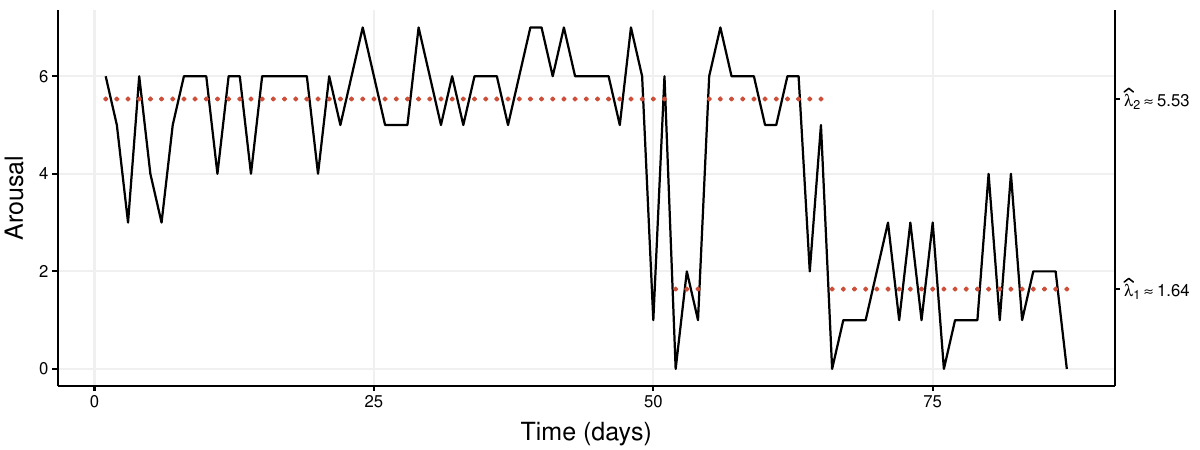} 

}

\caption{Plot of a two-state Poisson HMM fitted to the TYT data (of size 87). The colored dots correspond to the conditional mean of the inferred state at each time. Model estimates are displayed in \autoref{tab:estimates-cis-aic-bic-real-data-part1}.}\label{fig:data-plot-tinn}
\end{figure}

\end{knitrout}

We estimated Poisson HMMs with varying number of states by \texttt{nlminb} with {\tt{TMB}}'s gradient and Hessian functions passed as arguments (this approach was chosen for all estimated models in our examples). The preferred HMM in terms of AIC and BIC is a two-state model, see \autoref{tab:estimates-cis-aic-bic-real-data-part1} in \autoref{sec:appendix-estmod}. 

\autoref{fig:smoothing-tinn} displays the corresponding smoothing probabilities with 95\% Wald-type CIs, constructed using the standard error provided by {\tt{TMB}}. Intuitively, one might expect the uncertainty to be low when a smoothing probability takes values close to zero or one, whereas higher uncertainty should be inherent to smoothing probabilities further away from these bounds. The CIs illustrated in \autoref{fig:smoothing-tinn} follow this pattern. However, as pointed out by \citet{bacri}, this data set is atypically short for fitting HMMs. As we will see in the following, different patterns will emerge for longer sequences of observations.

\begin{knitrout}
\definecolor{shadecolor}{rgb}{0.969, 0.969, 0.969}\color{fgcolor}\begin{figure}[!htb]

{\centering \includegraphics[width=\maxwidth]{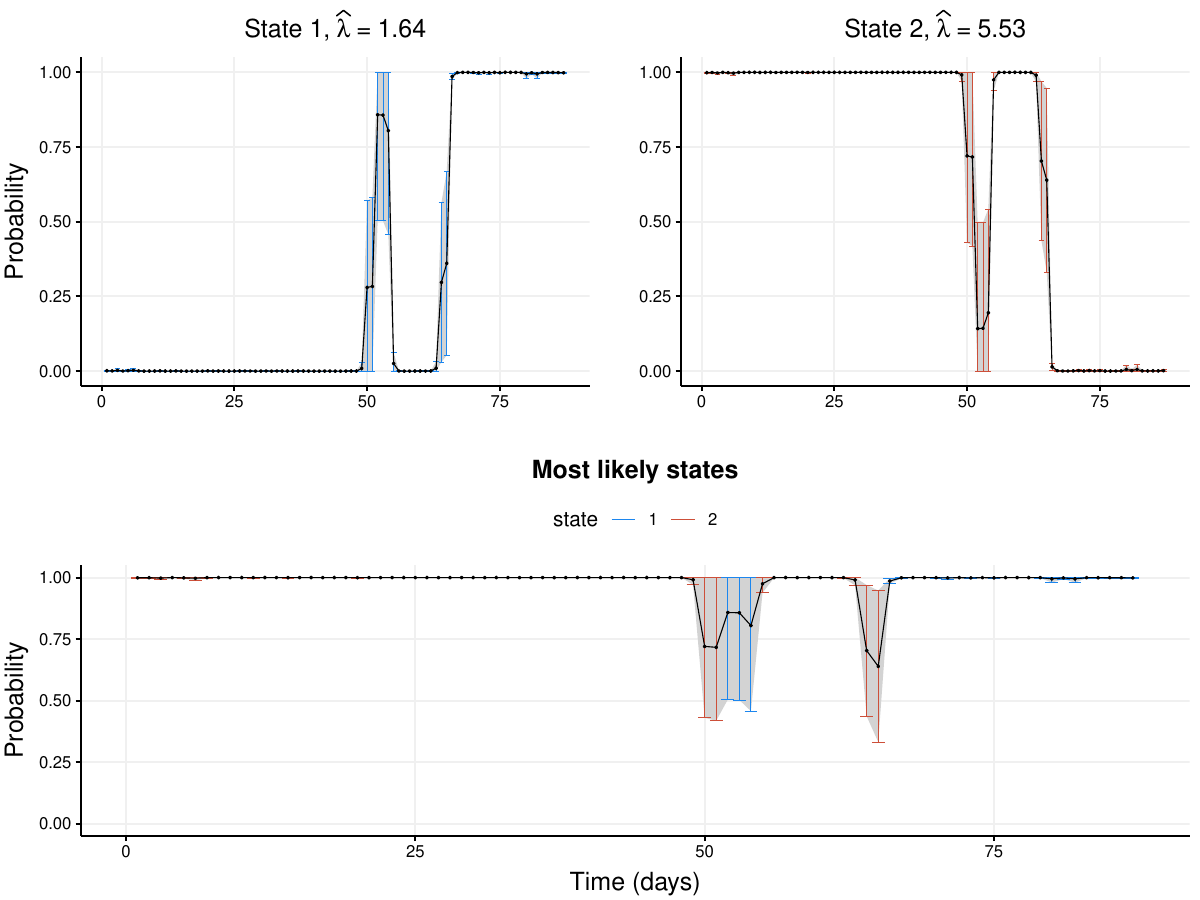} 

}

\caption{Smoothing probabilities and confidence intervals of a two-state Poisson HMM fitted to the TYT data set. The solid line shows the smoothing probability estimates and the 95\% CIs are represented by vertical lines. The bottom graph displays smoothing probabilities for the most likely state estimated for each data point. The vertical confidence interval lines are colored differently per hidden state. Further details on the model estimates are available in \autoref{tab:estimates-cis-aic-bic-real-data-part1}.}\label{fig:smoothing-tinn}
\end{figure}

\end{knitrout}

%%%%%%%%%%%%%%%%%%%%%%%%%%%%%%%%%%%%%%%%%%%%%%%%%%%%%%
\subsection{Soap}
\label{sec:sm-soap}
%%%%%%%%%%%%%%%%%%%%%%%%%%%%%%%%%%%%%%%%%%%%%%%%%%%%%%

In the following, we consider a data set of weekly sales of a soap in a supermarket. The data were provided by the Kilts Center for Marketing, Graduate School of Business of the University of Chicago and are available in the database at \url{https://research.chicagobooth.edu/kilts/marketing-databases/dominicks}.
The data are displayed in \autoref{fig:data-plot-soap}.
%The product was ``Zest White Water 15 oz.'', with code 3700031165 and store number 67.
Similarly to the previous section, we fitted multiple Poisson HMMs, and the preferred HMM in terms of AIC and BIC is a two-state model as well (see \autoref{tab:estimates-cis-aic-bic-real-data-part1} in \autoref{sec:appendix-estmod}). \autoref{fig:smoothing-soap} displays the smoothing probabilities resulting from this model.
The patterns of the CIs described in the previous section are, in principle, also present here.
Nevertheless, an exception is highlighted in the bottom panel of \autoref{fig:smoothing-soap}:
at $t = 152$, the smoothing probability of State 2 is closer to the upper boundary of one than the corresponding probability at $t = 147$. Yet, the uncertainty is higher at $t = 152$.
%We also note that many smoothing probability estimates not in the immediate neighborhood of 0.5 possess CIs that go through 0.5, showing that many state classifications are uncertain. This effect can be noticed at a larger degree with the S\&P 500 data set in \autoref{sec:sm-weekly}, where the thresholds are $1/3$ and $2/3$ due to the model having three hidden states.

\begin{knitrout}
\definecolor{shadecolor}{rgb}{0.969, 0.969, 0.969}\color{fgcolor}\begin{figure}[htb]

{\centering \includegraphics[width=\maxwidth]{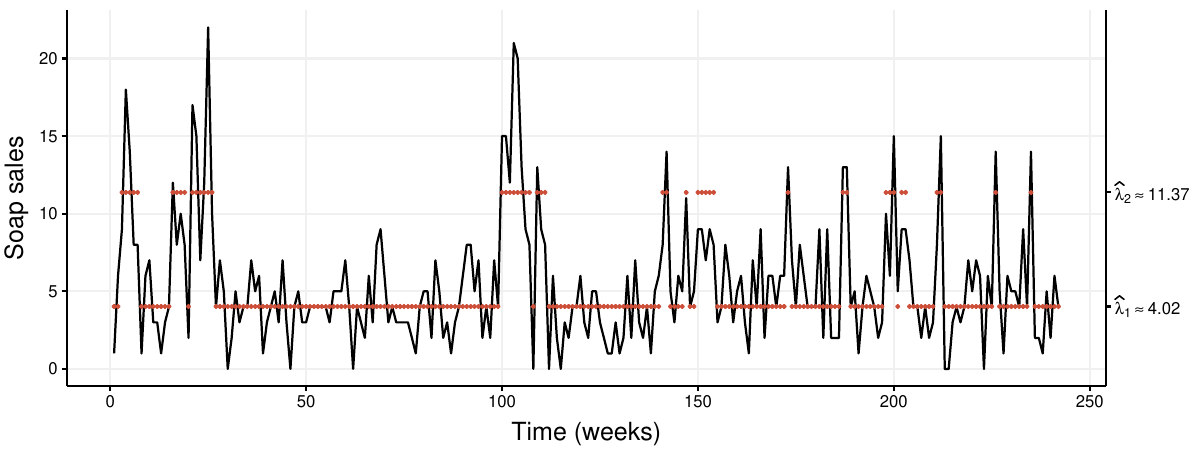} 

}

\caption{Plot of the soap data (of size 242), with fitted two-state Poisson HMM. The colored dots correspond to the conditional mean of the inferred state at each time. Model estimates are displayed in \autoref{tab:estimates-cis-aic-bic-real-data-part1}.}\label{fig:data-plot-soap}
\end{figure}

\end{knitrout}

\begin{knitrout}
\definecolor{shadecolor}{rgb}{0.969, 0.969, 0.969}\color{fgcolor}\begin{figure}[!htb]

{\centering \includegraphics[width=\maxwidth]{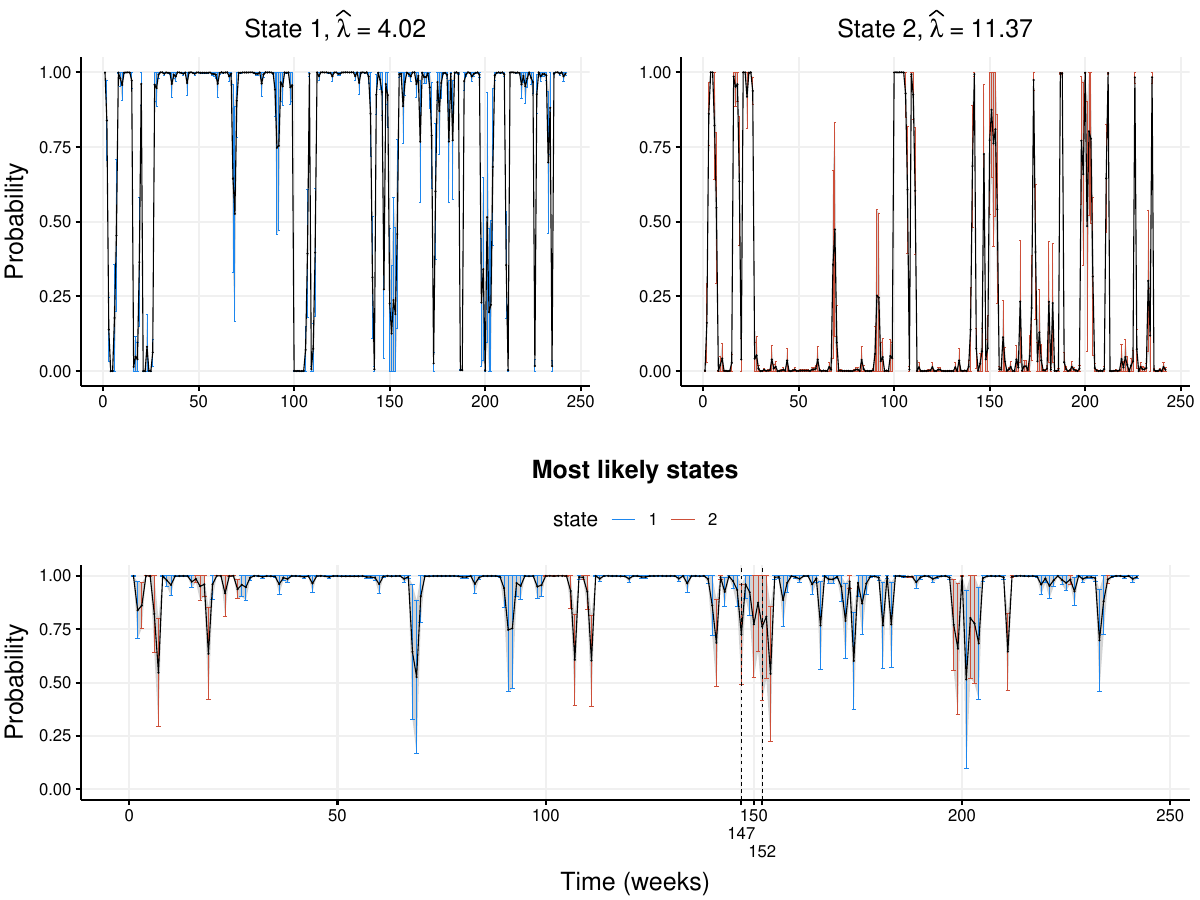} 

}

\caption{Smoothing probabilities and confidence intervals of a two-state Poisson HMM fitted to the soap data set. The solid line shows the smoothing probability estimates and the 95\% CIs are represented by vertical lines. The bottom left graph displays smoothing probabilities for the most likely state estimated for each data point. The vertical confidence interval lines are colored differently per hidden state. Further details on the model estimates are available in \autoref{tab:estimates-cis-aic-bic-real-data-part1}. }\label{fig:smoothing-soap}
\end{figure}

\end{knitrout}

%%%%%%%%%%%%%%%%%%%%%%%%%%%%%%%%%%%%%%%%%%%%%%%%%%%%%%
\subsection{Weekly returns}
\label{sec:sm-weekly}
%%%%%%%%%%%%%%%%%%%%%%%%%%%%%%%%%%%%%%%%%%%%%%%%%%%%%

The data set considered in this section are 2230 weekly log-returns based on the adjusted closing share price of the S\&P 500 stock market index retrieved from Yahoo Finance, between January $1^{st}$ 1980 and September $30^{th}$ 2022.
The returns are expressed in per cent to facilitate reading and interpreting the estimates.
As shown, e.g., by \citet{rydenb}, Gaussian HMMs reproduce well certain stylized facts of financial returns. We thus estimated such models with varying number of states.
A three-state model is preferred by the BIC, whereas the AIC is almost identical for three and four states (see \autoref{tab:estimates-cis-aic-bic-real-data-part1}). The estimates show a decrease in conditional standard deviation with increasing conditional mean.
This is a well-known property of many financial returns (see, e.g., \citet{schwerta,maheua,guidolina}) related to the behavior of market participants in crisis and calm periods. \autoref{fig:data-plot-sp500} shows the first 200 data points plotted along with conditional means from the preferred model.

\begin{knitrout}
\definecolor{shadecolor}{rgb}{0.969, 0.969, 0.969}\color{fgcolor}\begin{figure}[htb]

{\centering \includegraphics[width=\maxwidth]{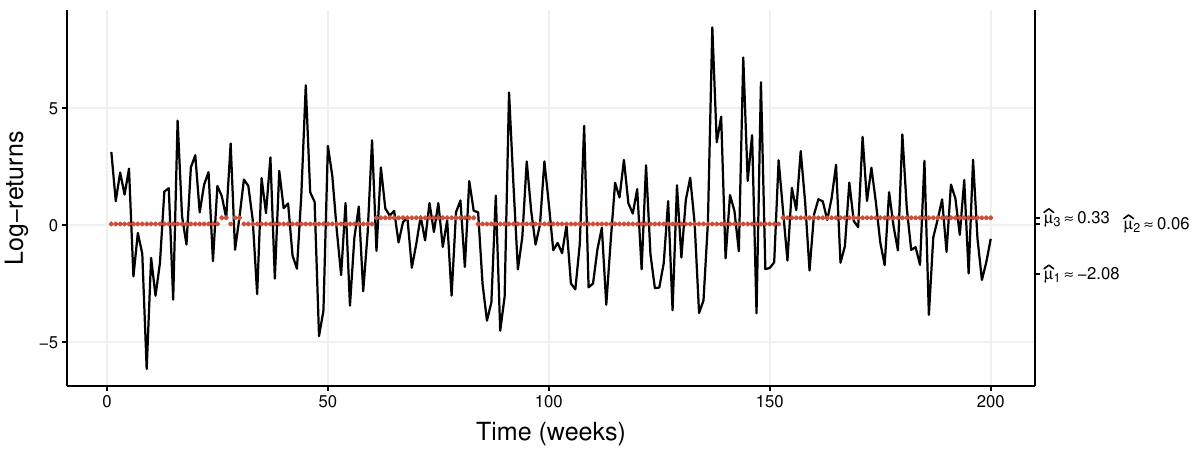} 

}

\caption{Plot of the weekly returns (of size 2230). The colored dots correspond to the conditional mean of the inferred state from a fitted 3-state Gaussian HMM (see \autoref{tab:estimates-cis-aic-bic-real-data-part1}). For readability, only the first 200 data are plotted.}\label{fig:data-plot-sp500}
\end{figure}

\end{knitrout}

\begin{knitrout}
\definecolor{shadecolor}{rgb}{0.969, 0.969, 0.969}\color{fgcolor}\begin{figure}[!htb]

{\centering \includegraphics[width=\maxwidth]{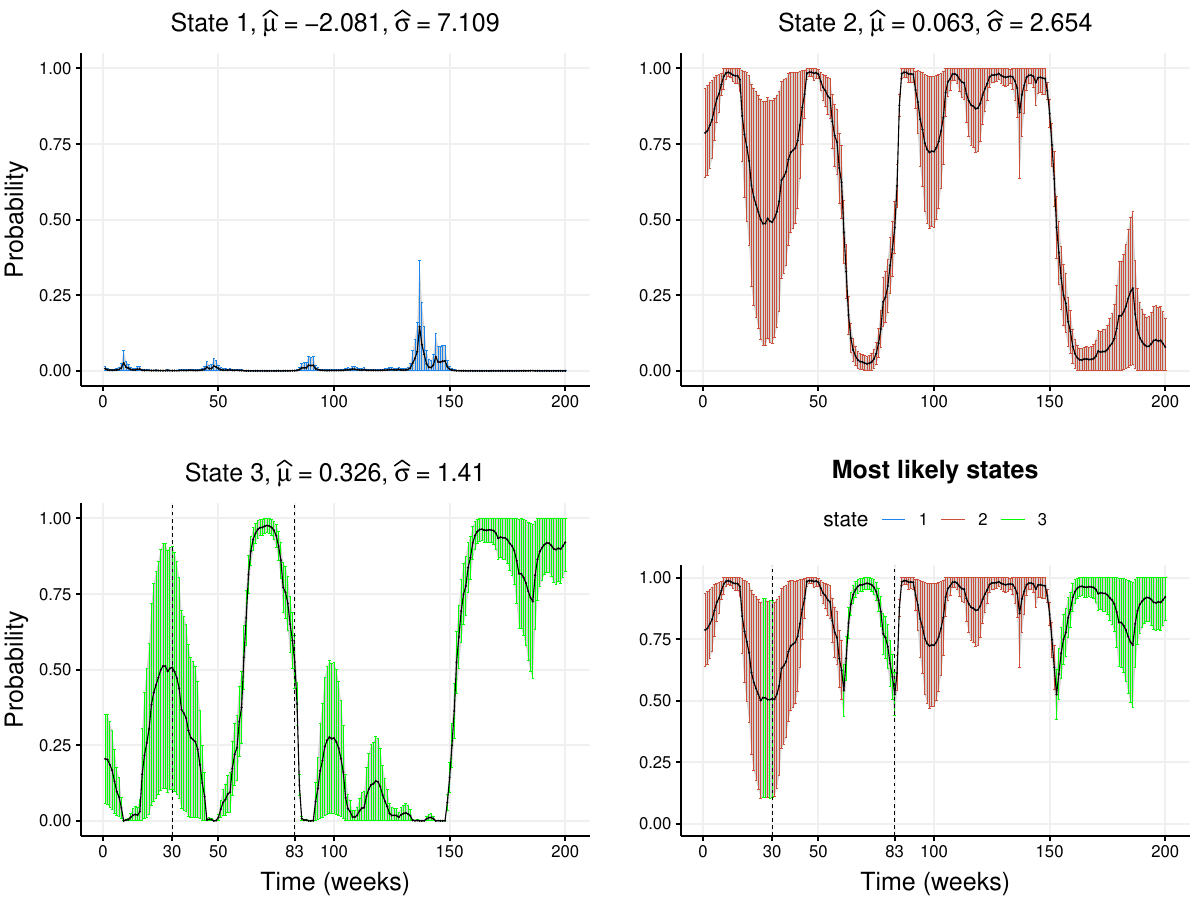} 

}

\caption{Smoothing probabilities and confidence intervals of a three-state Gaussian HMM fitted to the weekly returns data. The solid line shows the smoothing probability estimates and the 95\% CIs are represented by vertical lines. The bottom right graph displays smoothing probabilities for the most likely state estimated for each data point. The vertical confidence interval lines are colored differently per hidden state. Only the first 200 out of the 2230 values are shown for readability purposes. Further details on the model estimates are available in \autoref{tab:estimates-cis-aic-bic-real-data-part1}.}\label{fig:smoothing-sp500}
\end{figure}

\end{knitrout}

The subsequent \autoref{fig:smoothing-sp500} shows inferred smoothing probabilities together with their CIs. In addition to the previous applications, where smoothing probabilities close to the boundary seemed to be linked to lower uncertainty, this data set shows also some clear exceptions from this pattern. For example, the smoothing probabilities of State 3 inferred at $t = 30$ and $t = 83$ take the relatively close values 0.51 and 0.53, respectively. However, the associated uncertainty at $t = 30$ is visibly higher than the corresponding uncertainty at $t = 83$ (the estimated standard errors are 0.205 and 0.045). This may be explained by the fact that the inferred states closely around $t=30$ oscillated between State 2 and 3, thus indicating a high uncertainty of the state classification during this period. On the contrary, around $t=83$ a quick and persistent change from State 2 to 3 takes place.

%%%%%%%%%%%%%%%%%%%%%%%%%%%%%%%%%%%%%%%%%%%%%%%%%%%%%%
\subsection{Hospital}
\label{sec:sm-hospital}
%%%%%%%%%%%%%%%%%%%%%%%%%%%%%%%%%%%%%%%%%%%%%%%%%%%%%%

Basis for our last example is a data set issued by the hospital "H\^{o}pital Lariboisi\`ere" from Assistance Publique -- H\^{o}pitaux de Paris (a french hospital trust).
This data set consists of the hourly number of patient arrivals to the emergency ward during a period of roughly 10 years.
A subset of the data (over one week) is displayed in \autoref{fig:data-plot-hosp}.
We examine this due to several reasons.
First, with 87648 observations, this data set is much larger than the ones examined previously.
Secondly, the medical staff noticed differences between the rates of patient arrivals at day and night, respectively, which motivates the use of, e.g., an HMM.
Last, a Poisson HMM is a natural candidate for the observed count-type data. The preferred model by both AIC and BIC has five states (see \autoref{tab:estimates-cis-aic-bic-real-data-part2} in \autoref{sec:appendix-estmod})
and confirms the impression of the hospital employees: State 5 is mainly visited during core operating hours during day time.
The fourth and third states mainly occur late afternoon, early evening, and early morning.
Last, State 1 and 2 correspond to night time observations.

\begin{knitrout}
\definecolor{shadecolor}{rgb}{0.969, 0.969, 0.969}\color{fgcolor}\begin{figure}[htb]

{\centering \includegraphics[width=\maxwidth]{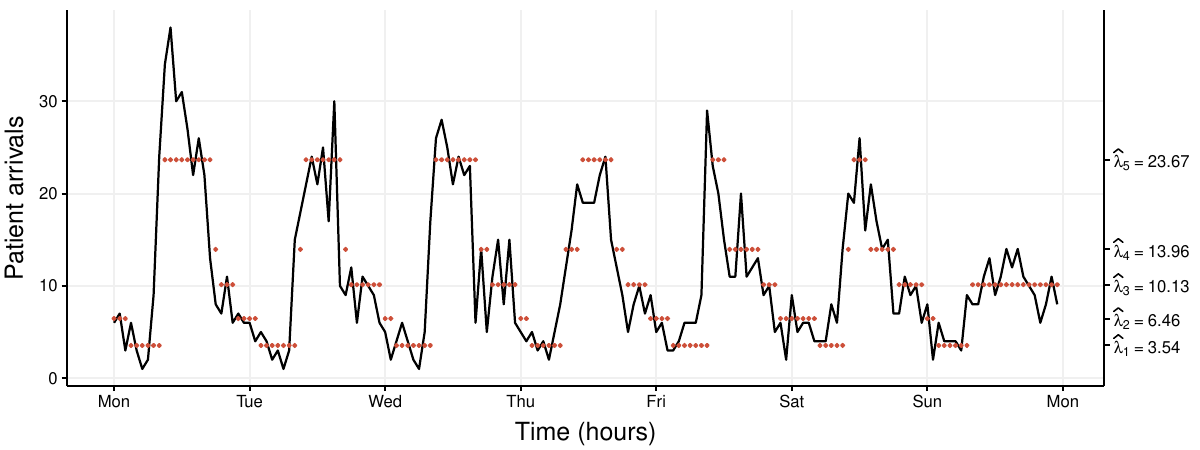} 

}

\caption{Plot of the hospital data (of size 87648). The colored dots correspond to the conditional mean of the inferred state at each time. For readability, only the first week starting on Monday (from Monday January $4^{th}$ 2010 at 00:00 to Sunday January $10^{th}$ 2010 at 23:59) is plotted instead of the entire 10 years. Model estimates are displayed in \autoref{tab:estimates-cis-aic-bic-real-data-part2}.}\label{fig:data-plot-hosp}
\end{figure}

\end{knitrout}

Even for a data set of this size, the smoothing probabilities and corresponding CIs can be derived with moderate computational effort and time: \autoref{fig:smoothing-hosp} displays the results.
Overall, the inferred CIs are relatively small, in particular in comparison with those obtained for the stock return data discussed in the previous example.
The low uncertainty in the state classification may most likely result from the clear, close to periodic transition patterns.

\begin{knitrout}
\definecolor{shadecolor}{rgb}{0.969, 0.969, 0.969}\color{fgcolor}\begin{figure}[!htb]

{\centering \includegraphics[width=\maxwidth]{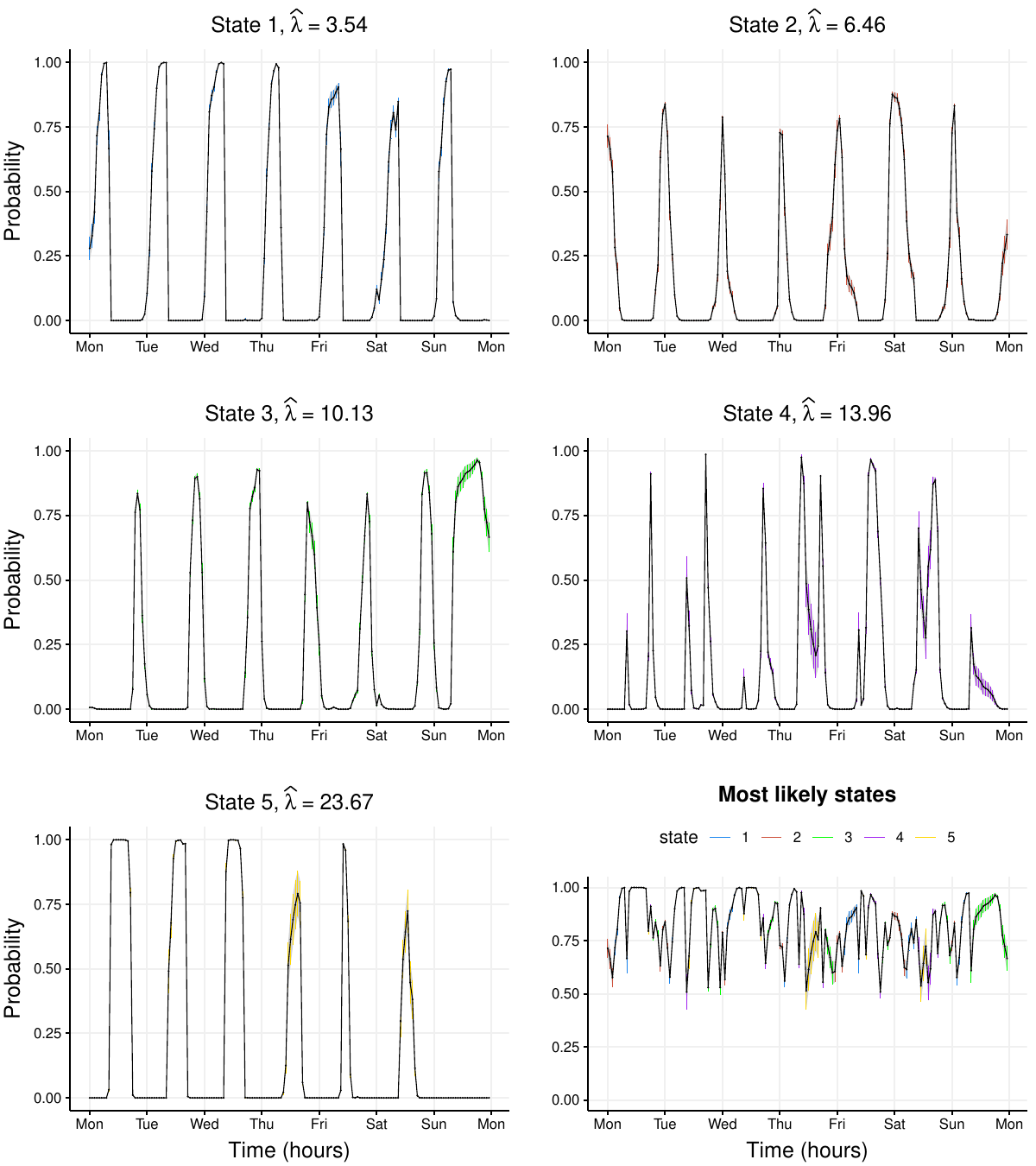} 

}

\caption{Smoothing probabilities and confidence intervals of a five-state Gaussian HMM fitted to the hospital data set. The solid line shows the smoothing probability estimates and the 95\% CIs are represented by vertical lines. The bottom right graph displays smoothing probabilities for the most likely state estimated for each data point. The vertical confidence interval lines are colored differently per hidden state. For readability, only the first week starting on Monday (from Monday January $4^{th}$ 2010 at 00:00 to Sunday January $10^{th}$ 2010 at 23:59) is plotted instead of the entire 10 years. Further details on the model estimates are available in \autoref{tab:estimates-cis-aic-bic-real-data-part2}.}\label{fig:smoothing-hosp}
\end{figure}

\end{knitrout}

%%%%%%%%%%%%%%%%%%%%%%%%%%%%%%%%%%%%%%%%%%%%%%%%%%%%%%%%%%%%%%%%%
\section{Performance and accuracy of different optimizers}
\label{sec:optimization}
%%%%%%%%%%%%%%%%%%%%%%%%%%%%%%%%%%%%%%%%%%%%%%%%%%%%%%%%%%%%%%%%%

This section compares the speed and accuracy of different optimizers in R using {\tt{TMB}} in a HMM estimation setting with DNM by several simulation studies with different settings.

In the first setting, we simulate time series consisting of 87 observations from a two-state Poisson HMM fitted to the TYT data, as we want to examine the different optimizers performance on data sets with relatively few observations.
In the second setting, we simulate time series of length 200 from a two-state Poisson HMM, and in the third setting, time series of length 200 from a two-state Gaussian HMM are simulated.
The specific choice of parameters is described in more detail in the forthcoming sections. 

%An additional design is available in Appendix \ref{sec:appendix-perf}, where simulated data sets of length DATA_SIZE_SIMU2 from a Poisson HMM, with parameters chosen closer together, to make estimation less straightforward, and to examine the impact of increased sample size. \textcolor{red}{BS: we can remove this design - as it basically do not add much to the findings we already have - moved out}

The comparisons that focus on computational speed and iterations of the different optimizers, are based on 200 replications from the different models, while the studies focusing on the accuracy of the optimizers are based on 1000 replications from the different models. 
Hence, the Monte-Carlo simulation setup in this paper closely resembles the setup of \citet{bacri}.
%\newline
%\newline
%\textcolor{red}{TB: Both. The speeds and iterations are compared on 200 samples, and the estimates are compared on 1000 samples}

%%%%%%%%%%%%%%%%%%%%%%%%%%%%%%%%%%%%%%%%%%%%%%%%%%
\subsection{Computational setup}
%%%%%%%%%%%%%%%%%%%%%%%%%%%%%%%%%%%%%%%%%%%%%%%%%%
%\textcolor{red}{BS: Should we maybe move this before the smoothing probs section - or maybe it is ok to leave the section here?}

The results presented in this paper required installing and using the {\tt{R}}-package {\tt{TMB}} and the software {\tt{Rtools}}, where the latter is needed to compile {\tt{C++}} code.
Scripts were coded in the interpreted programming language {\tt{R}} \citep{rcoreteam} using its eponymous software environment {\tt{RStudio}}.
For the purpose of making our results reproducible, the generation of random numbers is handled with the function \texttt{set.seed} under {\tt{R}} version number 3.6.0.
A seed is set multiple times in our scripts, ensuring that smaller parts can be easily duplicated without executing lengthy prior code.
A workstation with 4 Intel(R) Xeon(R) Gold 6134 processors (3.7 GHz) running under the Linux distribution Ubuntu 18.04.6 LTS (Bionic Beaver) with 384 GB RAM was used to execute our scripts. 

%%%%%%%%%%%%%%%%%%%%%%%%%%%%%%%%%%%%%%%%%%%%%%%%%%
\subsection{Selected optimizers and additional study design}
\label{sec:design}
%%%%%%%%%%%%%%%%%%%%%%%%%%%%%%%%%%%%%%%%%%%%%%%%%%

%\textcolor{red}{
%\textbf{\textit{BÅRD TEXT: Of course, here you need to expand quite a bit more. Explaining the simulation setup (processes etc.) (not enough to refer to paper 1), results etc.}}}

%{\hl BS: would it be better just to state that we selected the following optimizers based on their use and by some pre-tests of ours?}
%\newline
%{\hl TB: If we do that, wouldn't it open us to being asked by the reviewer if we can provide some evidence that they are used enough? I agree that your way reads better though.}

There exist a large number of optimizers for use within the {\tt{R}} ecosystem, see e.g. \citet{schwendinger}.
In a preliminary phase, multiple optimization routines were attempted, and the ones failing to converge on a few selected test data sets from the designs mentioned above were excluded (details available from the author upon request).
Hence, the following optimizers were selected for the simulation studies:

\begin{itemize}
\item \texttt{optim} from the {\tt{stats}} package to test the BFGS \citep{broyden, fletchera, goldfarb, shanno}, L-BFGS-B \citep{byrd}, CG \citep{fletcher}, and Nelder-Mead \citep{nelder} algorithms. 
\item \texttt{nlm} \citep{dennisa} and \texttt{nlminb} \citep{gay} from the same package to test popular non-linear unconstrained optimization algorithms.
\item \texttt{hjn} from the {\tt{optimr}} package \citep{R-optimr} to test the so-called Hooke and Jeeves Pattern Search Optimization \citep{hooke}.
\item \texttt{marqLevAlg} from the {\tt{marqLevAlg}} package \citep{R-marqLevAlg} to test a general-purpose optimization based on the Marquardt-Levenberg algorithm \citep{levenberg, marquardt}.
\item \texttt{ucminf} from the {\tt{ucminf}} package \citep{R-ucminf} to test a general-purpose optimization based on the Marquardt-Levenberg algorithm \citep{nielsen}.
\item \texttt{BBoptim} from the {\tt{BB}} package \citep{R-BB} to test a spectral projected gradient method for large scale optimization with simple constraints \citep{varadhan}.
\item \texttt{newuoa} from the {\tt{minqa}} package \citep{R-minqa} to test a large-scale nonlinear optimization algorithm \citet{powell}.
\end{itemize}

%A similar Monte-Carlo simulation strategy to \citet{bacri} was used to generate the simulated data sets and benchmark samples.

By also providing the Hessian and/or gradient to the optimizers that support such inputs, 22 optimization procedures are examined for each study design.
For the study of the computational speed and number of iterations, we applied the  {\tt{microbenchmark}} package \citep{R-microbenchmark} that allows us to obtain precise durations for each of the simulated data sets. Furthermore, HMM ML estimates were reordered by increasing Poisson rates or Gaussian means to avoid a random order of the states.
% Note that \citet{bacri} points out that HMM ML estimation with {\tt{TMB}} has a duration that varies little when repeated exactly.
% In other words, the normal background tasks of the computer do not noticeably impact speeds.

%Similarly, a bootstrap was executed on some of the data sets.
%More precisely, we used the ML estimates to generate sequences of states and random data from the states' respective conditional distributions.
%Then, the models of interest were re-estimated on each of the sampled data sets.

% Alternatively, an order could have been set from the start by estimating non-negative increments $\lambda_j - \lambda_{j-1}$ with $\lambda_0 \equiv 0$, as explained by \cite[Section 7.1.1 p.112]{zucchini} and similarly in the Gaussian case.
% But as \cite[Section 3.2 p.7]{bulla} points out, this can impose optimization issues: "over all series, the simplest parameterization, i.e., the use of log-transformed state-dependent parameters, leads to the best results as regards the number of failures and the convergence to the global maximum".
% Therefore, this alternative was not pursued.

It is important to note that we discarded all those samples for which the simulated state sequence did not sojourn at least once in each state.
This was to avoid identifiability and convergence issues when trying to estimate a $m$-state HMM on a data set where only $m - 1$ states are visited.
Further, samples where at least one optimizer failed to converge were also discarded to ensure comparability.
% The reason is that optimizers do not typically behave in the same way when failing to converge.
% Some may reach their limit of iterations, whereas others may have a tolerance too small.
% This failure can be caused by multiple reasons, many of which are not common with all optimizers in the list.
% Timing convergence failures would make durations impossible to meaningfully compare.
For the same reason, we set the maximum number of iterations to \ensuremath{10^{4}} whenever possible.
Note that some optimizers (\texttt{BFGS, L-BFGS-B, CG, Nelder-Mead, hjn, newuoa}) do not report a number of iterations and are therefore missing from our comparisons of iterations.
Finally, we use the true values as initial parameters.
For the TYT data, initial values are calculated as medians from the fitted models resulting from all optimizers listed above.

%(among all optimizers listed above) of estimates based on the TYT data are used for the first design. 

%On the real data set, optimizers seem to agree on one or rarely two possible values for each estimate (with negligible differences).
%In the latter case, an often large majority prefers one value.
%We prefer the estimate picked by majority and select the third quartile.
%The mode would be preferred but is unavailable due to small variations between estimates, likely due to computer approximations.

%%%%%%%%%%%%%%%%%%%%%%%%%%%%%%%%%%%%%%%%%%%%%%%%%%
\subsection{Results}
\label{results}
%%%%%%%%%%%%%%%%%%%%%%%%%%%%%%%%%%%%%%%%%%%%%%%%%%

%This section intends to compare the performance of various optimizer routines by using a couple of example data sets.
%They aim to be diverse in the number of observations and model complexity.
%These data sets include one of tinnitus ``arousal'', one of hospital patient arrivals, and simulated ones, as 
%described in \cref{sec:data_sets}.

In this section, we report the results from the simulation studies focusing on the performance of different optimizers in terms of the computational speed and the lack or presence of unreasonable estimates.
This section uses {\tt{R}} scripts that may be of interest to users investigating their own HMM settings, and are available in the supporting information.
We begin by reporting the speed of all optimization routines and thereafter the accuracy. 
The routine names contain the optimizer names and may be followed by either ``\_gr'' to indicate that the routine uses {\tt{TMB}}'s provided gradient function, ``\_he'' to indicate the use of {\tt{TMB}}'s provided Hessian function, or ``\_grhe'' to indicate the use of both gradient and Hessian functions provided by {\tt{TMB}}.
Note that some optimizers cannot make use of these functions and hence bear no extension at the end of their names.

%The same infrastructure and setup as in \citet{bacri} is used, therefore a preliminary check of its reliability is unnecessary.

%%%%%%%%%%%%%%%%%%%%%%%%%%%%%%%%%%%%%%%%%%%%%%%%%%
\subsubsection{Simulation study: TYT setting}
\label{sec:results-tinn}
%%%%%%%%%%%%%%%%%%%%%%%%%%%%%%%%%%%%%%%%%%%%%%%%%%

\begin{knitrout}
\definecolor{shadecolor}{rgb}{0.969, 0.969, 0.969}\color{fgcolor}\begin{figure}[htb]

{\centering \includegraphics[width=\maxwidth]{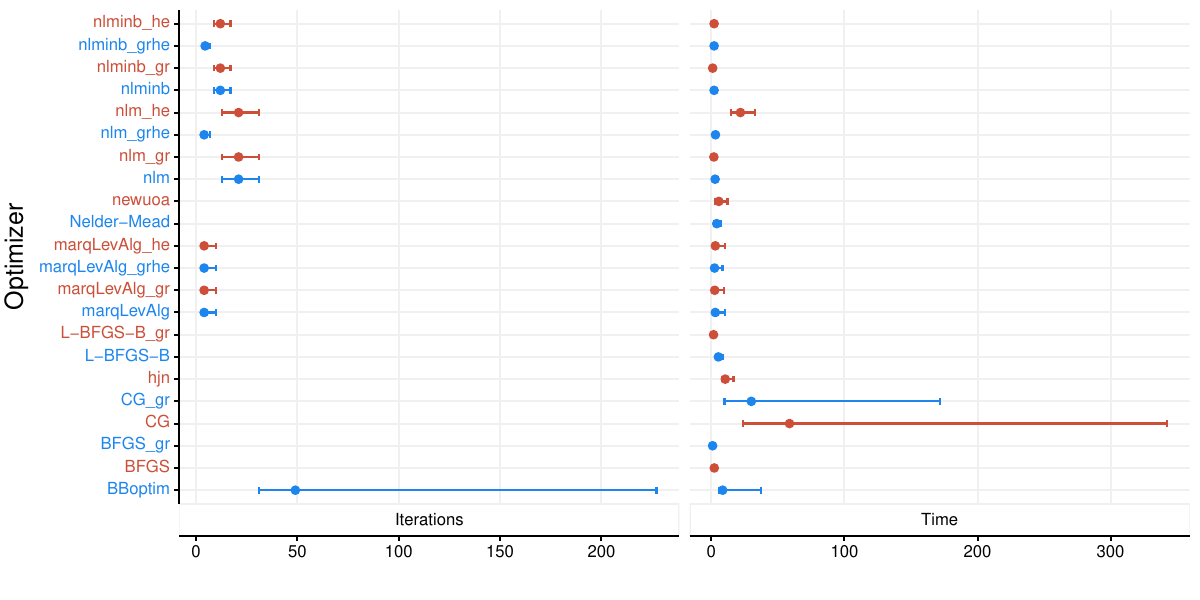} 

}

\caption[Median duration (in milliseconds) and median number of iterations together with 2.5\%- and 97.5\%-quantile when fitting two-state Poisson HMMs to 200 replications in the first setting]{Median duration (in milliseconds) and median number of iterations together with 2.5\%- and 97.5\%-quantile when fitting two-state Poisson HMMs to 200 replications in the first setting.}\label{fig:optimizers-acceleration-tinn-plot}
\end{figure}

\end{knitrout}

\autoref{fig:optimizers-acceleration-tinn-plot} shows the time required by each optimization routine, measured over the 200 replication. The number of iterations required by each optimization routine is also reported where available. Times range from 1 to 1340 milliseconds.
As shown by the large medians and wide CIs, \texttt{CG} and \texttt{CG\_gr} require substantially more time to estimate a HMM, compared to the other optimizers.
Furthermore, optimizers that take a long time to estimate HMMs require a high amount of iterations as well, as expected.
Notably, optimization routines almost always take longer when the Hessian is passed.
This likely results from a higher computational burden due to handling and evaluating Hessian matrices via {\tt{TMB}}.

Looking at the details, the optimizer \texttt{BFGS\_gr} is among the fastest optimizers, and this is partly explained by a relatively low number of iterations, and by the fact that it does not make use of the Hessian. 
However, we note that \texttt{nlm} and \texttt{nlmimb} families of optimizers are slower when the gradient and Hessian are not passed as argument.
The optimizer \texttt{BBoptim} is comparatively slow, and \texttt{nlminb\_grhe} fails to converge comparably often.

In addition to time durations, we investigate the accuracy of the optimizers, based on another 1000 replications from the two-state Poisson HMM in the first setting.
The main conclusion from this simulation is that the medians and empirical quantiles calculated  are in fact almost identical across all optimizers, and very close to the true parameter values used in the simulations. Results are reported in \autoref{fig:bootstrap-graph-tinn} in \autoref{sec:appendix-perf}.
These result imply that the different optimizers are equally accurate. Note, however, that the variation is quite high for some of the estimated parameters and for the nll, which results most likely from the limited number of observations.
   
%will all provide almost identical estimation results for the HMM.

%As shown by their broad CIs, nlls are fairly spread, with values ranging from a minimum of round(min(subset(bootstrap_tinn_results_with_nll, subset = Parameter == "nll", select = -Parameter))) to a maximum of round(max(subset(bootstrap_tinn_results_with_nll, subset = Parameter == "nll", select = -Parameter))).
%This is caused by some optimizers sometimes finding very different estimates on the same data, resulting in large differences in nll. It also explains the broad CIs on some estimates. Furthermore, the bootstrap estimates shown graphically in \autoref{fig:bootstrap-graph-tinn} shows little variations between nlls and parameter estimate medians and CIs accross the different optimizers.

%As with the smoothing probabilities, the width of the TPM's CIs grows with the distance to the boundaries zero and one.
%An intuitive interpretation is that their uncertainty should be low when the corresponding event's probability is near one or zero (in other words when the event is certain to happen or not), and high when its probability is far from zero or one (in other words when it is unsure whether the event happens or not).

%%%%%%%%%%%%%%%%%%%%%%%%%%%%%%%%%%%%%%%%%%%%%%%%%%
\subsubsection{Simulation study: two-state Poisson HMM}
\label{sec:results-simu1}
%%%%%%%%%%%%%%%%%%%%%%%%%%%%%%%%%%%%%%%%%%%%%%%%%%

\noindent In this second simulation setting, the parameters of our two-state Poisson HMM are given by
\begin{equation*}
\bm{\lambda} = (1, 7), \quad
\bm{\Gamma} = \begin{pmatrix} 0.95 & 0.05 \\ 0.15 & 0.85 \end{pmatrix},
\end{equation*}
and the sample size is fixed at 200.
 \autoref{fig:optimizers-acceleration-simu1-plot} illustrates our results, which are in line with those obtained in the first setting. More precisely, \texttt{BBoptim} still dominates the number of iterations, and \texttt{CG} and \texttt{CG\_gr} require more computational time than the others.
Overall, most optimizers need more time compared to the previous setting as there is more data to process, but the time increase for \texttt{nlm\_he} and \texttt{hjn} is much more substantial than for other optimizers.
%Surprisingly, the change in the number of iterations does not follow the same principle.
Moreover, although the data set is larger, some optimizers (e.g. \texttt{BBoptim} and \texttt{BFGS\_gr}) perform as fast as or faster than in the previous setting.
Furthermore, passing {\tt{TMB}}'s Hessian to optimizers does not always slow down the optimization. This can be seen, e.g., from \texttt{nlminb\_he} which exhibits a lower median time than \texttt{nlminb}. A last notable pattern concerns the popular \texttt{nlm} optimizer. When only the Hessian but not the gradient is supplied from \texttt{TMB}, the computational time increases substantially compared to \texttt{nlm\_gr} and \texttt{nlm\_grhe}. Interestingly, the equally popular optimizer \texttt{nlminb} is not affected in the same way. Thus, one should be careful when only supplying the Hessian from \texttt{TMB} while an optimizer carries out the gradient approximation. In any case, the low number of iterations required by both \texttt{nlminb\_grhe} and \texttt{nlm\_grhe} indicate a preferable convergence behavior when supplying both quantities.

In terms of accuracy of the parameter estimates and obtained nlls, our results are highly similar to those described in the first setting (for details, see \autoref{fig:bootstrap-graph-simu1} in \autoref{sec:appendix-perf}).

\begin{knitrout}
\definecolor{shadecolor}{rgb}{0.969, 0.969, 0.969}\color{fgcolor}\begin{figure}[htb]

{\centering \includegraphics[width=\maxwidth]{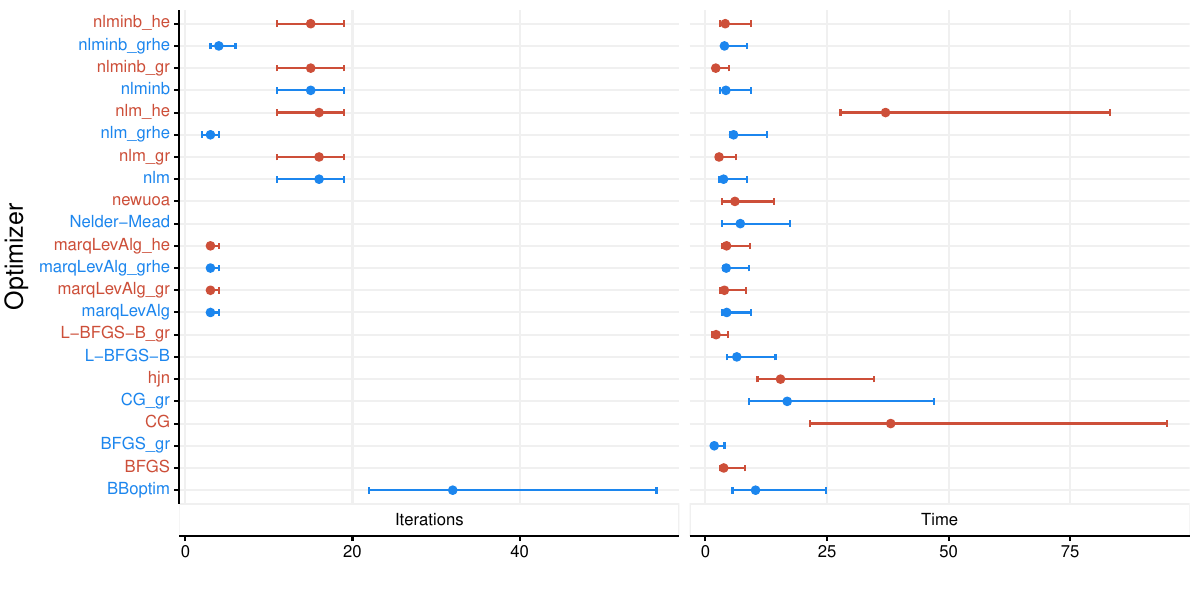} 

}

\caption[Median duration (in milliseconds) and median number of iterations together with 2.5\%- and 97.5\%-quantile when fitting two-state Poisson HMMs to 200 replications in the second setting]{Median duration (in milliseconds) and median number of iterations together with 2.5\%- and 97.5\%-quantile when fitting two-state Poisson HMMs to 200 replications in the second setting.}\label{fig:optimizers-acceleration-simu1-plot}
\end{figure}

\end{knitrout}

%\restoregeometry

%%%%%%%%%%%%%%%%%%%%%%%%%%%%%%%%%%%%%%%%%%%%%%%%%%
\subsubsection{Simulation study: two-state Gaussian HMM} %3 data}
\label{results-simu3}
%%%%%%%%%%%%%%%%%%%%%%%%%%%%%%%%%%%%%%%%%%%%%%%%%%

\noindent In our third and last simulation setting, the parameters of a two-state Gaussian HMM are given by
\begin{equation*}
\bm{\Gamma} = \begin{pmatrix} 0.95 & 0.05 \\ 0.15 & 0.85 \end{pmatrix}, \quad
\bm{\mu} = (-5, 5), \quad
\bm{\sigma} = (1, 5), 
\end{equation*}
and the sample size is fixed at 200. In terms of computational time and the number of iterations, the results observed for this setup are similar to the previously studied Poisson HMMs (see \autoref{fig:optimizers-acceleration-simu3-plot}). A minor aspect is an increase in both iterations and computational time, which is not surprising given the higher number of parameters.

Regarding estimation accuracy of the parameter estimates and obtained nlls, our results are again highly similar to those obtained in the previous two settings (illustrated graphically by \autoref{fig:bootstrap-graph-simu3} in \autoref{sec:appendix-perf}). 
  
\begin{knitrout}
\definecolor{shadecolor}{rgb}{0.969, 0.969, 0.969}\color{fgcolor}\begin{figure}[htb]

{\centering \includegraphics[width=\maxwidth]{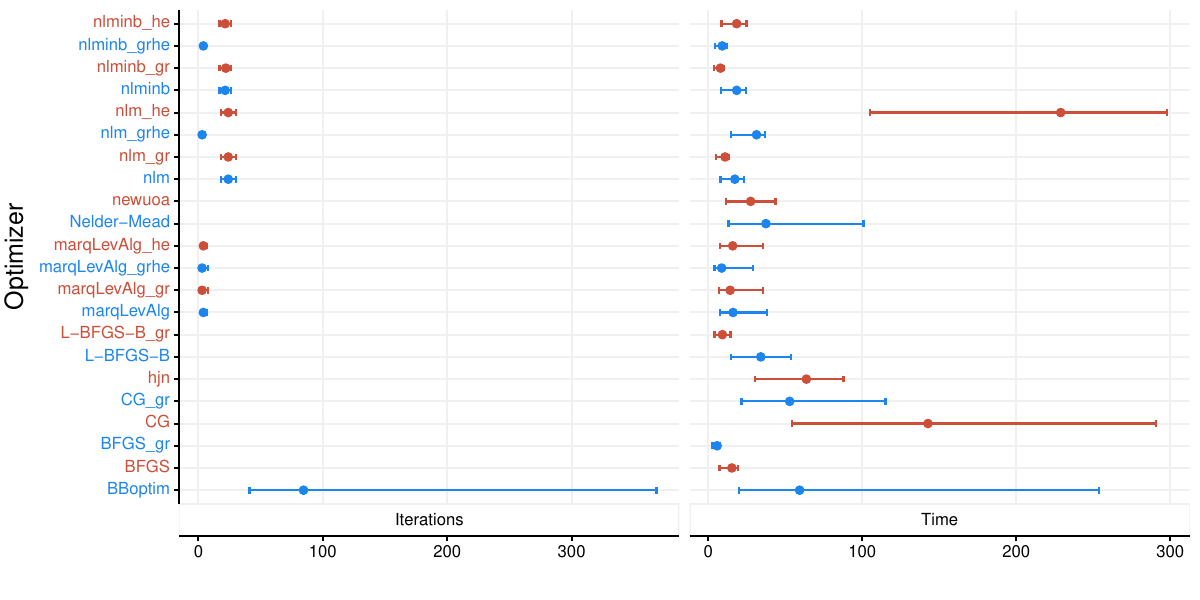} 

}

\caption[Median duration (in milliseconds) and median number of iterations together with 2.5\%- and 97.5\%-quantile when fitting two-state Gaussian HMMs to 200 replications in the third setting]{Median duration (in milliseconds) and median number of iterations together with 2.5\%- and 97.5\%-quantile when fitting two-state Gaussian HMMs to 200 replications in the third setting.}\label{fig:optimizers-acceleration-simu3-plot}
\end{figure}

\end{knitrout}

%\restoregeometry

%%%%%%%%%%%%%%%%%%%%%%%%%%%%%%%%%%%%%%%%%%%%%%%%%%%%%%%%%%%%%%%%%
\section{Robustness to initial value selection}
\label{sec:robustness}
%%%%%%%%%%%%%%%%%%%%%%%%%%%%%%%%%%%%%%%%%%%%%%%%%%%%%%%%%%%%%%%%%

In this section, we examine the impact of initial values of the parameters - including poorly selected ones - on the convergence behavior and accuracy of different optimizers. We investigate one real and several simulated data sets. In order to ``challenge'' the optimizers, all data sets are of comparable small size.

%%%%%%%%%%%%%%%%%%%%%%%%%%%%%%%%%%%%%%%%%%%%%%%%%%
\subsection{Study design}
\label{sec:robustness-design}
%%%%%%%%%%%%%%%%%%%%%%%%%%%%%%%%%%%%%%%%%%%%%%%%%%

We again consider three settings in the following. In the first, we investigate the TYT data set. In the second and third setting, we focus on simulated data from Poisson and Gaussian HMMs, respectively. More specifically, for the Poisson HMMs we generate one data set of sample size 200 for each of the HMMs considered. These HMMs are defined by all possible combinations of the following Poisson rates and TPMs:
\begin{align*}
& \bm{\lambda} \in \left\{ (1, 4), (5, 7) \right\} \text{and}\\
& \bm{\Gamma} \in \left\{ \begin{pmatrix} 0.9 & 0.1 \\ 0.2 & 0.8 \end{pmatrix}, \begin{pmatrix} 0.7 & 0.3 \\ 0.8 & 0.2 \end{pmatrix}, \begin{pmatrix} 0.1 & 0.9 \\ 0.8 & 0.2 \end{pmatrix}, \begin{pmatrix} 0.55 & 0.45 \\ 0.45 & 0.55 \end{pmatrix} \right\}.
\end{align*}
%For example, the second data set in this first simulation setting is generated by a two-state HMM based on the Poisson rates $\bm{\lambda} = (1,4)$ and the TPM $\bm{\Gamma} = matrix.to.LaTeX(true_gamma_2_simu1)$.
This setup leads to the generation of 2 x 4 data sets. For the Gaussian HMMs, we follow the same approach, %$200$
where the Gaussian means and standard deviations are selected from the sets
\begin{align*}
\bm{\mu} &\in \left\{ (-2, 2), (-1, 4) \right\} \text{and}\\
\bm{\sigma} &\in \left\{ (1.5, 2.5), (0.5, 1.5) \right\}.
\end{align*}
This setting thus generates 2 x 2 x 4 data sets.

For each data set, we pass a large range of initial values to the same optimizers considered in the previous section. In this way, we investigate the resistance of each optimizer to potentially poor initial values when fitting HMMs.
To generate sets of initial values, we consider the following potential candidates for Poisson rates $\bm{\lambda}$, Gaussian means $\bm{\mu}$, Gaussian standard deviations $\bm{\sigma}$, and TPMs $\bm{\Gamma}$: 
\begin{align*}
\lambda_1, \lambda_2 \in & \left\{ M, M + 0.5, M + 1, 1.5, \ldots, x_{max} \right\} ,\text{ where } \lambda_1 < \lambda_2,\\
\mu_1, \mu_2 \in & \left\{ x_{min}, x_{min} + 0.5, x_{min} + 1, \ldots, x_{max} - 0.5, x_{max} \right\}, \text{ where } \mu_1 < \mu_2,\\
\sigma_1, \sigma_2 \in & \left\{ 10 \text{ equidistant points going from } \sqrt{\frac{(x_{max} - x_{min})^2}{2T}} \text{ to } \sqrt{(x_{max} - \widehat{\mu})(\widehat{\mu} - x_{min})} \right\},\\
\text{and } \bm{\Gamma} \in
& \biggl\{
  \begin{pmatrix}
    0.1 & 0.9\\
    0.9 & 0.1
  \end{pmatrix},
  \begin{pmatrix}
    0.1 & 0.9\\
    0.8 & 0.2
  \end{pmatrix},
  \ldots,
  \begin{pmatrix}
    0.1 & 0.9\\
    0.1 & 0.9
  \end{pmatrix},
\\
& \begin{pmatrix}
    0.2 & 0.8\\
    0.9 & 0.1
  \end{pmatrix},
  \begin{pmatrix}
    0.2 & 0.8\\
    0.8 & 0.2
  \end{pmatrix},
  \ldots,
  \begin{pmatrix}
    0.2 & 0.8\\
    0.1 & 0.9
  \end{pmatrix},
\\
& \qquad \qquad \qquad \qquad \qquad \qquad, \ldots
  \begin{pmatrix}
    0.9 & 0.1\\
    0.1 & 0.9
  \end{pmatrix}
\biggl\},
\end{align*}
where $x_{min} = \min(x_1, x_2, \ldots, x_T)$, $x_{max} = \max(x_1, x_2, \ldots, x_T)$, $\widehat{\mu} = \frac{1}{T}$ $\sum_{i=1}^T x_i$ and $M = \max(0.5, x_{min})$. The motivation for this selection is as follows. First, Poisson means have to be greater than zero, so we set their lower boundary to $M$. Secondly, the $\widehat{\lambda}_i$'s have to belong to the interval $(x_{min}, x_{max})$, cfr. \citep{bohning}. This applies to the Gaussian means as well. Thirdly, the upper and lower limit of ${\sigma}_1, {\sigma}_2 $ is motivated by the Bhatia-Davis \citep{bhatia} and the von Szokefalvi Nagy inequalities \citep{nagy}.
%The simulation respects the assumption of independence required by the latter inequality.

%Also note that we impose $\lambda_1 < \lambda_2$. By allowing all possibilities, many settings would be equivalent and lead to the same model being estimated, the only difference being the order of the states. The inequality constraint prevents this situation. With more than two states, the constraint would become $\lambda_1 < \lambda_2 < \ldots < \lambda_m$. For the same reason, we impose $\mu_1 < \mu_2$.

For the Poisson HMMs fitted, we consider all possible combination of the parameters described above. For the Gaussian HMMs, however, we sample 12500 initial parameters for each of the 16   data sets to reduce computation time (the total amount reduces from 39066300 to 200000). As in the previous \autoref{sec:optimization}, we limit the maximal number of iterations to \ensuremath{10^{4}} whenever possible. Other settings were left as default. 

To evaluate the performance of the optimizers considered, we rely on two criteria. First, we register whether the optimizer converges (all errors are registered regardless of their type). Secondly, when an optimizer converges, we also register if the nll is equal to the "true" nll (with a gentle margin of $\pm$ 5\%). We calculate this "true" nll as the median of all nlls derived with all optimizers initiated with the true parameter values. 

% For the TYT data, we use all available sets of initial values, for the simulated data sets we starting the optimization from the true initial values only.
%This design yields tens of millions sets of initial parameters, across all study designs, from which we can infer convergence failure rates for different optimizers and the rate at which they find the correct global minimum nll.

%%%%%%%%%%%%%%%%%%%%%%%%%%%%%%%%%%%%%%%%%%%%%%%%%%
\subsection{Results}
%%%%%%%%%%%%%%%%%%%%%%%%%%%%%%%%%%%%%%%%%%%%%%%%%%

\autoref{tab:performance-rates-tinn} shows the results for the two-state Poisson HMMs fitted to the TYT data. The four \texttt{marqLevAlg} follow closely. We note that \texttt{hjn} has the best performance: it basically does not fail and converges almost always. Moreover, \texttt{nlminb\_grhe} fails to converge much more often than other optimizers, but is the most likely to find the global maximum when it converges.
For the other optimizers, the failure rate is generally low (less than 5\%) and the global maximum is found in more than 65\% of the cases.

% latex table generated in R 4.2.2 by xtable 1.8-4 package
% Mon Jan 30 17:38:27 2023
\begin{table}[ht]
\centering
\caption{Performance of multiple optimizers estimating Poisson HMMs from the TYT data (first setting) over 7371 different sets of initial parameters. The first column lists all optimizers. The second column shows how often optimizers fail to converge successfully. The third column displays how often optimizers successfully found the global maximum of the nll when converging.} 
\label{tab:performance-rates-tinn}
\begin{tabular}{rrr}
  \hline
 & Failures $(\%)$ & Global maximum found $(\%)$ \\ 
  \hline
BFGS & 0.00 & 66.83 \\ 
  BFGS\_gr & 0.00 & 66.83 \\ 
  L-BFGS-B & 3.43 & 71.38 \\ 
  L-BFGS-B\_gr & 3.46 & 71.22 \\ 
  CG & 4.90 & 71.48 \\ 
  CG\_gr & 4.91 & 71.49 \\ 
  Nelder-Mead & 0.00 & 69.88 \\ 
  nlm & 0.00 & 65.19 \\ 
  nlm\_gr & 0.00 & 65.22 \\ 
  nlm\_he & 0.00 & 65.19 \\ 
  nlm\_grhe & 0.00 & 72.99 \\ 
  nlminb & 0.00 & 74.96 \\ 
  nlminb\_gr & 0.00 & 74.94 \\ 
  nlminb\_he & 0.00 & 74.96 \\ 
  nlminb\_grhe & 24.41 & 99.96 \\ 
  hjn & 0.00 & 97.40 \\ 
  marqLevAlg & 1.64 & 88.43 \\ 
  marqLevAlg\_gr & 1.99 & 88.64 \\ 
  marqLevAlg\_he & 1.64 & 88.43 \\ 
  marqLevAlg\_grhe & 1.94 & 88.60 \\ 
  newuoa & 0.00 & 71.67 \\ 
  BBoptim & 0.00 & 72.11 \\ 
   \hline
\end{tabular}
\end{table}

\autoref{tab:performance-rates-simu1} reports the results from the second setting, also with two-state Poisson HMMs. The results show that all optimizers find the global maximum in the majority of cases ($> 96\%$) when converging. Moreover, the failure rates are relatively low for all optimizers with the exception of \texttt{CG}, which stands out with comparably high failure rates. %Furthermore, we note that the \texttt{nlminb}, with either the gradient or Hessian passed, and \texttt{hjn} have both excellent performances.  

Finally, \autoref{tab:performance-rates-simu3} presents the results from the third setting with Gaussian HMMs. The results regarding the failure rates point in the same direction as those from the Poisson HMM setting: \texttt{CG} attains the highes failure rate, followed by \texttt{newuoa} and \texttt{L-BFGS-B}. The remaining optimizers perform satisfactorily. Note, however, that \texttt{nlminb} does not benefit from being supplied with the gradient and Hessian from \texttt{TMB}. Regarding the global maximum found, many algorithms reach success rates of about 95\% or more. Notable exceptions are \texttt{Nelder-Mead} and \texttt{nlm} when not both gradient and Hessian are provided by \texttt{TMB}. As observed previously for the TYT data, \texttt{nlm} benefits strongly from the full use of \texttt{TMB}.

% latex table generated in R 4.2.2 by xtable 1.8-4 package
% Mon Jan 30 17:38:27 2023
\begin{table}[ht]
\centering
\caption{Performance of multiple optimizers estimating Poisson HMMs from the second setting over 194481 different sets of initial parameters. The first column lists all optimizers. The second column shows how often optimizers fail to converge successfully. The third column displays how often optimizers successfully found the global maximum of the nll when converging.} 
\label{tab:performance-rates-simu1}
\begin{tabular}{rrr}
  \hline
 & Failures $(\%)$ & Global maximum found $(\%)$ \\ 
  \hline
BFGS & 0.00 & 96.68 \\ 
  BFGS\_gr & 0.00 & 96.68 \\ 
  L-BFGS-B & 0.30 & 100.00 \\ 
  L-BFGS-B\_gr & 0.29 & 100.00 \\ 
  CG & 14.23 & 99.49 \\ 
  CG\_gr & 14.47 & 99.53 \\ 
  Nelder-Mead & 0.00 & 99.91 \\ 
  nlm & 0.07 & 98.42 \\ 
  nlm\_gr & 0.02 & 98.42 \\ 
  nlm\_he & 0.07 & 98.42 \\ 
  nlm\_grhe & 0.00 & 99.50 \\ 
  nlminb & 0.00 & 100.00 \\ 
  nlminb\_gr & 0.00 & 100.00 \\ 
  nlminb\_he & 0.00 & 100.00 \\ 
  nlminb\_grhe & 4.62 & 100.00 \\ 
  hjn & 0.00 & 99.99 \\ 
  marqLevAlg & 3.19 & 99.98 \\ 
  marqLevAlg\_gr & 3.20 & 100.00 \\ 
  marqLevAlg\_he & 3.19 & 99.98 \\ 
  marqLevAlg\_grhe & 3.20 & 100.00 \\ 
  newuoa & 0.34 & 99.77 \\ 
  BBoptim & 4.20 & 100.00 \\ 
   \hline
\end{tabular}
\end{table}

% latex table generated in R 4.2.2 by xtable 1.8-4 package
% Mon Jan 30 17:38:29 2023
\begin{table}[ht]
\centering
\caption{Performance of multiple optimizers estimating Gaussian HMMs from the third setting over 200000 different sets of initial parameters randomly drawn from 39066300 different candidate sets of initial values. The first column lists all optimizers. The second column shows how often optimizers fail to converge successfully. The third column displays how often optimizers successfully found the global maximum of the nll when converging.} 
\label{tab:performance-rates-simu3}
\begin{tabular}{rrr}
  \hline
 & Failures $(\%)$ & Global maximum found $(\%)$ \\ 
  \hline
BFGS & 0.00 & 84.24 \\ 
  BFGS\_gr & 0.00 & 84.24 \\ 
  L-BFGS-B & 6.58 & 94.40 \\ 
  L-BFGS-B\_gr & 6.75 & 94.45 \\ 
  CG & 8.62 & 98.68 \\ 
  CG\_gr & 8.82 & 98.81 \\ 
  Nelder-Mead & 0.03 & 80.55 \\ 
  nlm & 0.39 & 87.77 \\ 
  nlm\_gr & 0.11 & 87.73 \\ 
  nlm\_he & 0.39 & 87.77 \\ 
  nlm\_grhe & 0.32 & 97.21 \\ 
  nlminb & 0.22 & 97.18 \\ 
  nlminb\_gr & 0.26 & 97.20 \\ 
  nlminb\_he & 0.22 & 97.18 \\ 
  nlminb\_grhe & 5.69 & 97.97 \\ 
  hjn & 0.53 & 95.21 \\ 
  marqLevAlg & 2.55 & 98.18 \\ 
  marqLevAlg\_gr & 2.66 & 98.34 \\ 
  marqLevAlg\_he & 2.55 & 98.18 \\ 
  marqLevAlg\_grhe & 2.64 & 98.33 \\ 
  newuoa & 7.58 & 97.44 \\ 
  BBoptim & 1.13 & 98.61 \\ 
   \hline
\end{tabular}
\end{table}

\clearpage

%%%%%%%%%%%%%%%%%%%%%%%%%%%%%%%%%%%%%%%%%%%%%%%%%%%%%%%%%%%%%
\subsection{Hybrid algorithm}
%%%%%%%%%%%%%%%%%%%%%%%%%%%%%%%%%%%%%%%%%%%%%%%%%%%%%%%%%%%%%

Last, we ran a small simulation study investigating a hybrid algorithm within the {\tt{TMB}} framework in the spirit of \citet{bulla}. The hybrid algorithm starts with the \texttt{Nelder-Mead} optimizer and switches to \texttt{nlminb} after a certain number of iterations. The idea is to benefit from both \texttt{Nelder-Mead}'s rapid departure from poor initial values  and from the high convergence speed of \texttt{nlminb}. Our study is performed on 1000 random sets of initial values in a similar fashion to \autoref{sec:robustness}, using the weekly returns data set. For every set of initial values, we sequentially increase the number of iterations carried out by \texttt{Nelder-Mead} by ten (starting at one) until convergence or until $10 000$ iterations has been reached, in which case we classify the attempt as a failure. For comparison, we also run an optimization only with \texttt{nlminb} starting directly with the same sets of inital values. In both cases, \texttt{nlminb} benefits from the gradient and Hessian provided by {\tt{TMB}}. 

 \autoref{fig:Nelder-Mead-convergence-per-iteration} reports the results of our study. The left most columns show that only one iteration was required to converge in 827 of the 1000 sets of initial values with the hybrid algorithm, while \texttt{nlminb} converges in only 698 out of these 827 sets. Furthermore, for 39 additional sets of initial values, the hybrid algorithm converges with 10 iterations carried out by the \texttt{Nelder-Mead} optimizer, whereas \texttt{nlminb} fails 26 times on these sets. Increasing the number of initial iterations carried out by \texttt{Nelder-Mead} for the hybrid algorithm leads to similar patterns. Finally, both optimization routines fail to converge for 18 sets of initial values. Thus, in total, the hybrid algorithm with one or more \texttt{Nelder-Mead} iterations failed in only $1.8 \%$ of the cases. In comparison, direct use of \texttt{nlminb} failed in around $12 \%$ of the cases. This indicates that the hybrid algorithm requires rather few iterations of \texttt{Nelder-Mead} to improve the convergence rate substantially.

\begin{knitrout}
\definecolor{shadecolor}{rgb}{0.969, 0.969, 0.969}\color{fgcolor}\begin{figure}[htb]

{\centering \includegraphics[width=\maxwidth]{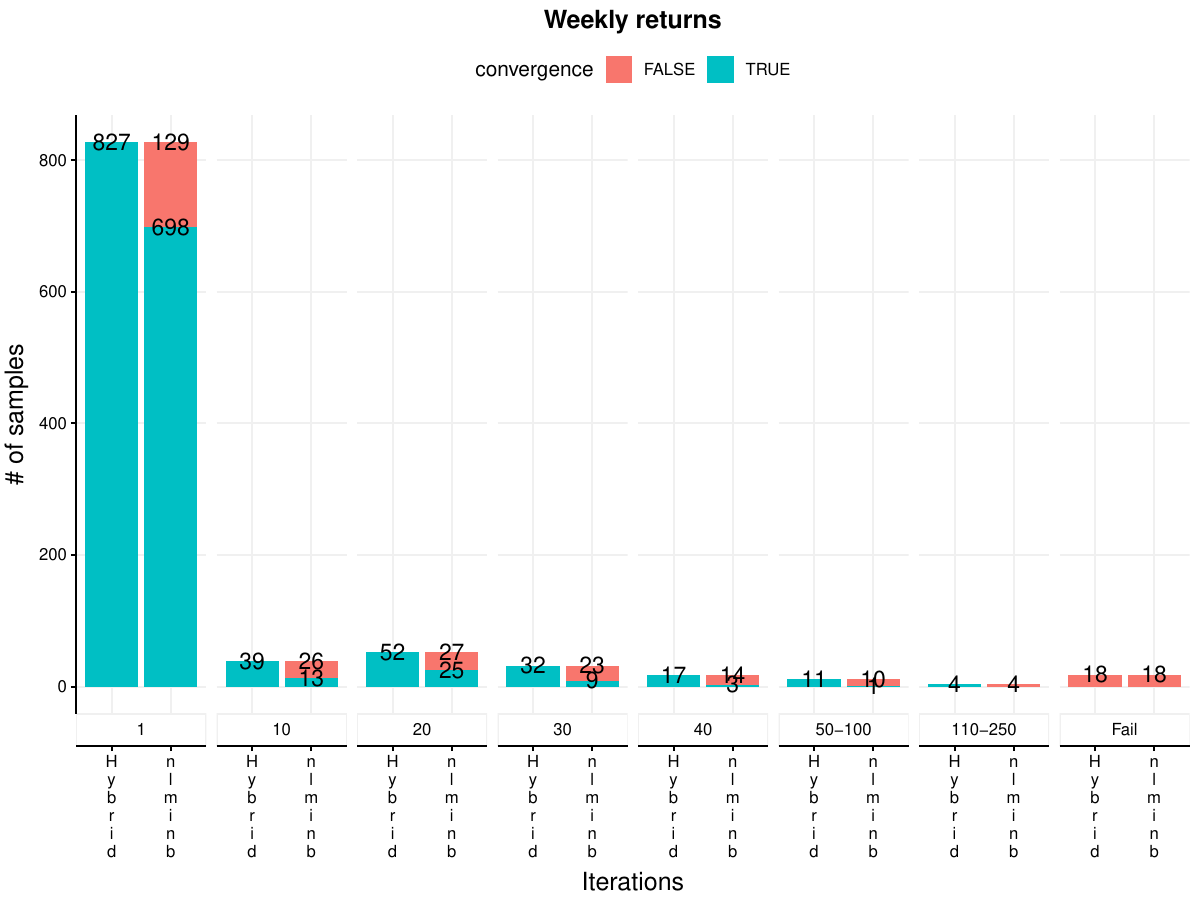} 

}

\caption{Convergence counts of \texttt{nlminb} and a hybrid algorithm (first \texttt{Nelder-Mead}, then \texttt{nlminb}). The hybrid algorithm uses 1, 10, 20,$\ldots$ iterations for the \texttt{Nelder-Mead}, then passes the estimates to \texttt{nlminb} as initial values. We randomly picked 1000 sets of initial values out of 1458000 potential candidates.}\label{fig:Nelder-Mead-convergence-per-iteration}
\end{figure}

\end{knitrout}

\section{Concluding remarks}
\label{sec:conclusion}
%%%%%%%%%%%%%%%%%%%%%%%%%%%%%%%%%%%%%%%%%%%%%%%%%%%%%%%%%%%%%%%%%

This paper addresses a couple of aspects concerning parameter estimation for HMMs via {\tt{TMB}} and {\tt{R}} using direct numerical maximization (DNM) of the likelihood. The advantage of {\tt{TMB}} is that it permits to quantify the uncertainty of any quantity depending on the estimated parameters. This is achieved by combining the delta method with automatic differentiation.
Moreover, {\tt{TMB}} provides exact calculations of first- and second-order derivatives of the (log-)likelihood of a model by automatic differentiation. This allows for efficient gradient- and/or Hessian-based optimization of the likelihood.   

In the first part of the paper, we propose a straightforward technique to quantify the uncertainty of smoothing probabilities via the calculation of Wald-type CIs. The advantage of this approach is that one avoids computationally intensive bootstrap methods. By means of several examples, we illustrate how such CIs provide additional insight into the commonly used state classification via smoothing probabilities. For a practitioner working with HMMs, the presented uncertainty quantification constitutes a new tool for obtaining a better understanding of the dynamics of the hidden process.

Subsequently, we examine speed and accuracy of the different optimizers in three simulation settings. Our results show a slight variation in both number of iterations and computational speed. In particular, the optimizers {\tt{nlminb}}, {\tt{nlm}}, {\tt{marqLevAlg}}, and {\tt{BFGS}} usually possess the highest computational speed. The computing time required by all optimizers reduces when using the gradient provided by {\tt{TMB}}. The number of iterations reduces in particular for {\tt{nlminb}} and {\tt{nlm}} when employing both gradient and Hessian from {\tt{TMB}}. This suggests a more direct path towards the (global) likelihood maximum compared to optimizer-internal approximation routines.\\ 
Regarding the accuracy, measured in terms of attained negative log-likelihood and parameter estimates (and their variability) the results show little variations across the different optimizers. This is indeed a  positive finding, indicating that the performance of the optimizers is relatively equal in terms of accuracy - however, this statement only holds true when using optimal initial value.       
Then, we examine robustness towards initial values across different optimizers. In three settings, we measure a) how often optimizers fail to converge and b) how often they successfully reach the global maximum of the log-likelihood function when starting from a wide range of sets of initial values. In particular, {\tt{nlm}}, {\tt{nlminb}}, and {\tt{hjn}} show an overall good performance. Notably, {\tt{nlm}} benefits strongly from employing both gradient and Hessian provided by {\tt{TMB}}, which is not the case for {\tt{nlminb}}. Altogether, we observe a trade-off between failure rates and convergence to the global maximum. Nevertheless, none of the optimizers shows an exceptionally bad performance.

Finally, we illustrate that a hybrid algorithm starting with the \texttt{Nelder-Mead} optimizer and switching to \texttt{nlminb} after a certain number of iterations converges more often than {\tt{nlminb}}. Notably an effect is visible even with a single initial iteration carried out by \texttt{Nelder-Mead}. 

 \clearpage

% \vspace*{1pc}
% \footnotesize{
% \noindent {\bf{Acknowledgement}}

\section*{Acknowledgements}
% \fontsize{9pt}{11pt} % 2nd parameter = 1.2 times the font size, https://tex.stackexchange.com/a/48277
% \selectfont
%BLINDED
We thank the University of Regensburg and European School for Interdisciplinary Tinnitus Research (ESIT) for providing access to the TYT data. Special thanks go to J.~Sim\~{o}es for data preparation, and W.~Schlee and B.~Langguth for helpful comments and suggestions. We also thank Anthony Chauvin and Bertrand Galichon for providing access to the hospital data set along with the necessary authorizations. J.~Bulla's profound thanks for the continued moral support go to the highly skilled, very helpful, and always friendly employees of Vestbrygg AS (Org.~ID 912105954).
\vspace*{1pc}

% \noindent {\bf{Conflict of Interest}}

\section*{Disclosure statement}

The authors have declared no conflict of interest.

\section*{Funding}

%BLINDED
The work of J. Bulla was supported by the GENDER-Net Co-Plus Fund (GNP-182). This work was also supported by the Financial Market Fund (Norwegian Research Council project no. 309218).

%%%%%%%%%%%%%%%%%%%%%%%%%%%%%%%%%%%%%%%%%%%%%%%%%%%%%%%%%%%%%%%%%
%\clearpage
% \bibliographystyle{plainnat}
\bibliographystyle{apalike}
\bibliography{paper2,packages}

\begin{thebibliography}{}

\bibitem[Ailliot et~al., 2015]{ailliot}
Ailliot, P., Allard, D., Monbet, V., and Naveau, P. (2015).
\newblock {Stochastic weather generators: an overview of weather type models}.
\newblock {\em Journal de la soci\'et\'e fran\c{c}aise de statistique}, 156(1):101--113.

\bibitem[Altman and Petkau, 2005]{altmana}
Altman, R.~M. and Petkau, A.~J. (2005).
\newblock {Application of hidden Markov models to multiple sclerosis lesion count data}.
\newblock {\em Statistics in Medicine}, 24(15):2335--2344.

\bibitem[Bacri et~al., 2022]{bacri}
Bacri, T., Berentsen, G.~D., Bulla, J., and H{\o}lleland, S. (2022).
\newblock A gentle tutorial on accelerated parameter and confidence interval estimation for hidden {{Markov}} models using {{Template Model Builder}}.
\newblock {\em Biometrical Journal}.

\bibitem[Bates et~al., 2022]{R-minqa}
Bates, D., Mullen, K.~M., Nash, J.~C., and Varadhan, R. (2022).
\newblock {\em minqa: Derivative-Free Optimization Algorithms by Quadratic Approximation}.
\newblock R package version 1.2.5.

\bibitem[Baum and Petrie, 1966]{baum}
Baum, L.~E. and Petrie, T. (1966).
\newblock Statistical {{Inference}} for {{Probabilistic Functions}} of {{Finite State Markov Chains}}.
\newblock {\em The Annals of Mathematical Statistics}, 37(6):1554--1563.

\bibitem[Baum et~al., 1970]{bauma}
Baum, L.~E., Petrie, T., Soules, G., and Weiss, N. (1970).
\newblock A maximization technique occurring in the statistical analysis of probabilistic functions of {{Markov}} chains.
\newblock {\em The Annals of Mathematical Statistics}, 41(1):164--171.

\bibitem[Bhatia and Davis, 2000]{bhatia}
Bhatia, R. and Davis, C. (2000).
\newblock A {{Better Bound}} on the {{Variance}}.
\newblock {\em The American Mathematical Monthly}, 107(4):353--357.

\bibitem[B{\"o}hning, 1999]{bohning}
B{\"o}hning, D. (1999).
\newblock {\em Computer-Assisted Analysis of Mixtures and Applications: Meta-Analysis, Disease Mapping and Others}, volume~81.
\newblock {CRC press}.

\bibitem[Broyden, 1970]{broyden}
Broyden, C.~G. (1970).
\newblock The {{Convergence}} of a {{Class}} of {{Double-rank Minimization Algorithms}} 1. {{General Considerations}}.
\newblock {\em IMA Journal of Applied Mathematics}, 6(1):76--90.

\bibitem[Bulla and Berzel, 2008]{bulla}
Bulla, J. and Berzel, A. (2008).
\newblock Computational issues in parameter estimation for stationary hidden {{Markov}} models.
\newblock {\em Computational Statistics}, 23(1):1--18.

\bibitem[Byrd et~al., 1995]{byrd}
Byrd, R.~H., Lu, P., Nocedal, J., and Zhu, C. (1995).
\newblock A {{Limited Memory Algorithm}} for {{Bound Constrained Optimization}}.
\newblock {\em SIAM Journal on Scientific Computing}, 16(5):1190--1208.

\bibitem[Capp{\'e} et~al., 2006]{cappe}
Capp{\'e}, O., Moulines, E., and Ryden, T. (2006).
\newblock {\em Inference in {{Hidden Markov Models}}}.
\newblock {Springer Science \& Business Media}.

\bibitem[Dempster et~al., 1977]{dempster}
Dempster, A.~P., Laird, N.~M., and Rubin, D.~B. (1977).
\newblock Maximum {{Likelihood}} from {{Incomplete Data Via}} the {{EM Algorithm}}.
\newblock {\em Journal of the Royal Statistical Society: Series B (Methodological)}, 39(1):1--22.

\bibitem[Dennis and Mor{\'e}, 1977]{dennis}
Dennis, John~E., J. and Mor{\'e}, J.~J. (1977).
\newblock Quasi-{{Newton Methods}}, {{Motivation}} and {{Theory}}.
\newblock {\em SIAM Review}, 19(1):46--89.

\bibitem[Dennis and Schnabel, 1996]{dennisa}
Dennis, J.~E. and Schnabel, R.~B. (1996).
\newblock {\em Numerical {{Methods}} for {{Unconstrained Optimization}} and {{Nonlinear Equations}}}.
\newblock Classics in {{Applied Mathematics}}. {Society for Industrial and Applied Mathematics}.

\bibitem[Durbin, 1998]{durbin}
Durbin, R. (1998).
\newblock {\em Biological {{Sequence Analysis}}: {{Probabilistic Models}} of {{Proteins}} and {{Nucleic Acids}}}.
\newblock {Cambridge University Press}.

\bibitem[Eddy, 1998]{eddy}
Eddy, S.~R. (1998).
\newblock Profile hidden {{Markov}} models.
\newblock {\em Bioinformatics}, 14(9):755--763.

\bibitem[Efron and Tibshirani, 1993]{efron}
Efron, B. and Tibshirani, R.~J. (1993).
\newblock {\em An Introduction to the Bootstrap}.
\newblock {Chapman \& Hall}, {New York, N.Y.; London}.

\bibitem[Feller, 1968]{Feller}
Feller, W. (1968).
\newblock {\em An Introduction to Probability Theory and Its Applications}.
\newblock {Wiley}.

\bibitem[Fletcher, 1970]{fletchera}
Fletcher, R. (1970).
\newblock A new approach to variable metric algorithms.
\newblock {\em The Computer Journal}, 13(3):317--322.

\bibitem[Fletcher, 2013]{fletcher}
Fletcher, R. (2013).
\newblock {\em Practical {{Methods}} of {{Optimization}}}.
\newblock {John Wiley \& Sons}.

\bibitem[Fredkin and Rice, 1992]{fredkin}
Fredkin, D.~R. and Rice, J.~A. (1992).
\newblock Bayesian {{Restoration}} of {{Single-Channel Patch Clamp Recordings}}.
\newblock {\em Biometrics}, 48(2):427--448.

\bibitem[Gay, 1990]{gay}
Gay, D.~M. (1990).
\newblock Usage summary for selected optimization routines.
\newblock {\em Computing science technical report}, 153:1--21.

\bibitem[Goldfarb, 1970]{goldfarb}
Goldfarb, D. (1970).
\newblock A family of variable-metric methods derived by variational means.
\newblock {\em Mathematics of Computation}, 24(109):23--26.

\bibitem[Guidolin and Timmermann, 2005]{guidolina}
Guidolin, M. and Timmermann, A. (2005).
\newblock Economic {{Implications}} of {{Bull}} and {{Bear Regimes}} in {{UK Stock}} and {{Bond Returns}}.
\newblock {\em The Economic Journal}, 115(500):111--143.

\bibitem[Hamilton, 1989]{hamilton}
Hamilton, J.~D. (1989).
\newblock A {{New Approach}} to the {{Economic Analysis}} of {{Nonstationary Time Series}} and the {{Business Cycle}}.
\newblock {\em Econometrica}, 57(2):357--384.

\bibitem[H{\"a}rdle et~al., 2003]{hardle}
H{\"a}rdle, W., Horowitz, J., and Kreiss, J.-P. (2003).
\newblock Bootstrap {{Methods}} for {{Time Series}}.
\newblock {\em International Statistical Review}, 71(2):435--459.

\bibitem[Hassan et~al., 2007]{hassan}
Hassan, M.~R., Nath, B., and Kirley, M. (2007).
\newblock A fusion model of {{HMM}}, {{ANN}} and {{GA}} for stock market forecasting.
\newblock {\em Expert Systems with Applications}, 33(1):171--180.

\bibitem[Hooke and Jeeves, 1961]{hooke}
Hooke, R. and Jeeves, T.~A. (1961).
\newblock `` {{Direct Search}}'' {{Solution}} of {{Numerical}} and {{Statistical Problems}}.
\newblock {\em Journal of the ACM}, 8(2):212--229.

\bibitem[Juang and Rabiner, 1991]{juang}
Juang, B.~H. and Rabiner, L.~R. (1991).
\newblock Hidden {{Markov Models}} for {{Speech Recognition}}.
\newblock {\em Technometrics}, 33(3):251--272.

\bibitem[Kato et~al., 2002]{kato}
Kato, J., Watanabe, T., Joga, S., Rittscher, J., and Blake, A. (2002).
\newblock An {{HMM-based}} segmentation method for traffic monitoring movies.
\newblock {\em IEEE Transactions on Pattern Analysis and Machine Intelligence}, 24(9):1291--1296.

\bibitem[Kristensen et~al., 2016]{kristensen}
Kristensen, K., Nielsen, A., Berg, C., Skaug, H., and Bell, B. (2016).
\newblock {{TMB}}: {{Automatic}} differentiation and laplace approximation.
\newblock {\em Journal of Statistical Software, Articles}, 70(5):1--21.

\bibitem[Lange and Weeks, 1989]{lange}
Lange, K. and Weeks, D.~E. (1989).
\newblock Efficient computation of lod scores: Genotype elimination, genotype redefinition, and hybrid maximum likelihood algorithms.
\newblock {\em Annals of Human Genetics}, 53(1):67--83.

\bibitem[Levenberg, 1944]{levenberg}
Levenberg, K. (1944).
\newblock A method for the solution of certain non-linear problems in least squares.
\newblock {\em Quarterly of Applied Mathematics}, 2(2):164--168.

\bibitem[Liporace, 1982]{liporace}
Liporace, L. (1982).
\newblock Maximum likelihood estimation for multivariate observations of {{Markov}} sources.
\newblock {\em IEEE Transactions on Information Theory}, 28(5):729--734.

\bibitem[Lystig and Hughes, 2002]{lystig}
Lystig, T.~C. and Hughes, J.~P. (2002).
\newblock Exact {{Computation}} of the {{Observed Information Matrix}} for {{Hidden Markov Models}}.
\newblock {\em Journal of Computational and Graphical Statistics}, 11(3):678--689.

\bibitem[MacDonald and Zucchini, 1997]{macdonald}
MacDonald, I.~L. and Zucchini, W. (1997).
\newblock {\em Hidden {{Markov}} and Other Models for Discrete-Valued Time Series}, volume~70 of {\em Monographs on Statistics and Applied Probability}.
\newblock {Chapman \& Hall}, {London}.

\bibitem[{maheu} and McCurdy, 2000]{maheua}
{maheu}, J.~M. and McCurdy, T.~H. (2000).
\newblock Identifying {{Bull}} and {{Bear Markets}} in {{Stock Returns}}.
\newblock {\em Journal of Business \& Economic Statistics}, 18(1):100--112.

\bibitem[Marquardt, 1963]{marquardt}
Marquardt, D.~W. (1963).
\newblock An {{Algorithm}} for {{Least-Squares Estimation}} of {{Nonlinear Parameters}}.
\newblock {\em Journal of the Society for Industrial and Applied Mathematics}, 11(2):431--441.

\bibitem[McClintock et~al., 2020]{mcclintock}
McClintock, B.~T., Langrock, R., Gimenez, O., Cam, E., Borchers, D.~L., Glennie, R., and Patterson, T.~A. (2020).
\newblock Uncovering ecological state dynamics with hidden {{Markov}} models.
\newblock {\em Ecology Letters}, 23(12):1878--1903.

\bibitem[Mersmann, 2021]{R-microbenchmark}
Mersmann, O. (2021).
\newblock {\em microbenchmark: Accurate Timing Functions}.
\newblock R package version 1.4.9.

\bibitem[Mor et~al., 2021]{mor}
Mor, B., Garhwal, S., and Kumar, A. (2021).
\newblock A {{Systematic Review}} of {{Hidden Markov Models}} and {{Their Applications}}.
\newblock {\em Archives of Computational Methods in Engineering}, 28(3):1429--1448.

\bibitem[Nagy, 1918]{nagy}
Nagy, J. (1918).
\newblock {\"Uber algebraische Gleichungen mit lauter reellen Wurzeln.}
\newblock {\em Jahresbericht der Deutschen Mathematiker-Vereinigung}, 27:37--43.

\bibitem[Nash and Varadhan, 2019]{R-optimr}
Nash, J.~C. and Varadhan, R. (2019).
\newblock {\em optimr: A Replacement and Extension of the optim Function}.
\newblock R package version 2019-12.16.

\bibitem[Nelder and Mead, 1965]{nelder}
Nelder, J.~A. and Mead, R. (1965).
\newblock A {{Simplex Method}} for {{Function Minimization}}.
\newblock {\em The Computer Journal}, 7(4):308--313.

\bibitem[Nielsen, 2000]{nielsen}
Nielsen, H.~B. (2000).
\newblock {\em {{UCMINF}} - an Algorithm for Unconstrained, Nonlinear Optimization}.
\newblock {Informatics and Mathematical Modelling, Technical University of Denmark, DTU}.

\bibitem[Nielsen and Mortensen, 2022]{R-ucminf}
Nielsen, H.~B. and Mortensen, S.~B. (2022).
\newblock {\em ucminf: General-Purpose Unconstrained Non-Linear Optimization}.
\newblock R package version 1.1-4.1.

\bibitem[Oudelha and Ainon, 2010]{oudelha}
Oudelha, M. and Ainon, R.~N. (2010).
\newblock {{HMM}} parameters estimation using hybrid {{Baum-Welch}} genetic algorithm.
\newblock In {\em 2010 {{International Symposium}} on {{Information Technology}}}, volume~2, pages 542--545.

\bibitem[Philipps et~al., 2022]{R-marqLevAlg}
Philipps, V., Proust-Lima, C., Prague, M., Hejblum, B., Commenges, D., and Diakite, A. (2022).
\newblock {\em marqLevAlg: A Parallelized General-Purpose Optimization Based on Marquardt-Levenberg Algorithm}.
\newblock R package version 2.0.7.

\bibitem[Powell, 2006]{powell}
Powell, M. J.~D. (2006).
\newblock The {{NEWUOA}} software for unconstrained optimization without derivatives.
\newblock In Di~Pillo, G. and Roma, M., editors, {\em Large-{{Scale Nonlinear Optimization}}}, Nonconvex {{Optimization}} and {{Its Applications}}, pages 255--297. {Springer US}, {Boston, MA}.

\bibitem[Probst et~al., 2016]{probst}
Probst, T., Pryss, R., Langguth, B., and Schlee, W. (2016).
\newblock Emotion dynamics and tinnitus: {{Daily}} life data from the ``{{TrackYourTinnitus}}'' application.
\newblock {\em Scientific Reports}, 6(1):31166.

\bibitem[Probst et~al., 2017]{probsta}
Probst, T., Pryss, R.~C., Langguth, B., Rauschecker, J.~P., Schobel, J., Reichert, M., Spiliopoulou, M., Schlee, W., and Zimmermann, J. (2017).
\newblock Does {{Tinnitus Depend}} on {{Time-of-Day}}? {{An Ecological Momentary Assessment Study}} with the ``{{TrackYourTinnitus}}'' {{Application}}.
\newblock {\em Frontiers in Aging Neuroscience}, 9:253.

\bibitem[Pryss et~al., 2015a]{pryssa}
Pryss, R., Reichert, M., Herrmann, J., Langguth, B., and Schlee, W. (2015a).
\newblock Mobile {{Crowd Sensing}} in {{Clinical}} and {{Psychological Trials}} \textendash{} {{A Case Study}}.
\newblock In {\em 2015 {{IEEE}} 28th {{International Symposium}} on {{Computer-Based Medical Systems}}}, pages 23--24. {IEEE}.

\bibitem[Pryss et~al., 2015b]{pryss}
Pryss, R., Reichert, M., Langguth, B., and Schlee, W. (2015b).
\newblock Mobile {{Crowd Sensing Services}} for {{Tinnitus Assessment}}, {{Therapy}}, and {{Research}}.
\newblock In {\em 2015 {{IEEE International Conference}} on {{Mobile Services}}}, pages 352--359. {IEEE}.

\bibitem[{R Core Team}, 2021]{rcoreteam}
{R Core Team} (2021).
\newblock {\em R: {{A}} Language and Environment for Statistical Computing}.
\newblock {R Foundation for Statistical Computing}, {Vienna, Austria}.

\bibitem[Rabiner, 1989]{rabiner}
Rabiner, L. (1989).
\newblock A tutorial on hidden {{Markov}} models and selected applications in speech recognition.
\newblock {\em Proceedings of the IEEE}, 77(2):257--286.

\bibitem[Rabiner and Juang, 1986]{rabinera}
Rabiner, L. and Juang, B. (1986).
\newblock An introduction to hidden {{Markov}} models.
\newblock {\em IEEE ASSP Magazine}, 3(1):4--16.

\bibitem[Redner and Walker, 1984]{redner}
Redner, R.~A. and Walker, H.~F. (1984).
\newblock Mixture {{Densities}}, {{Maximum Likelihood}} and the {{EM Algorithm}}.
\newblock {\em SIAM Review}, 26(2):195--239.

\bibitem[Ryd{\'e}n et~al., 1998]{rydenb}
Ryd{\'e}n, T., Ter{\"a}svirta, T., and {\AA}sbrink, S. (1998).
\newblock Stylized {{Facts}} of {{Daily Return Series}} and the {{Hidden Markov Model}}.
\newblock {\em Journal of Applied Econometrics}, 13(3):217--244.

\bibitem[Schadt et~al., 1998]{schadt}
Schadt, E.~E., Sinsheimer, J.~S., and Lange, K. (1998).
\newblock Computational {{Advances}} in {{Maximum Likelihood Methods}} for {{Molecular Phylogeny}}.
\newblock {\em Genome Research}, 8(3):222--233.

\bibitem[Schnabel et~al., 1985]{schnabel}
Schnabel, R.~B., Koonatz, J.~E., and Weiss, B.~E. (1985).
\newblock A modular system of algorithms for unconstrained minimization.
\newblock {\em ACM Transactions on Mathematical Software}, 11(4):419--440.

\bibitem[Schwendinger and Borchers, 2022]{schwendinger}
Schwendinger, F. and Borchers, H.~W. (2022).
\newblock {{CRAN Task View}}: {{Optimization}} and {{Mathematical Programming}}.
\newblock https://CRAN.R-project.org/view=Optimization.

\bibitem[Schwert, 1989]{schwerta}
Schwert, G.~W. (1989).
\newblock Why does stock market volatility change over time.
\newblock {\em Journal of Finance}, 44(5):1115--1153.

\bibitem[Shanno, 1970]{shanno}
Shanno, D.~F. (1970).
\newblock Conditioning of quasi-{{Newton}} methods for function minimization.
\newblock {\em Mathematics of Computation}, 24(111):647--656.

\bibitem[Stolcke and Omohundro, 1993]{stolcke}
Stolcke, A. and Omohundro, S. (1993).
\newblock Hidden {{Markov}} model induction by {{Bayesian}} model merging.
\newblock {\em Advances in neural information processing systems}, pages 11--11.

\bibitem[Turner, 2008]{turner}
Turner, R. (2008).
\newblock Direct maximization of the likelihood of a hidden {{Markov}} model.
\newblock {\em Computational Statistics \& Data Analysis}, 52(9):4147--4160.

\bibitem[Varadhan and Gilbert, 2010]{varadhan}
Varadhan, R. and Gilbert, P. (2010).
\newblock {{BB}}: {{An R Package}} for {{Solving}} a {{Large System}} of {{Nonlinear Equations}} and for {{Optimizing}} a {{High-Dimensional Nonlinear Objective Function}}.
\newblock {\em Journal of Statistical Software}, 32:1--26.

\bibitem[Varadhan and Gilbert, 2019]{R-BB}
Varadhan, R. and Gilbert, P. (2019).
\newblock {\em BB: Solving and Optimizing Large-Scale Nonlinear Systems}.
\newblock R package version 2019.10-1.

\bibitem[Visser et~al., 2000]{visser}
Visser, I., Raijmakers, M. E.~J., and Molenaar, P. C.~M. (2000).
\newblock Confidence intervals for hidden {{Markov}} model parameters.
\newblock {\em British Journal of Mathematical and Statistical Psychology}, 53(2):317--327.

\bibitem[Wu, 1983]{wu}
Wu, C. F.~J. (1983).
\newblock On the {{Convergence Properties}} of the {{EM Algorithm}}.
\newblock {\em The Annals of Statistics}, 11(1):95--103.

\bibitem[Zucchini et~al., 2016]{zucchini}
Zucchini, W., MacDonald, I., and Langrock, R. (2016).
\newblock {\em Hidden Markov Models for Time Series: An Introduction Using r, Second Edition}.
\newblock Chapman \& {{Hall}}/{{CRC}} Monographs on Statistics \& Applied Probability. {CRC Press}.

\end{thebibliography}
% \addcontentsline{toc}{chapter}{References}
%%%%%%%%%%%%%%%%%%%%%%%%%%%%%%%%%%%%%%%%%%%%%%%%%%%%%%%%%%%%%%%%%

\clearpage
%%%%%%%%%%%%%%%%%%%%%%%%%%%%%%%%%%%%%%%%%%%%%%%%%%%%%%%%%%%%%%%%%
\appendix

\begin{center}
  {\bf \Huge Appendix}
\end{center}
%%%%%%%%%%%%%%%%%%%%%%%%%%%%%%%%%%%%%%%%%%%%%%%%%%%%%%%%%%%%%%%%%

%The supplemental material provides some additional tables and figures related to the results from the main paper, and also provide a few additional studies. 

%%%%%%%%%%%%%%%%%%%%%%%%%%%%%%%%%%%%%%%%%%%%%%%%%
\section{Forward algorithm and backward algorithm}
\label{sec:appendix-hmm_fwbw}
%%%%%%%%%%%%%%%%%%%%%%%%%%%%%%%%%%%%%%%%%%%%%%%%%

The pass through observations can be efficiently computed piece by piece, from left (i.e., time $s=1$) to right (i.e., time $s=t$) or right to left by the so-called ``forward algorithm'' and ``backward algorithm'' respectively. With these two algorithms, the likelihood can be computed recursively. To formulate the forward algorithm, let us define the vector $\alpha_t$ for $t = 1, 2, \ldots, T$ so that
\begin{align*}
\bm{\alpha}_t &= \bm{\delta} \bm{P}(x_1)\bm{\Gamma} \bm{P}(x_2) \bm{\Gamma} \bm{P}(x_3) \ldots \bm{\Gamma} \bm{P}(x_t)\\
&= \bm{\delta} \bm{P}(x_1) \prod_{s=2}^{t}\bm{\Gamma} \bm{P}(x_s)\\
&= \left( \alpha_t(1), \ldots, \alpha_t(m) \right)
\end{align*}
for $t = 1, 2, \ldots, T$.
The name of the algorithm comes from computing it via the recursion
\begin{gather*}
\bm{\alpha}_0 = \bm{\delta} \bm{P}(x_1),\\
\bm{\alpha}_t = \bm{\alpha}_{t-1} \bm{\Gamma} \bm{P}(x_t) \text{ for } t = 1, 2, \ldots, T.
\end{gather*}
After passing through all observations until time $T$, the likelihood is derived from
\begin{gather*}
L(\bm{\theta}) = \bm{\alpha}_T \bm{1}'.
\end{gather*}

The backward algorithm is very similar to the forward algorithm. The sole difference lies in starting from the last instead of the first observation. To specify the backward algorithm, let us define the vector $\bm{\beta}_t$ for $t = 1, 2, \ldots, T$ so that
\begin{align*}
\bm{\beta}'_t &= \bm{\Gamma} \bm{P}(x_{t+1}) \bm{\Gamma} \bm{P}(x_{t+2}) \ldots \bm{\Gamma} \bm{P}(x_T) \ldots \bm{1}'\\
&= \left(\prod_{s=t+1}^{T}\bm{\Gamma} \bm{P}(x_s) \right) \bm{1}'\\
&= \left( \beta_t(1), \ldots, \beta_t(m) \right).
\end{align*}
The name backward algorithm results from the way of recursively calculating $\bm{\beta}_t$ by
\begin{gather*}
\bm{\beta}_T = \bm{1}'\\
\bm{\beta}_t = \bm{\Gamma} \bm{P}(x_{t+1}) \bm{\beta}_{t+1} \text{ for } t = T-1, T-2, \ldots, 1.
\end{gather*}
As before, the likelihood can be obtained after a pass through all observations by
\begin{gather*}
L(\bm{\theta}) = \bm{\delta} \bm{\beta}_1.
\end{gather*}

Usually, only the forward algorithm is used for parameter estimation.
Nonetheless, either or both algorithms may be necessary to derive certain quantities of interest, e.g., smoothing probabilities (as we explore in \autoref{sec:smoothing-uncertainty}) or forecast probabilities. For practical implementation, attention needs to be paid to underflow errors which can quickly occur in both algorithms (due to the elements of the TPM and / or conditional probabilties taking values in $[0,1]$). Therefore, we implemented a scaled version of the forward algorithm as suggested by \cite[p.~48]{zucchini}. This version directly provides the negative log-likelihood (nll) as result.

%%%%%%%%%%%%%%%%%%%%%%%%%%%%%%%%%%%%%%%%%%%%%%%%%%%%%%%%%%%%
\section{Reparameterization of the likelihood function}
\label{sec:appendix-reparameterization}
%%%%%%%%%%%%%%%%%%%%%%%%%%%%%%%%%%%%%%%%%%%%%%%%%%%%%%%%%%%%

Along with the data, $\bm{\theta}$ serves as input for calculating the likelihood by the previously illustrated forward algorithms. Some constraints need to be imposed on $\bm{\theta}$:
\begin{enumerate}
\item In the case of the Poisson HMM, the parameter vector $\bm{\lambda}\ = (\lambda_1, \dots, \lambda_m)$ necessary for computing the conditional distribution has to consist of strictly positive elements.
\item The elements $\gamma_{ij}$ of the TPM $\bm{\Gamma}$ must be non-negative, and each row of $\bm{\Gamma}$ needs to sum up to to one.
\end{enumerate}
Tacking these constraints by means of constrained optimization routines when maximizing the likelihood can lead to several difficulties. For example, constrained optimization routines may require a significantly amount of fine-tuning. Moreover, the number of available optimization routines is significantly reduced. A common alternative works by reparametrizing the log-likelihood as a function of unconstrained, so-called ``working'' parameters $\{\bf{T}, \bm{\eta}\}= g^{-1}(\bm{\Gamma}, \bm{\lambda})$, as follows.
One possibility to reparametrize $\bm{\Gamma}$ is given by
\begin{equation*}
\tau_{ij} = \log\left(\frac{\gamma_{ij}}{1 - \sum_{k \neq i} \gamma_{ik}}\right) = \log(\gamma_{ij}/\gamma_{ii}) \text{ for } i \neq j.
\end{equation*}
where $\tau_{ij} \in \mathbb{R}$ are $m(m-1)$ unconstrained elements of an $m \times m$ matrix $\bf{T}$ with missing diagonal elements. The matching reverse transformation is
\begin{equation*}
\gamma_{ij} = \frac{\exp(\tau_{ij})}{1 + \sum_{k \neq i} \exp(\tau_{ik})} \text{ for } i \neq j\text{,}
\end{equation*}
and the diagonal elements of $\bm{\Gamma}$ follow from $\sum_j \gamma_{ij} = 1$, $i = 1,...,m$ (see \citet{zucchini}, p.~51).

For a Poisson HMM, a simple method to reparametrize the conditional Poisson means $\bm{\lambda}$ into $\bm{\eta}$ is given by $\eta_i = \log(\lambda_i)$ for $i = 1, \ldots, m$.
Consequently, the constrained ``natural'' parameters are obtained via $\lambda_i = \exp(\eta_i)$.
Estimation of the natural parameters $\{\bm{\Gamma}, \bm{\lambda}\}$ can therefore be obtained by maximization of the reparametrized likelihood with respect to $\{\bf{T}, \bm{\eta}\}$ then by a transformation of the estimated working parameters back to natural parameters via the above transformations, i.e. $\{\hat{\bm{\Gamma}}, \hat{\bm{\lambda}}\} = g(\hat{\bm{P}}, \hat{\bm{\eta}})$.
In general, the above procedure requires that the function $g$ is one-to-one.

With a Gaussian HMM, the means are already unconstrained and do not require any transformation.
However, the standard deviations can be transformed similarly to the Poisson rates via $\eta_i = \log(\sigma_i)$ for $i = 1, \ldots, m$, and the ``natural'' parameters are then obtained via the corresponding reverse transformation $\sigma_i = \exp(\eta_i)$.

%%%%%%%%%%%%%%%%%%%%%%%%%%%%%%%%%%%%%%%%%%%%%%%%%%%%%%%%%%%%
\section{Confidence intervals}
\label{sec:appendix-cis}
%%%%%%%%%%%%%%%%%%%%%%%%%%%%%%%%%%%%%%%%%%%%%%%%%%%%%%%%%%%%

When handling statistical models, confidence intervals (CIs) for the estimated parameters can be derived via various approaches. One common such approach bases on finite-difference approximations of the Hessian. However, as \citet{visser} points out, there are better alternatives when dealing with most HMMs. Preferred are profile likelihood-based CIs or bootstrap-based CIs, where the latter are now widely used despite the potentially large computational load \citep{bulla, zucchini}. With common Poisson HMMs, \citet{bacri} shows that {\tt{TMB}} can yield valid Wald-type CIs in a fraction of the time required by classical bootstrap-based methods.

In this section, we describe the Wald-type and the bootstrap-based approach. The likelihood profile method frequently fails to yield CIs (see \citep{bacri}) and is therefore not detailed.

%%%%%%%%%%%%%%%%%%%%%%%%%%%%%%%%%%%%%%%%%%%%%%%%%%%%%%%%%%%%%%%%%
\subsection{Wald-type confidence intervals}
\label{sec:appendix-wald}
%%%%%%%%%%%%%%%%%%%%%%%%%%%%%%%%%%%%%%%%%%%%%%%%%%%%%%%%%%%%%%%%%

Evaluating the Hessian at the optimum found by numerical optimization procedures is basis for Wald-tye CIs. As illustrated above, the optimized negative log-likelihood function of a HMM typically depends on the working parameters. However, drawing inference on the natural parameters is by far more relevant than inference on the working parameters. Since the Hessian $\nabla^2\log L(\{\hat{\bm{T}}, \hat{\bm{\eta}}\})$ relies on the working parameters $\hat{\bm{\eta}}$, an estimate of the covariance matrix of $\{\hat{\bm{\Gamma}}, \hat{\bm{\lambda}}\}$ can be obtained via the delta method through
\begin{equation}
\Sigma_{\hat{\bm{\Gamma}}, \hat{\bm{\lambda}}} = - \nabla g(\hat{\bm{T}}, \hat{\bm{\eta}})\left(\nabla^2\log L(\hat{\bm{T}}, \hat{\bf{\eta}})\right)^{-1}\nabla g(\hat{\bm{T}}, \hat{\bm{\eta}})^\prime,
\label{eq:deltamethod}
\end{equation}
with $\{\hat{\bm{\Gamma}}, \hat{\bm{\lambda}}\} = g(\hat{\bm{T}}, \hat{\bm{\eta}})$ as defined in \cite[][Ch.~3.3.1]{zucchini} and \citet{bacri}.
From this, {\tt{TMB}} deduces the standard errors for the working and natural parameters, from which 95\% CIs can be formed via $\hat{\bm{\lambda}} \pm q_{0.975} \cdot \text{SE}(\hat{\bm{\lambda}})$ where $q_{0.975} \approx 1.96$ is the $97.5^{\text{th}}$ percentile from the standard normal distribution, and $\text{SE}(\hat{\bm{\lambda}})$ denotes the standard error of $\hat{\bm{\lambda}}$.

% {\hl REPHRASE AFTER}
% 
% From a practical perspective, however, inference about the natural parameters usually is of interest.
% As the Hessian $\nabla^2\log L(\{\hat{\bm{T}}, \hat{\bm{\eta}}\})$ refers to the working parameters $\{\bm{T}, \bm{\eta}\}$, the delta method is suitable to obtain an estimate of the covariance matrix of $\{\hat{\bm{\Gamma}}, \hat{\bm{\lambda}}\}$ by
% \begin{equation}
% \Sigma_{\hat{\bm{\Gamma}}, \hat{\bm{\lambda}}} = - \nabla g(\hat{\bm{T}}, \hat{\bm{\eta}})\left(\nabla^2\log L(\hat{\bm{T}}, \hat{\bf{\eta}})\right)^{-1}\nabla g(\hat{\bm{T}}, \hat{\bm{\eta}})^\prime,
% \label{eq:deltamethod}
% \end{equation}
% with $\{\hat{\bm{\Gamma}}, \hat{\bm{\lambda}}\} = g(\hat{\bm{T}}, \hat{\bm{\eta}})$ as defined in \autoref{sec:reparameterization}.
% 
% {\hl REPHRASE BEFORE}

%%%%%%%%%%%%%%%%%%%%%%%%%%%%%%%%%%%%%%%%%%%%%%%%%%%%%%%%%%%%%%%%%
\subsection{Bootstrap-based confidence intervals}
\label{sec:appendix-boostrap}
%%%%%%%%%%%%%%%%%%%%%%%%%%%%%%%%%%%%%%%%%%%%%%%%%%%%%%%%%%%%%%%%%

The bootstrap method was described by \citet{efron} in their seminal article, and is widely used by many practitioners. Many bootstrap techniques have been developed since then, and have been extensively applied in the scientific literature. This paper will not review these techniques, but will instead use one of them: the so-called parametric bootstrap. As \citet{hardle} points out, the parametric bootstrap is suitable for time series, and hence motivates this choice. For details on the parametric bootstrap implementation in {\tt{R}}, we refer to \cite[][Ch.~3.6.2, pp.~58-60]{zucchini}. 

At its heart, bootstrapping requires some form of re-sampling to be carried out. Then, the chosen model is re-estimated on each new sample, and eventually some form of aggregation of the results takes place.
We make use of {\tt{TMB}} to accelerate the re-estimation procedure, then look at the resulting median along with the 2.5th and 97.5th empirical percentile to infer a 95\% CI.

%%%%%%%%%%%%%%%%%%%%%%%%%%%%%%%%%%%%%%%%%%%%%%%%%%%%%%
\section{Estimated models for real data sets}
\label{sec:appendix-estmod}
%%%%%%%%%%%%%%%%%%%%%%%%%%%%%%%%%%%%%%%%%%%%%%%%%%%%%%

% SUMMARY TABLE PART 1

% latex table generated in R 4.2.2 by xtable 1.8-4 package
% Mon Jan 30 17:43:50 2023
\begin{table}[ht]
\centering
\caption{Estimates, Wald-type 95\% CIs, for a two-state Poisson HMM fitted to the TYT data, a two-state Poisson HMM fitted to the soap data, and a three-state Gaussian HMM fitted to the weekly returns data. From left to right, the columns contain: the parameter name, parameter estimate, and lower (L.) and upper (U.) bound of the corresponding 95\% CI derived via the Hessian provided by TMB, repeated for the three HMMs. As standard deviations and Poisson rates must be non-negative, the CIs are adjusted when necessary. Similarly, the TPM elements and the stationary distribution must take values between 0 and 1 and the CIs are adjusted accordingly when necessary. AIC and BIC scores are displayed under for corresponding models with two, three, four, and five states (when the model estimation converges). Model estimates are derived by \texttt{nlminb} with {\tt{TMB}}'s gradient and Hessian functions passed as arguments.} 
\label{tab:estimates-cis-aic-bic-real-data-part1}
\begin{tabular}{cccc|cccc|cccc}
  \multicolumn{4}{c}{TYT data set} & \multicolumn{4}{c}{Soap data set} & \multicolumn{4}{c}{Weekly returns data set} \\ \hline
Par. & Est. & L. & U. & Par. & Est. & L. & U. & Par. & Est. & L. & U. \\ 
  \hline
$\lambda_{1}$ & 1.64 & 1.09 & 2.18 & $\lambda_{1}$ & 4.02 & 3.66 & 4.38 & $\mu_{1}$ & -2.08 & -4.3 & 0.14 \\ 
  $\lambda_{2}$ & 5.53 & 4.91 & 6.16 & $\lambda_{2}$ & 11.37 & 9.94 & 12.80 & $\mu_{2}$ & 0.06 & -0.13 & 0.25 \\ 
  $\gamma_{11}$ & 0.95 & 0.86 & 1.00 & $\gamma_{11}$ & 0.91 & 0.86 & 0.96 & $\mu_{3}$ & 0.33 & 0.24 & 0.41 \\ 
  $\gamma_{12}$ & 0.05 & 0 & 0.14 & $\gamma_{12}$ & 0.09 & 0.04 & 0.14 & $\sigma_{1}$ & 7.11 & 5.25 & 8.97 \\ 
  $\gamma_{21}$ & 0.03 & 0 & 0.07 & $\gamma_{21}$ & 0.37 & 0.2 & 0.54 & $\sigma_{2}$ & 2.65 & 2.47 & 2.84 \\ 
  $\gamma_{22}$ & 0.97 & 0.93 & 1.00 & $\gamma_{22}$ & 0.63 & 0.46 & 0.80 & $\sigma_{3}$ & 1.41 & 1.34 & 1.48 \\ 
  $\delta_{1}$ & 0.34 & 0 & 0.79 & $\delta_{1}$ & 0.81 & 0.71 & 0.91 & $\gamma_{11}$ & 0.98 & 0.97 & 0.99 \\ 
  $\delta_{2}$ & 0.66 & 0.21 & 1.00 & $\delta_{2}$ & 0.19 & 0.09 & 0.29 & $\gamma_{12}$ & 0 & 0 & 0.00 \\ 
   &  &  &  &  &  &  &  & $\gamma_{13}$ & 0.02 & 0.01 & 0.03 \\ 
   &  &  &  &  &  &  &  & $\gamma_{21}$ & 0 & 0 & 0.00 \\ 
   &  &  &  &  &  &  &  & $\gamma_{22}$ & 0.81 & 0.63 & 0.99 \\ 
   &  &  &  &  &  &  &  & $\gamma_{23}$ & 0.19 & 0.01 & 0.37 \\ 
   &  &  &  &  &  &  &  & $\gamma_{31}$ & 0.03 & 0.01 & 0.04 \\ 
   &  &  &  &  &  &  &  & $\gamma_{32}$ & 0.01 & 0 & 0.02 \\ 
   &  &  &  &  &  &  &  & $\gamma_{33}$ & 0.96 & 0.94 & 0.98 \\ 
   &  &  &  &  &  &  &  & $\delta_{1}$ & 0.42 & 0.4 & 0.45 \\ 
   &  &  &  &  &  &  &  & $\delta_{2}$ & 0.55 & 0.42 & 0.69 \\ 
   &  &  &  &  &  &  &  & $\delta_{3}$ & 0.02 & 0 & 0.17 \\ 
   \hline
m & AIC & BIC &  & m & AIC & BIC &  & m & AIC & BIC &  \\ 
   \hline
2 & 345 & 355 &  & 2 & 1245 & 1259 &  & 2 & 9563 & 9597 &  \\ 
  3 & 355 & 377 &  & 3 & 1239 & 1270 &  & 3 & 9476 & 9544 &  \\ 
   &  &  &  & 4 & 1241 & 1297 &  & 4 & 9475 & 9590 &  \\ 
   \hline
\end{tabular}
\end{table}

% SUMMARY TABLE PART 2

  \providecommand{\huxb}[2]{\arrayrulecolor[RGB]{#1}\global\arrayrulewidth=#2pt}
  \providecommand{\huxvb}[2]{\color[RGB]{#1}\vrule width #2pt}
  \providecommand{\huxtpad}[1]{\rule{0pt}{#1}}
  \providecommand{\huxbpad}[1]{\rule[-#1]{0pt}{#1}}

\begin{table}[ht]
\begin{centerbox}
\begin{threeparttable}
\captionsetup{justification=centering,singlelinecheck=off}
\caption{Estimates, Wald-type 95\% CIs, for a five-state Poisson HMM fitted to the hospital data. From left to right, the columns contain: the parameter name, parameter estimate, and lower (L.) and upper (U.) bound of the corresponding 95\% CI derived via the Hessian provided by TMB, repeated for the three HMMs. As standard deviations and Poisson rates must be non-negative, the CIs are adjusted when necessary. Similarly, the TPM elements and the stationary distribution must take values between 0 and 1 and the CIs are adjusted accordingly when necessary. AIC and BIC scores are displayed under for corresponding models with two, three, four, and five states (when the model estimation converges). Model estimates are derived by \texttt{nlminb} with {\tt{TMB}}'s gradient and Hessian functions passed as arguments.}
 \label{tab:estimates-cis-aic-bic-real-data-part2}
\setlength{\tabcolsep}{0pt}
\begin{tabular}{l l l l l l l l}

\hhline{>{\huxb{0, 0, 0}{0.4}}|}
\arrayrulecolor{black}

\multicolumn{4}{!{\huxvb{0, 0, 0}{0}}c!{\huxvb{0, 0, 0}{0.4}}}{\huxtpad{0.5pt + 1em}\centering \hspace{0.5pt} Hospital data set (1) \hspace{0.5pt}\huxbpad{0.5pt}} &
\multicolumn{4}{c!{\huxvb{0, 0, 0}{0}}}{\huxtpad{0.5pt + 1em}\centering \hspace{0.5pt} Hospital data set (2) \hspace{0.5pt}\huxbpad{0.5pt}} \tabularnewline[-0.5pt]

\hhline{>{\huxb{0, 0, 0}{0.4}}->{\huxb{0, 0, 0}{0.4}}->{\huxb{0, 0, 0}{0.4}}->{\huxb{0, 0, 0}{0.4}}->{\huxb{0, 0, 0}{0.4}}->{\huxb{0, 0, 0}{0.4}}->{\huxb{0, 0, 0}{0.4}}->{\huxb{0, 0, 0}{0.4}}-}
\arrayrulecolor{black}

\multicolumn{1}{!{\huxvb{0, 0, 0}{0}}c!{\huxvb{0, 0, 0}{0}}}{\huxtpad{0.5pt + 1em}\centering \hspace{0.5pt} Par. \hspace{0.5pt}\huxbpad{0.5pt}} &
\multicolumn{1}{c!{\huxvb{0, 0, 0}{0}}}{\huxtpad{0.5pt + 1em}\centering \hspace{0.5pt} Est. \hspace{0.5pt}\huxbpad{0.5pt}} &
\multicolumn{1}{c!{\huxvb{0, 0, 0}{0}}}{\huxtpad{0.5pt + 1em}\centering \hspace{0.5pt} L. \hspace{0.5pt}\huxbpad{0.5pt}} &
\multicolumn{1}{c!{\huxvb{0, 0, 0}{0.4}}}{\huxtpad{0.5pt + 1em}\centering \hspace{0.5pt} U. \hspace{0.5pt}\huxbpad{0.5pt}} &
\multicolumn{1}{c!{\huxvb{0, 0, 0}{0}}}{\huxtpad{0.5pt + 1em}\centering \hspace{0.5pt} Par. \hspace{0.5pt}\huxbpad{0.5pt}} &
\multicolumn{1}{c!{\huxvb{0, 0, 0}{0}}}{\huxtpad{0.5pt + 1em}\centering \hspace{0.5pt} Est. \hspace{0.5pt}\huxbpad{0.5pt}} &
\multicolumn{1}{c!{\huxvb{0, 0, 0}{0}}}{\huxtpad{0.5pt + 1em}\centering \hspace{0.5pt} L. \hspace{0.5pt}\huxbpad{0.5pt}} &
\multicolumn{1}{c!{\huxvb{0, 0, 0}{0}}}{\huxtpad{0.5pt + 1em}\centering \hspace{0.5pt} U. \hspace{0.5pt}\huxbpad{0.5pt}} \tabularnewline[-0.5pt]

\hhline{>{\huxb{0, 0, 0}{0.4}}->{\huxb{0, 0, 0}{0.4}}->{\huxb{0, 0, 0}{0.4}}->{\huxb{0, 0, 0}{0.4}}->{\huxb{0, 0, 0}{0.4}}->{\huxb{0, 0, 0}{0.4}}->{\huxb{0, 0, 0}{0.4}}->{\huxb{0, 0, 0}{0.4}}-}
\arrayrulecolor{black}

\multicolumn{1}{!{\huxvb{0, 0, 0}{0}}c!{\huxvb{0, 0, 0}{0}}}{\huxtpad{0.5pt + 1em}\centering \hspace{0.5pt} $\lambda_{1}$ \hspace{0.5pt}\huxbpad{0.5pt}} &
\multicolumn{1}{c!{\huxvb{0, 0, 0}{0}}}{\huxtpad{0.5pt + 1em}\centering \hspace{0.5pt} 3.54 \hspace{0.5pt}\huxbpad{0.5pt}} &
\multicolumn{1}{c!{\huxvb{0, 0, 0}{0}}}{\huxtpad{0.5pt + 1em}\centering \hspace{0.5pt} 3.51 \hspace{0.5pt}\huxbpad{0.5pt}} &
\multicolumn{1}{c!{\huxvb{0, 0, 0}{0.4}}}{\huxtpad{0.5pt + 1em}\centering \hspace{0.5pt} 3.58 \hspace{0.5pt}\huxbpad{0.5pt}} &
\multicolumn{1}{c!{\huxvb{0, 0, 0}{0}}}{\huxtpad{0.5pt + 1em}\centering \hspace{0.5pt} $\gamma_{34}$ \hspace{0.5pt}\huxbpad{0.5pt}} &
\multicolumn{1}{c!{\huxvb{0, 0, 0}{0}}}{\huxtpad{0.5pt + 1em}\centering \hspace{0.5pt} 0 \hspace{0.5pt}\huxbpad{0.5pt}} &
\multicolumn{1}{c!{\huxvb{0, 0, 0}{0}}}{\huxtpad{0.5pt + 1em}\centering \hspace{0.5pt} 0 \hspace{0.5pt}\huxbpad{0.5pt}} &
\multicolumn{1}{c!{\huxvb{0, 0, 0}{0}}}{\huxtpad{0.5pt + 1em}\centering \hspace{0.5pt} 0 \hspace{0.5pt}\huxbpad{0.5pt}} \tabularnewline[-0.5pt]

\hhline{>{\huxb{0, 0, 0}{0.4}}|}
\arrayrulecolor{black}

\multicolumn{1}{!{\huxvb{0, 0, 0}{0}}c!{\huxvb{0, 0, 0}{0}}}{\huxtpad{0.5pt + 1em}\centering \hspace{0.5pt} $\lambda_{2}$ \hspace{0.5pt}\huxbpad{0.5pt}} &
\multicolumn{1}{c!{\huxvb{0, 0, 0}{0}}}{\huxtpad{0.5pt + 1em}\centering \hspace{0.5pt} 6.46 \hspace{0.5pt}\huxbpad{0.5pt}} &
\multicolumn{1}{c!{\huxvb{0, 0, 0}{0}}}{\huxtpad{0.5pt + 1em}\centering \hspace{0.5pt} 6.33 \hspace{0.5pt}\huxbpad{0.5pt}} &
\multicolumn{1}{c!{\huxvb{0, 0, 0}{0.4}}}{\huxtpad{0.5pt + 1em}\centering \hspace{0.5pt} 6.59 \hspace{0.5pt}\huxbpad{0.5pt}} &
\multicolumn{1}{c!{\huxvb{0, 0, 0}{0}}}{\huxtpad{0.5pt + 1em}\centering \hspace{0.5pt} $\gamma_{35}$ \hspace{0.5pt}\huxbpad{0.5pt}} &
\multicolumn{1}{c!{\huxvb{0, 0, 0}{0}}}{\huxtpad{0.5pt + 1em}\centering \hspace{0.5pt} 0 \hspace{0.5pt}\huxbpad{0.5pt}} &
\multicolumn{1}{c!{\huxvb{0, 0, 0}{0}}}{\huxtpad{0.5pt + 1em}\centering \hspace{0.5pt} 0 \hspace{0.5pt}\huxbpad{0.5pt}} &
\multicolumn{1}{c!{\huxvb{0, 0, 0}{0}}}{\huxtpad{0.5pt + 1em}\centering \hspace{0.5pt} 0 \hspace{0.5pt}\huxbpad{0.5pt}} \tabularnewline[-0.5pt]

\hhline{>{\huxb{0, 0, 0}{0.4}}|}
\arrayrulecolor{black}

\multicolumn{1}{!{\huxvb{0, 0, 0}{0}}c!{\huxvb{0, 0, 0}{0}}}{\huxtpad{0.5pt + 1em}\centering \hspace{0.5pt} $\lambda_{3}$ \hspace{0.5pt}\huxbpad{0.5pt}} &
\multicolumn{1}{c!{\huxvb{0, 0, 0}{0}}}{\huxtpad{0.5pt + 1em}\centering \hspace{0.5pt} 10.13 \hspace{0.5pt}\huxbpad{0.5pt}} &
\multicolumn{1}{c!{\huxvb{0, 0, 0}{0}}}{\huxtpad{0.5pt + 1em}\centering \hspace{0.5pt} 10.03 \hspace{0.5pt}\huxbpad{0.5pt}} &
\multicolumn{1}{c!{\huxvb{0, 0, 0}{0.4}}}{\huxtpad{0.5pt + 1em}\centering \hspace{0.5pt} 10.24 \hspace{0.5pt}\huxbpad{0.5pt}} &
\multicolumn{1}{c!{\huxvb{0, 0, 0}{0}}}{\huxtpad{0.5pt + 1em}\centering \hspace{0.5pt} $\gamma_{41}$ \hspace{0.5pt}\huxbpad{0.5pt}} &
\multicolumn{1}{c!{\huxvb{0, 0, 0}{0}}}{\huxtpad{0.5pt + 1em}\centering \hspace{0.5pt} 0 \hspace{0.5pt}\huxbpad{0.5pt}} &
\multicolumn{1}{c!{\huxvb{0, 0, 0}{0}}}{\huxtpad{0.5pt + 1em}\centering \hspace{0.5pt} 0 \hspace{0.5pt}\huxbpad{0.5pt}} &
\multicolumn{1}{c!{\huxvb{0, 0, 0}{0}}}{\huxtpad{0.5pt + 1em}\centering \hspace{0.5pt} 0 \hspace{0.5pt}\huxbpad{0.5pt}} \tabularnewline[-0.5pt]

\hhline{>{\huxb{0, 0, 0}{0.4}}|}
\arrayrulecolor{black}

\multicolumn{1}{!{\huxvb{0, 0, 0}{0}}c!{\huxvb{0, 0, 0}{0}}}{\huxtpad{0.5pt + 1em}\centering \hspace{0.5pt} $\lambda_{4}$ \hspace{0.5pt}\huxbpad{0.5pt}} &
\multicolumn{1}{c!{\huxvb{0, 0, 0}{0}}}{\huxtpad{0.5pt + 1em}\centering \hspace{0.5pt} 13.96 \hspace{0.5pt}\huxbpad{0.5pt}} &
\multicolumn{1}{c!{\huxvb{0, 0, 0}{0}}}{\huxtpad{0.5pt + 1em}\centering \hspace{0.5pt} 13.84 \hspace{0.5pt}\huxbpad{0.5pt}} &
\multicolumn{1}{c!{\huxvb{0, 0, 0}{0.4}}}{\huxtpad{0.5pt + 1em}\centering \hspace{0.5pt} 14.07 \hspace{0.5pt}\huxbpad{0.5pt}} &
\multicolumn{1}{c!{\huxvb{0, 0, 0}{0}}}{\huxtpad{0.5pt + 1em}\centering \hspace{0.5pt} $\gamma_{42}$ \hspace{0.5pt}\huxbpad{0.5pt}} &
\multicolumn{1}{c!{\huxvb{0, 0, 0}{0}}}{\huxtpad{0.5pt + 1em}\centering \hspace{0.5pt} 0 \hspace{0.5pt}\huxbpad{0.5pt}} &
\multicolumn{1}{c!{\huxvb{0, 0, 0}{0}}}{\huxtpad{0.5pt + 1em}\centering \hspace{0.5pt} 0 \hspace{0.5pt}\huxbpad{0.5pt}} &
\multicolumn{1}{c!{\huxvb{0, 0, 0}{0}}}{\huxtpad{0.5pt + 1em}\centering \hspace{0.5pt} 0 \hspace{0.5pt}\huxbpad{0.5pt}} \tabularnewline[-0.5pt]

\hhline{>{\huxb{0, 0, 0}{0.4}}|}
\arrayrulecolor{black}

\multicolumn{1}{!{\huxvb{0, 0, 0}{0}}c!{\huxvb{0, 0, 0}{0}}}{\huxtpad{0.5pt + 1em}\centering \hspace{0.5pt} $\lambda_{5}$ \hspace{0.5pt}\huxbpad{0.5pt}} &
\multicolumn{1}{c!{\huxvb{0, 0, 0}{0}}}{\huxtpad{0.5pt + 1em}\centering \hspace{0.5pt} 23.67 \hspace{0.5pt}\huxbpad{0.5pt}} &
\multicolumn{1}{c!{\huxvb{0, 0, 0}{0}}}{\huxtpad{0.5pt + 1em}\centering \hspace{0.5pt} 23.38 \hspace{0.5pt}\huxbpad{0.5pt}} &
\multicolumn{1}{c!{\huxvb{0, 0, 0}{0.4}}}{\huxtpad{0.5pt + 1em}\centering \hspace{0.5pt} 23.95 \hspace{0.5pt}\huxbpad{0.5pt}} &
\multicolumn{1}{c!{\huxvb{0, 0, 0}{0}}}{\huxtpad{0.5pt + 1em}\centering \hspace{0.5pt} $\gamma_{43}$ \hspace{0.5pt}\huxbpad{0.5pt}} &
\multicolumn{1}{c!{\huxvb{0, 0, 0}{0}}}{\huxtpad{0.5pt + 1em}\centering \hspace{0.5pt} 0.13 \hspace{0.5pt}\huxbpad{0.5pt}} &
\multicolumn{1}{c!{\huxvb{0, 0, 0}{0}}}{\huxtpad{0.5pt + 1em}\centering \hspace{0.5pt} 0.12 \hspace{0.5pt}\huxbpad{0.5pt}} &
\multicolumn{1}{c!{\huxvb{0, 0, 0}{0}}}{\huxtpad{0.5pt + 1em}\centering \hspace{0.5pt} 0.13 \hspace{0.5pt}\huxbpad{0.5pt}} \tabularnewline[-0.5pt]

\hhline{>{\huxb{0, 0, 0}{0.4}}|}
\arrayrulecolor{black}

\multicolumn{1}{!{\huxvb{0, 0, 0}{0}}c!{\huxvb{0, 0, 0}{0}}}{\huxtpad{0.5pt + 1em}\centering \hspace{0.5pt} $\gamma_{11}$ \hspace{0.5pt}\huxbpad{0.5pt}} &
\multicolumn{1}{c!{\huxvb{0, 0, 0}{0}}}{\huxtpad{0.5pt + 1em}\centering \hspace{0.5pt} 0.82 \hspace{0.5pt}\huxbpad{0.5pt}} &
\multicolumn{1}{c!{\huxvb{0, 0, 0}{0}}}{\huxtpad{0.5pt + 1em}\centering \hspace{0.5pt} 0.81 \hspace{0.5pt}\huxbpad{0.5pt}} &
\multicolumn{1}{c!{\huxvb{0, 0, 0}{0.4}}}{\huxtpad{0.5pt + 1em}\centering \hspace{0.5pt} 0.83 \hspace{0.5pt}\huxbpad{0.5pt}} &
\multicolumn{1}{c!{\huxvb{0, 0, 0}{0}}}{\huxtpad{0.5pt + 1em}\centering \hspace{0.5pt} $\gamma_{44}$ \hspace{0.5pt}\huxbpad{0.5pt}} &
\multicolumn{1}{c!{\huxvb{0, 0, 0}{0}}}{\huxtpad{0.5pt + 1em}\centering \hspace{0.5pt} 0.87 \hspace{0.5pt}\huxbpad{0.5pt}} &
\multicolumn{1}{c!{\huxvb{0, 0, 0}{0}}}{\huxtpad{0.5pt + 1em}\centering \hspace{0.5pt} 0.86 \hspace{0.5pt}\huxbpad{0.5pt}} &
\multicolumn{1}{c!{\huxvb{0, 0, 0}{0}}}{\huxtpad{0.5pt + 1em}\centering \hspace{0.5pt} 0.87 \hspace{0.5pt}\huxbpad{0.5pt}} \tabularnewline[-0.5pt]

\hhline{>{\huxb{0, 0, 0}{0.4}}|}
\arrayrulecolor{black}

\multicolumn{1}{!{\huxvb{0, 0, 0}{0}}c!{\huxvb{0, 0, 0}{0}}}{\huxtpad{0.5pt + 1em}\centering \hspace{0.5pt} $\gamma_{12}$ \hspace{0.5pt}\huxbpad{0.5pt}} &
\multicolumn{1}{c!{\huxvb{0, 0, 0}{0}}}{\huxtpad{0.5pt + 1em}\centering \hspace{0.5pt} 0 \hspace{0.5pt}\huxbpad{0.5pt}} &
\multicolumn{1}{c!{\huxvb{0, 0, 0}{0}}}{\huxtpad{0.5pt + 1em}\centering \hspace{0.5pt} 0 \hspace{0.5pt}\huxbpad{0.5pt}} &
\multicolumn{1}{c!{\huxvb{0, 0, 0}{0.4}}}{\huxtpad{0.5pt + 1em}\centering \hspace{0.5pt} 0 \hspace{0.5pt}\huxbpad{0.5pt}} &
\multicolumn{1}{c!{\huxvb{0, 0, 0}{0}}}{\huxtpad{0.5pt + 1em}\centering \hspace{0.5pt} $\gamma_{45}$ \hspace{0.5pt}\huxbpad{0.5pt}} &
\multicolumn{1}{c!{\huxvb{0, 0, 0}{0}}}{\huxtpad{0.5pt + 1em}\centering \hspace{0.5pt} 0.01 \hspace{0.5pt}\huxbpad{0.5pt}} &
\multicolumn{1}{c!{\huxvb{0, 0, 0}{0}}}{\huxtpad{0.5pt + 1em}\centering \hspace{0.5pt} 0.01 \hspace{0.5pt}\huxbpad{0.5pt}} &
\multicolumn{1}{c!{\huxvb{0, 0, 0}{0}}}{\huxtpad{0.5pt + 1em}\centering \hspace{0.5pt} 0.01 \hspace{0.5pt}\huxbpad{0.5pt}} \tabularnewline[-0.5pt]

\hhline{>{\huxb{0, 0, 0}{0.4}}|}
\arrayrulecolor{black}

\multicolumn{1}{!{\huxvb{0, 0, 0}{0}}c!{\huxvb{0, 0, 0}{0}}}{\huxtpad{0.5pt + 1em}\centering \hspace{0.5pt} $\gamma_{13}$ \hspace{0.5pt}\huxbpad{0.5pt}} &
\multicolumn{1}{c!{\huxvb{0, 0, 0}{0}}}{\huxtpad{0.5pt + 1em}\centering \hspace{0.5pt} 0.03 \hspace{0.5pt}\huxbpad{0.5pt}} &
\multicolumn{1}{c!{\huxvb{0, 0, 0}{0}}}{\huxtpad{0.5pt + 1em}\centering \hspace{0.5pt} 0.02 \hspace{0.5pt}\huxbpad{0.5pt}} &
\multicolumn{1}{c!{\huxvb{0, 0, 0}{0.4}}}{\huxtpad{0.5pt + 1em}\centering \hspace{0.5pt} 0.03 \hspace{0.5pt}\huxbpad{0.5pt}} &
\multicolumn{1}{c!{\huxvb{0, 0, 0}{0}}}{\huxtpad{0.5pt + 1em}\centering \hspace{0.5pt} $\gamma_{51}$ \hspace{0.5pt}\huxbpad{0.5pt}} &
\multicolumn{1}{c!{\huxvb{0, 0, 0}{0}}}{\huxtpad{0.5pt + 1em}\centering \hspace{0.5pt} 0 \hspace{0.5pt}\huxbpad{0.5pt}} &
\multicolumn{1}{c!{\huxvb{0, 0, 0}{0}}}{\huxtpad{0.5pt + 1em}\centering \hspace{0.5pt} 0 \hspace{0.5pt}\huxbpad{0.5pt}} &
\multicolumn{1}{c!{\huxvb{0, 0, 0}{0}}}{\huxtpad{0.5pt + 1em}\centering \hspace{0.5pt} 0 \hspace{0.5pt}\huxbpad{0.5pt}} \tabularnewline[-0.5pt]

\hhline{>{\huxb{0, 0, 0}{0.4}}|}
\arrayrulecolor{black}

\multicolumn{1}{!{\huxvb{0, 0, 0}{0}}c!{\huxvb{0, 0, 0}{0}}}{\huxtpad{0.5pt + 1em}\centering \hspace{0.5pt} $\gamma_{14}$ \hspace{0.5pt}\huxbpad{0.5pt}} &
\multicolumn{1}{c!{\huxvb{0, 0, 0}{0}}}{\huxtpad{0.5pt + 1em}\centering \hspace{0.5pt} 0.13 \hspace{0.5pt}\huxbpad{0.5pt}} &
\multicolumn{1}{c!{\huxvb{0, 0, 0}{0}}}{\huxtpad{0.5pt + 1em}\centering \hspace{0.5pt} 0.12 \hspace{0.5pt}\huxbpad{0.5pt}} &
\multicolumn{1}{c!{\huxvb{0, 0, 0}{0.4}}}{\huxtpad{0.5pt + 1em}\centering \hspace{0.5pt} 0.14 \hspace{0.5pt}\huxbpad{0.5pt}} &
\multicolumn{1}{c!{\huxvb{0, 0, 0}{0}}}{\huxtpad{0.5pt + 1em}\centering \hspace{0.5pt} $\gamma_{52}$ \hspace{0.5pt}\huxbpad{0.5pt}} &
\multicolumn{1}{c!{\huxvb{0, 0, 0}{0}}}{\huxtpad{0.5pt + 1em}\centering \hspace{0.5pt} 0 \hspace{0.5pt}\huxbpad{0.5pt}} &
\multicolumn{1}{c!{\huxvb{0, 0, 0}{0}}}{\huxtpad{0.5pt + 1em}\centering \hspace{0.5pt} 0 \hspace{0.5pt}\huxbpad{0.5pt}} &
\multicolumn{1}{c!{\huxvb{0, 0, 0}{0}}}{\huxtpad{0.5pt + 1em}\centering \hspace{0.5pt} 0 \hspace{0.5pt}\huxbpad{0.5pt}} \tabularnewline[-0.5pt]

\hhline{>{\huxb{0, 0, 0}{0.4}}|}
\arrayrulecolor{black}

\multicolumn{1}{!{\huxvb{0, 0, 0}{0}}c!{\huxvb{0, 0, 0}{0}}}{\huxtpad{0.5pt + 1em}\centering \hspace{0.5pt} $\gamma_{15}$ \hspace{0.5pt}\huxbpad{0.5pt}} &
\multicolumn{1}{c!{\huxvb{0, 0, 0}{0}}}{\huxtpad{0.5pt + 1em}\centering \hspace{0.5pt} 0.02 \hspace{0.5pt}\huxbpad{0.5pt}} &
\multicolumn{1}{c!{\huxvb{0, 0, 0}{0}}}{\huxtpad{0.5pt + 1em}\centering \hspace{0.5pt} 0.02 \hspace{0.5pt}\huxbpad{0.5pt}} &
\multicolumn{1}{c!{\huxvb{0, 0, 0}{0.4}}}{\huxtpad{0.5pt + 1em}\centering \hspace{0.5pt} 0.02 \hspace{0.5pt}\huxbpad{0.5pt}} &
\multicolumn{1}{c!{\huxvb{0, 0, 0}{0}}}{\huxtpad{0.5pt + 1em}\centering \hspace{0.5pt} $\gamma_{53}$ \hspace{0.5pt}\huxbpad{0.5pt}} &
\multicolumn{1}{c!{\huxvb{0, 0, 0}{0}}}{\huxtpad{0.5pt + 1em}\centering \hspace{0.5pt} 0 \hspace{0.5pt}\huxbpad{0.5pt}} &
\multicolumn{1}{c!{\huxvb{0, 0, 0}{0}}}{\huxtpad{0.5pt + 1em}\centering \hspace{0.5pt} 0 \hspace{0.5pt}\huxbpad{0.5pt}} &
\multicolumn{1}{c!{\huxvb{0, 0, 0}{0}}}{\huxtpad{0.5pt + 1em}\centering \hspace{0.5pt} 0 \hspace{0.5pt}\huxbpad{0.5pt}} \tabularnewline[-0.5pt]

\hhline{>{\huxb{0, 0, 0}{0.4}}|}
\arrayrulecolor{black}

\multicolumn{1}{!{\huxvb{0, 0, 0}{0}}c!{\huxvb{0, 0, 0}{0}}}{\huxtpad{0.5pt + 1em}\centering \hspace{0.5pt} $\gamma_{21}$ \hspace{0.5pt}\huxbpad{0.5pt}} &
\multicolumn{1}{c!{\huxvb{0, 0, 0}{0}}}{\huxtpad{0.5pt + 1em}\centering \hspace{0.5pt} 0.28 \hspace{0.5pt}\huxbpad{0.5pt}} &
\multicolumn{1}{c!{\huxvb{0, 0, 0}{0}}}{\huxtpad{0.5pt + 1em}\centering \hspace{0.5pt} 0.27 \hspace{0.5pt}\huxbpad{0.5pt}} &
\multicolumn{1}{c!{\huxvb{0, 0, 0}{0.4}}}{\huxtpad{0.5pt + 1em}\centering \hspace{0.5pt} 0.3 \hspace{0.5pt}\huxbpad{0.5pt}} &
\multicolumn{1}{c!{\huxvb{0, 0, 0}{0}}}{\huxtpad{0.5pt + 1em}\centering \hspace{0.5pt} $\gamma_{54}$ \hspace{0.5pt}\huxbpad{0.5pt}} &
\multicolumn{1}{c!{\huxvb{0, 0, 0}{0}}}{\huxtpad{0.5pt + 1em}\centering \hspace{0.5pt} 0.19 \hspace{0.5pt}\huxbpad{0.5pt}} &
\multicolumn{1}{c!{\huxvb{0, 0, 0}{0}}}{\huxtpad{0.5pt + 1em}\centering \hspace{0.5pt} 0.18 \hspace{0.5pt}\huxbpad{0.5pt}} &
\multicolumn{1}{c!{\huxvb{0, 0, 0}{0}}}{\huxtpad{0.5pt + 1em}\centering \hspace{0.5pt} 0.21 \hspace{0.5pt}\huxbpad{0.5pt}} \tabularnewline[-0.5pt]

\hhline{>{\huxb{0, 0, 0}{0.4}}|}
\arrayrulecolor{black}

\multicolumn{1}{!{\huxvb{0, 0, 0}{0}}c!{\huxvb{0, 0, 0}{0}}}{\huxtpad{0.5pt + 1em}\centering \hspace{0.5pt} $\gamma_{22}$ \hspace{0.5pt}\huxbpad{0.5pt}} &
\multicolumn{1}{c!{\huxvb{0, 0, 0}{0}}}{\huxtpad{0.5pt + 1em}\centering \hspace{0.5pt} 0.72 \hspace{0.5pt}\huxbpad{0.5pt}} &
\multicolumn{1}{c!{\huxvb{0, 0, 0}{0}}}{\huxtpad{0.5pt + 1em}\centering \hspace{0.5pt} 0.7 \hspace{0.5pt}\huxbpad{0.5pt}} &
\multicolumn{1}{c!{\huxvb{0, 0, 0}{0.4}}}{\huxtpad{0.5pt + 1em}\centering \hspace{0.5pt} 0.73 \hspace{0.5pt}\huxbpad{0.5pt}} &
\multicolumn{1}{c!{\huxvb{0, 0, 0}{0}}}{\huxtpad{0.5pt + 1em}\centering \hspace{0.5pt} $\gamma_{55}$ \hspace{0.5pt}\huxbpad{0.5pt}} &
\multicolumn{1}{c!{\huxvb{0, 0, 0}{0}}}{\huxtpad{0.5pt + 1em}\centering \hspace{0.5pt} 0.8 \hspace{0.5pt}\huxbpad{0.5pt}} &
\multicolumn{1}{c!{\huxvb{0, 0, 0}{0}}}{\huxtpad{0.5pt + 1em}\centering \hspace{0.5pt} 0.79 \hspace{0.5pt}\huxbpad{0.5pt}} &
\multicolumn{1}{c!{\huxvb{0, 0, 0}{0}}}{\huxtpad{0.5pt + 1em}\centering \hspace{0.5pt} 0.82 \hspace{0.5pt}\huxbpad{0.5pt}} \tabularnewline[-0.5pt]

\hhline{>{\huxb{0, 0, 0}{0.4}}|}
\arrayrulecolor{black}

\multicolumn{1}{!{\huxvb{0, 0, 0}{0}}c!{\huxvb{0, 0, 0}{0}}}{\huxtpad{0.5pt + 1em}\centering \hspace{0.5pt} $\gamma_{23}$ \hspace{0.5pt}\huxbpad{0.5pt}} &
\multicolumn{1}{c!{\huxvb{0, 0, 0}{0}}}{\huxtpad{0.5pt + 1em}\centering \hspace{0.5pt} 0 \hspace{0.5pt}\huxbpad{0.5pt}} &
\multicolumn{1}{c!{\huxvb{0, 0, 0}{0}}}{\huxtpad{0.5pt + 1em}\centering \hspace{0.5pt} 0 \hspace{0.5pt}\huxbpad{0.5pt}} &
\multicolumn{1}{c!{\huxvb{0, 0, 0}{0.4}}}{\huxtpad{0.5pt + 1em}\centering \hspace{0.5pt} 0 \hspace{0.5pt}\huxbpad{0.5pt}} &
\multicolumn{1}{c!{\huxvb{0, 0, 0}{0}}}{\huxtpad{0.5pt + 1em}\centering \hspace{0.5pt} $\delta_{1}$ \hspace{0.5pt}\huxbpad{0.5pt}} &
\multicolumn{1}{c!{\huxvb{0, 0, 0}{0}}}{\huxtpad{0.5pt + 1em}\centering \hspace{0.5pt} 0.24 \hspace{0.5pt}\huxbpad{0.5pt}} &
\multicolumn{1}{c!{\huxvb{0, 0, 0}{0}}}{\huxtpad{0.5pt + 1em}\centering \hspace{0.5pt} 0.23 \hspace{0.5pt}\huxbpad{0.5pt}} &
\multicolumn{1}{c!{\huxvb{0, 0, 0}{0}}}{\huxtpad{0.5pt + 1em}\centering \hspace{0.5pt} 0.25 \hspace{0.5pt}\huxbpad{0.5pt}} \tabularnewline[-0.5pt]

\hhline{>{\huxb{0, 0, 0}{0.4}}|}
\arrayrulecolor{black}

\multicolumn{1}{!{\huxvb{0, 0, 0}{0}}c!{\huxvb{0, 0, 0}{0}}}{\huxtpad{0.5pt + 1em}\centering \hspace{0.5pt} $\gamma_{24}$ \hspace{0.5pt}\huxbpad{0.5pt}} &
\multicolumn{1}{c!{\huxvb{0, 0, 0}{0}}}{\huxtpad{0.5pt + 1em}\centering \hspace{0.5pt} 0 \hspace{0.5pt}\huxbpad{0.5pt}} &
\multicolumn{1}{c!{\huxvb{0, 0, 0}{0}}}{\huxtpad{0.5pt + 1em}\centering \hspace{0.5pt} 0 \hspace{0.5pt}\huxbpad{0.5pt}} &
\multicolumn{1}{c!{\huxvb{0, 0, 0}{0.4}}}{\huxtpad{0.5pt + 1em}\centering \hspace{0.5pt} 0 \hspace{0.5pt}\huxbpad{0.5pt}} &
\multicolumn{1}{c!{\huxvb{0, 0, 0}{0}}}{\huxtpad{0.5pt + 1em}\centering \hspace{0.5pt} $\delta_{2}$ \hspace{0.5pt}\huxbpad{0.5pt}} &
\multicolumn{1}{c!{\huxvb{0, 0, 0}{0}}}{\huxtpad{0.5pt + 1em}\centering \hspace{0.5pt} 0.15 \hspace{0.5pt}\huxbpad{0.5pt}} &
\multicolumn{1}{c!{\huxvb{0, 0, 0}{0}}}{\huxtpad{0.5pt + 1em}\centering \hspace{0.5pt} 0.15 \hspace{0.5pt}\huxbpad{0.5pt}} &
\multicolumn{1}{c!{\huxvb{0, 0, 0}{0}}}{\huxtpad{0.5pt + 1em}\centering \hspace{0.5pt} 0.16 \hspace{0.5pt}\huxbpad{0.5pt}} \tabularnewline[-0.5pt]

\hhline{>{\huxb{0, 0, 0}{0.4}}|}
\arrayrulecolor{black}

\multicolumn{1}{!{\huxvb{0, 0, 0}{0}}c!{\huxvb{0, 0, 0}{0}}}{\huxtpad{0.5pt + 1em}\centering \hspace{0.5pt} $\gamma_{25}$ \hspace{0.5pt}\huxbpad{0.5pt}} &
\multicolumn{1}{c!{\huxvb{0, 0, 0}{0}}}{\huxtpad{0.5pt + 1em}\centering \hspace{0.5pt} 0 \hspace{0.5pt}\huxbpad{0.5pt}} &
\multicolumn{1}{c!{\huxvb{0, 0, 0}{0}}}{\huxtpad{0.5pt + 1em}\centering \hspace{0.5pt} 0 \hspace{0.5pt}\huxbpad{0.5pt}} &
\multicolumn{1}{c!{\huxvb{0, 0, 0}{0.4}}}{\huxtpad{0.5pt + 1em}\centering \hspace{0.5pt} 0 \hspace{0.5pt}\huxbpad{0.5pt}} &
\multicolumn{1}{c!{\huxvb{0, 0, 0}{0}}}{\huxtpad{0.5pt + 1em}\centering \hspace{0.5pt} $\delta_{3}$ \hspace{0.5pt}\huxbpad{0.5pt}} &
\multicolumn{1}{c!{\huxvb{0, 0, 0}{0}}}{\huxtpad{0.5pt + 1em}\centering \hspace{0.5pt} 0.28 \hspace{0.5pt}\huxbpad{0.5pt}} &
\multicolumn{1}{c!{\huxvb{0, 0, 0}{0}}}{\huxtpad{0.5pt + 1em}\centering \hspace{0.5pt} 0.27 \hspace{0.5pt}\huxbpad{0.5pt}} &
\multicolumn{1}{c!{\huxvb{0, 0, 0}{0}}}{\huxtpad{0.5pt + 1em}\centering \hspace{0.5pt} 0.29 \hspace{0.5pt}\huxbpad{0.5pt}} \tabularnewline[-0.5pt]

\hhline{>{\huxb{0, 0, 0}{0.4}}|}
\arrayrulecolor{black}

\multicolumn{1}{!{\huxvb{0, 0, 0}{0}}c!{\huxvb{0, 0, 0}{0}}}{\huxtpad{0.5pt + 1em}\centering \hspace{0.5pt} $\gamma_{31}$ \hspace{0.5pt}\huxbpad{0.5pt}} &
\multicolumn{1}{c!{\huxvb{0, 0, 0}{0}}}{\huxtpad{0.5pt + 1em}\centering \hspace{0.5pt} 0 \hspace{0.5pt}\huxbpad{0.5pt}} &
\multicolumn{1}{c!{\huxvb{0, 0, 0}{0}}}{\huxtpad{0.5pt + 1em}\centering \hspace{0.5pt} 0 \hspace{0.5pt}\huxbpad{0.5pt}} &
\multicolumn{1}{c!{\huxvb{0, 0, 0}{0.4}}}{\huxtpad{0.5pt + 1em}\centering \hspace{0.5pt} 0 \hspace{0.5pt}\huxbpad{0.5pt}} &
\multicolumn{1}{c!{\huxvb{0, 0, 0}{0}}}{\huxtpad{0.5pt + 1em}\centering \hspace{0.5pt} $\delta_{4}$ \hspace{0.5pt}\huxbpad{0.5pt}} &
\multicolumn{1}{c!{\huxvb{0, 0, 0}{0}}}{\huxtpad{0.5pt + 1em}\centering \hspace{0.5pt} 0.29 \hspace{0.5pt}\huxbpad{0.5pt}} &
\multicolumn{1}{c!{\huxvb{0, 0, 0}{0}}}{\huxtpad{0.5pt + 1em}\centering \hspace{0.5pt} 0.28 \hspace{0.5pt}\huxbpad{0.5pt}} &
\multicolumn{1}{c!{\huxvb{0, 0, 0}{0}}}{\huxtpad{0.5pt + 1em}\centering \hspace{0.5pt} 0.3 \hspace{0.5pt}\huxbpad{0.5pt}} \tabularnewline[-0.5pt]

\hhline{>{\huxb{0, 0, 0}{0.4}}|}
\arrayrulecolor{black}

\multicolumn{1}{!{\huxvb{0, 0, 0}{0}}c!{\huxvb{0, 0, 0}{0}}}{\huxtpad{0.5pt + 1em}\centering \hspace{0.5pt} $\gamma_{32}$ \hspace{0.5pt}\huxbpad{0.5pt}} &
\multicolumn{1}{c!{\huxvb{0, 0, 0}{0}}}{\huxtpad{0.5pt + 1em}\centering \hspace{0.5pt} 0.15 \hspace{0.5pt}\huxbpad{0.5pt}} &
\multicolumn{1}{c!{\huxvb{0, 0, 0}{0}}}{\huxtpad{0.5pt + 1em}\centering \hspace{0.5pt} 0.15 \hspace{0.5pt}\huxbpad{0.5pt}} &
\multicolumn{1}{c!{\huxvb{0, 0, 0}{0.4}}}{\huxtpad{0.5pt + 1em}\centering \hspace{0.5pt} 0.16 \hspace{0.5pt}\huxbpad{0.5pt}} &
\multicolumn{1}{c!{\huxvb{0, 0, 0}{0}}}{\huxtpad{0.5pt + 1em}\centering \hspace{0.5pt} $\delta_{5}$ \hspace{0.5pt}\huxbpad{0.5pt}} &
\multicolumn{1}{c!{\huxvb{0, 0, 0}{0}}}{\huxtpad{0.5pt + 1em}\centering \hspace{0.5pt} 0.04 \hspace{0.5pt}\huxbpad{0.5pt}} &
\multicolumn{1}{c!{\huxvb{0, 0, 0}{0}}}{\huxtpad{0.5pt + 1em}\centering \hspace{0.5pt} 0.03 \hspace{0.5pt}\huxbpad{0.5pt}} &
\multicolumn{1}{c!{\huxvb{0, 0, 0}{0}}}{\huxtpad{0.5pt + 1em}\centering \hspace{0.5pt} 0.04 \hspace{0.5pt}\huxbpad{0.5pt}} \tabularnewline[-0.5pt]

\hhline{>{\huxb{0, 0, 0}{0.4}}|}
\arrayrulecolor{black}

\multicolumn{1}{!{\huxvb{0, 0, 0}{0}}c!{\huxvb{0, 0, 0}{0}}}{\huxtpad{0.5pt + 1em}\centering \hspace{0.5pt} $\gamma_{33}$ \hspace{0.5pt}\huxbpad{0.5pt}} &
\multicolumn{1}{c!{\huxvb{0, 0, 0}{0}}}{\huxtpad{0.5pt + 1em}\centering \hspace{0.5pt} 0.85 \hspace{0.5pt}\huxbpad{0.5pt}} &
\multicolumn{1}{c!{\huxvb{0, 0, 0}{0}}}{\huxtpad{0.5pt + 1em}\centering \hspace{0.5pt} 0.84 \hspace{0.5pt}\huxbpad{0.5pt}} &
\multicolumn{1}{c!{\huxvb{0, 0, 0}{0.4}}}{\huxtpad{0.5pt + 1em}\centering \hspace{0.5pt} 0.85 \hspace{0.5pt}\huxbpad{0.5pt}} &
\multicolumn{1}{c!{\huxvb{0, 0, 0}{0}}}{\huxtpad{0.5pt + 1em}\centering \hspace{0.5pt}  \hspace{0.5pt}\huxbpad{0.5pt}} &
\multicolumn{1}{c!{\huxvb{0, 0, 0}{0}}}{\huxtpad{0.5pt + 1em}\centering \hspace{0.5pt}  \hspace{0.5pt}\huxbpad{0.5pt}} &
\multicolumn{1}{c!{\huxvb{0, 0, 0}{0}}}{\huxtpad{0.5pt + 1em}\centering \hspace{0.5pt}  \hspace{0.5pt}\huxbpad{0.5pt}} &
\multicolumn{1}{c!{\huxvb{0, 0, 0}{0}}}{\huxtpad{0.5pt + 1em}\centering \hspace{0.5pt}  \hspace{0.5pt}\huxbpad{0.5pt}} \tabularnewline[-0.5pt]

\hhline{>{\huxb{0, 0, 0}{0.4}}->{\huxb{0, 0, 0}{0.4}}->{\huxb{0, 0, 0}{0.4}}->{\huxb{0, 0, 0}{0.4}}->{\huxb{0, 0, 0}{0.4}}->{\huxb{0, 0, 0}{0.4}}->{\huxb{0, 0, 0}{0.4}}->{\huxb{0, 0, 0}{0.4}}-}
\arrayrulecolor{black}

\multicolumn{1}{!{\huxvb{0, 0, 0}{0}}c!{\huxvb{0, 0, 0}{0}}}{\huxtpad{0.5pt + 1em}\centering \hspace{0.5pt}  \hspace{0.5pt}\huxbpad{0.5pt}} &
\multicolumn{1}{c!{\huxvb{0, 0, 0}{0}}}{\huxtpad{0.5pt + 1em}\centering \hspace{0.5pt}  \hspace{0.5pt}\huxbpad{0.5pt}} &
\multicolumn{1}{c!{\huxvb{0, 0, 0}{0}}}{\huxtpad{0.5pt + 1em}\centering \hspace{0.5pt} m \hspace{0.5pt}\huxbpad{0.5pt}} &
\multicolumn{1}{c!{\huxvb{0, 0, 0}{0}}}{\huxtpad{0.5pt + 1em}\centering \hspace{0.5pt} AIC \hspace{0.5pt}\huxbpad{0.5pt}} &
\multicolumn{1}{c!{\huxvb{0, 0, 0}{0}}}{\huxtpad{0.5pt + 1em}\centering \hspace{0.5pt} BIC \hspace{0.5pt}\huxbpad{0.5pt}} &
\multicolumn{1}{c!{\huxvb{0, 0, 0}{0}}}{\huxtpad{0.5pt + 1em}\centering \hspace{0.5pt}  \hspace{0.5pt}\huxbpad{0.5pt}} &
\multicolumn{1}{c!{\huxvb{0, 0, 0}{0}}}{\huxtpad{0.5pt + 1em}\centering \hspace{0.5pt}  \hspace{0.5pt}\huxbpad{0.5pt}} &
\multicolumn{1}{c!{\huxvb{0, 0, 0}{0}}}{\huxtpad{0.5pt + 1em}\centering \hspace{0.5pt}  \hspace{0.5pt}\huxbpad{0.5pt}} \tabularnewline[-0.5pt]

\hhline{>{\huxb{255, 255, 255}{0.4}}->{\huxb{255, 255, 255}{0.4}}->{\huxb{0, 0, 0}{0.4}}->{\huxb{0, 0, 0}{0.4}}->{\huxb{0, 0, 0}{0.4}}->{\huxb{255, 255, 255}{0.4}}->{\huxb{255, 255, 255}{0.4}}->{\huxb{255, 255, 255}{0.4}}-}
\arrayrulecolor{black}

\multicolumn{1}{!{\huxvb{0, 0, 0}{0}}c!{\huxvb{0, 0, 0}{0}}}{\huxtpad{0.5pt + 1em}\centering \hspace{0.5pt}  \hspace{0.5pt}\huxbpad{0.5pt}} &
\multicolumn{1}{c!{\huxvb{0, 0, 0}{0}}}{\huxtpad{0.5pt + 1em}\centering \hspace{0.5pt}  \hspace{0.5pt}\huxbpad{0.5pt}} &
\multicolumn{1}{c!{\huxvb{0, 0, 0}{0}}}{\huxtpad{0.5pt + 1em}\centering \hspace{0.5pt} 2 \hspace{0.5pt}\huxbpad{0.5pt}} &
\multicolumn{1}{c!{\huxvb{0, 0, 0}{0}}}{\huxtpad{0.5pt + 1em}\centering \hspace{0.5pt} 518598 \hspace{0.5pt}\huxbpad{0.5pt}} &
\multicolumn{1}{c!{\huxvb{0, 0, 0}{0}}}{\huxtpad{0.5pt + 1em}\centering \hspace{0.5pt} 518636 \hspace{0.5pt}\huxbpad{0.5pt}} &
\multicolumn{1}{c!{\huxvb{0, 0, 0}{0}}}{\huxtpad{0.5pt + 1em}\centering \hspace{0.5pt}  \hspace{0.5pt}\huxbpad{0.5pt}} &
\multicolumn{1}{c!{\huxvb{0, 0, 0}{0}}}{\huxtpad{0.5pt + 1em}\centering \hspace{0.5pt}  \hspace{0.5pt}\huxbpad{0.5pt}} &
\multicolumn{1}{c!{\huxvb{0, 0, 0}{0}}}{\huxtpad{0.5pt + 1em}\centering \hspace{0.5pt}  \hspace{0.5pt}\huxbpad{0.5pt}} \tabularnewline[-0.5pt]

\hhline{}
\arrayrulecolor{black}

\multicolumn{1}{!{\huxvb{0, 0, 0}{0}}c!{\huxvb{0, 0, 0}{0}}}{\huxtpad{0.5pt + 1em}\centering \hspace{0.5pt}  \hspace{0.5pt}\huxbpad{0.5pt}} &
\multicolumn{1}{c!{\huxvb{0, 0, 0}{0}}}{\huxtpad{0.5pt + 1em}\centering \hspace{0.5pt}  \hspace{0.5pt}\huxbpad{0.5pt}} &
\multicolumn{1}{c!{\huxvb{0, 0, 0}{0}}}{\huxtpad{0.5pt + 1em}\centering \hspace{0.5pt} 3 \hspace{0.5pt}\huxbpad{0.5pt}} &
\multicolumn{1}{c!{\huxvb{0, 0, 0}{0}}}{\huxtpad{0.5pt + 1em}\centering \hspace{0.5pt} 494824 \hspace{0.5pt}\huxbpad{0.5pt}} &
\multicolumn{1}{c!{\huxvb{0, 0, 0}{0}}}{\huxtpad{0.5pt + 1em}\centering \hspace{0.5pt} 494908 \hspace{0.5pt}\huxbpad{0.5pt}} &
\multicolumn{1}{c!{\huxvb{0, 0, 0}{0}}}{\huxtpad{0.5pt + 1em}\centering \hspace{0.5pt}  \hspace{0.5pt}\huxbpad{0.5pt}} &
\multicolumn{1}{c!{\huxvb{0, 0, 0}{0}}}{\huxtpad{0.5pt + 1em}\centering \hspace{0.5pt}  \hspace{0.5pt}\huxbpad{0.5pt}} &
\multicolumn{1}{c!{\huxvb{0, 0, 0}{0}}}{\huxtpad{0.5pt + 1em}\centering \hspace{0.5pt}  \hspace{0.5pt}\huxbpad{0.5pt}} \tabularnewline[-0.5pt]

\hhline{}
\arrayrulecolor{black}

\multicolumn{1}{!{\huxvb{0, 0, 0}{0}}c!{\huxvb{0, 0, 0}{0}}}{\huxtpad{0.5pt + 1em}\centering \hspace{0.5pt}  \hspace{0.5pt}\huxbpad{0.5pt}} &
\multicolumn{1}{c!{\huxvb{0, 0, 0}{0}}}{\huxtpad{0.5pt + 1em}\centering \hspace{0.5pt}  \hspace{0.5pt}\huxbpad{0.5pt}} &
\multicolumn{1}{c!{\huxvb{0, 0, 0}{0}}}{\huxtpad{0.5pt + 1em}\centering \hspace{0.5pt} 4 \hspace{0.5pt}\huxbpad{0.5pt}} &
\multicolumn{1}{c!{\huxvb{0, 0, 0}{0}}}{\huxtpad{0.5pt + 1em}\centering \hspace{0.5pt} 485206 \hspace{0.5pt}\huxbpad{0.5pt}} &
\multicolumn{1}{c!{\huxvb{0, 0, 0}{0}}}{\huxtpad{0.5pt + 1em}\centering \hspace{0.5pt} 485356 \hspace{0.5pt}\huxbpad{0.5pt}} &
\multicolumn{1}{c!{\huxvb{0, 0, 0}{0}}}{\huxtpad{0.5pt + 1em}\centering \hspace{0.5pt}  \hspace{0.5pt}\huxbpad{0.5pt}} &
\multicolumn{1}{c!{\huxvb{0, 0, 0}{0}}}{\huxtpad{0.5pt + 1em}\centering \hspace{0.5pt}  \hspace{0.5pt}\huxbpad{0.5pt}} &
\multicolumn{1}{c!{\huxvb{0, 0, 0}{0}}}{\huxtpad{0.5pt + 1em}\centering \hspace{0.5pt}  \hspace{0.5pt}\huxbpad{0.5pt}} \tabularnewline[-0.5pt]

\hhline{}
\arrayrulecolor{black}

\multicolumn{1}{!{\huxvb{0, 0, 0}{0}}c!{\huxvb{0, 0, 0}{0}}}{\huxtpad{0.5pt + 1em}\centering \hspace{0.5pt}  \hspace{0.5pt}\huxbpad{0.5pt}} &
\multicolumn{1}{c!{\huxvb{0, 0, 0}{0}}}{\huxtpad{0.5pt + 1em}\centering \hspace{0.5pt}  \hspace{0.5pt}\huxbpad{0.5pt}} &
\multicolumn{1}{c!{\huxvb{0, 0, 0}{0}}}{\huxtpad{0.5pt + 1em}\centering \hspace{0.5pt} 5 \hspace{0.5pt}\huxbpad{0.5pt}} &
\multicolumn{1}{c!{\huxvb{0, 0, 0}{0}}}{\huxtpad{0.5pt + 1em}\centering \hspace{0.5pt} 481627 \hspace{0.5pt}\huxbpad{0.5pt}} &
\multicolumn{1}{c!{\huxvb{0, 0, 0}{0}}}{\huxtpad{0.5pt + 1em}\centering \hspace{0.5pt} 481862 \hspace{0.5pt}\huxbpad{0.5pt}} &
\multicolumn{1}{c!{\huxvb{0, 0, 0}{0}}}{\huxtpad{0.5pt + 1em}\centering \hspace{0.5pt}  \hspace{0.5pt}\huxbpad{0.5pt}} &
\multicolumn{1}{c!{\huxvb{0, 0, 0}{0}}}{\huxtpad{0.5pt + 1em}\centering \hspace{0.5pt}  \hspace{0.5pt}\huxbpad{0.5pt}} &
\multicolumn{1}{c!{\huxvb{0, 0, 0}{0}}}{\huxtpad{0.5pt + 1em}\centering \hspace{0.5pt}  \hspace{0.5pt}\huxbpad{0.5pt}} \tabularnewline[-0.5pt]

\hhline{>{\huxb{0, 0, 0}{0.4}}->{\huxb{0, 0, 0}{0.4}}->{\huxb{0, 0, 0}{0.4}}->{\huxb{0, 0, 0}{0.4}}->{\huxb{0, 0, 0}{0.4}}->{\huxb{0, 0, 0}{0.4}}->{\huxb{0, 0, 0}{0.4}}->{\huxb{0, 0, 0}{0.4}}-}
\arrayrulecolor{black}
\end{tabular}
\end{threeparttable}\par\end{centerbox}

\end{table}

\clearpage

%%%%%%%%%%%%%%%%%%%%%%%%%%%%%%%%%%%%%%%%%%%%%%%%%%%%%%%%%%%%%%%%%
\section{Performance and accuracy of different optimizers: additional figures and results}
\label{sec:appendix-perf}
%%%%%%%%%%%%%%%%%%%%%%%%%%%%%%%%%%%%%%%%%%%%%%%%%%%%%%%%%%%%%%%%%

%In this section we present a few corresponding figures to the results from Section \ref{sec:optimization}. %, and an additional simulation study.

\cref{fig:bootstrap-graph-tinn,fig:bootstrap-graph-simu1,fig:bootstrap-graph-simu3} show the median of the parameter estimates and the nll, along with 95\% CIs from the three different simulation designs described in \autoref{sec:optimization} across all optimizers studied, over 1000 realizations from the different models. All figures show that the different optimizers behave almost identically in terms of accuracy. However, we restate that the initial values used here are equal to the true parameter values.

\begin{knitrout}
\definecolor{shadecolor}{rgb}{0.969, 0.969, 0.969}\color{fgcolor}\begin{figure}[htb]

{\centering \includegraphics[width=\maxwidth]{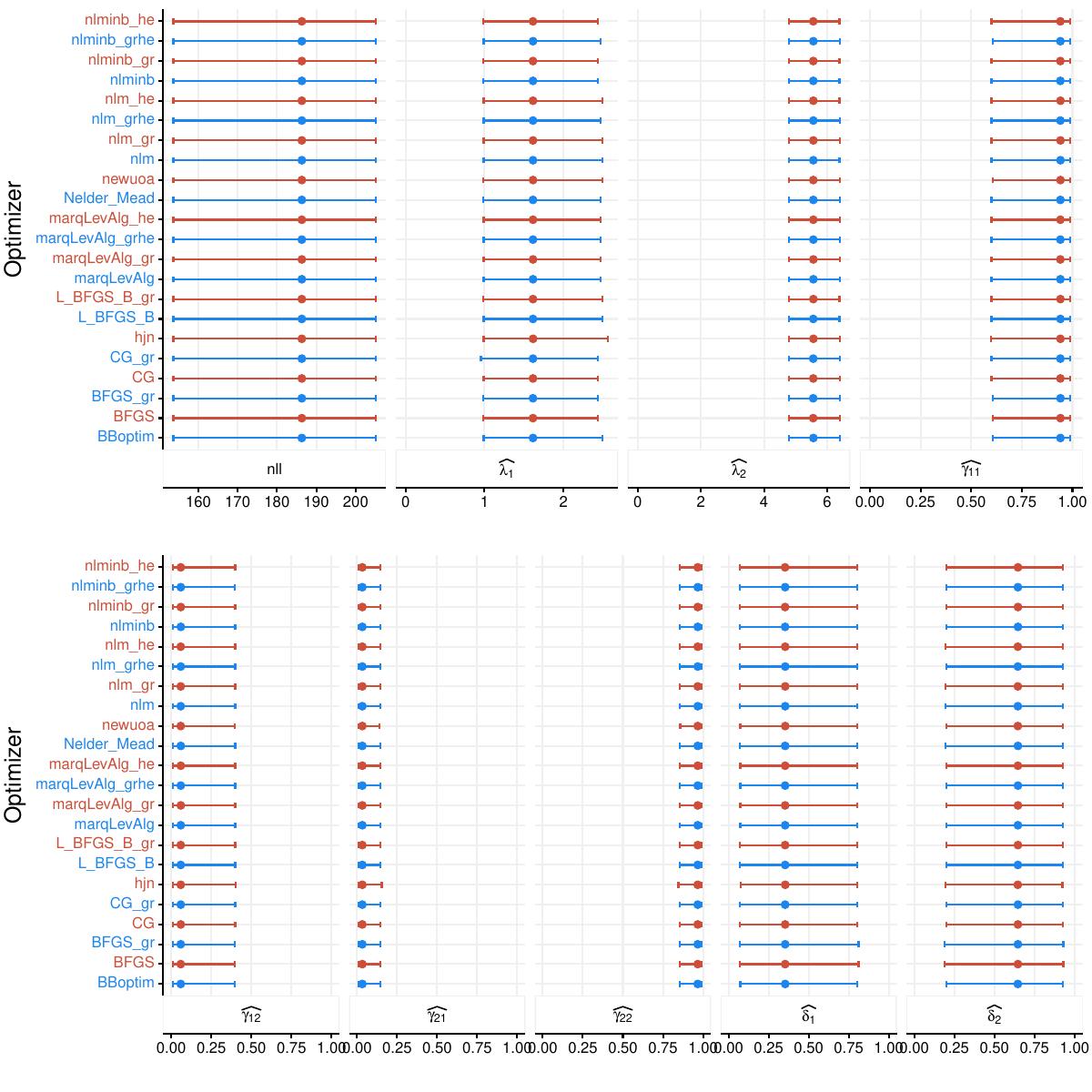} 

}

\caption[Plots of estimates and NLL when estimating a two-state Poisson HMM on the TYT data set (87 data), over 1000 realizations]{Plots of estimates and NLL when estimating a two-state Poisson HMM on the TYT data set (87 data), over 1000 realizations. The columns display in order the NLL, Poisson rates, TPM elements, and the stationary distribution. The dots represent the medians, and the lines display the 95\% percentile CIs.}\label{fig:bootstrap-graph-tinn}
\end{figure}

\end{knitrout}

\begin{knitrout}
\definecolor{shadecolor}{rgb}{0.969, 0.969, 0.969}\color{fgcolor}\begin{figure}[htb]

{\centering \includegraphics[width=\maxwidth]{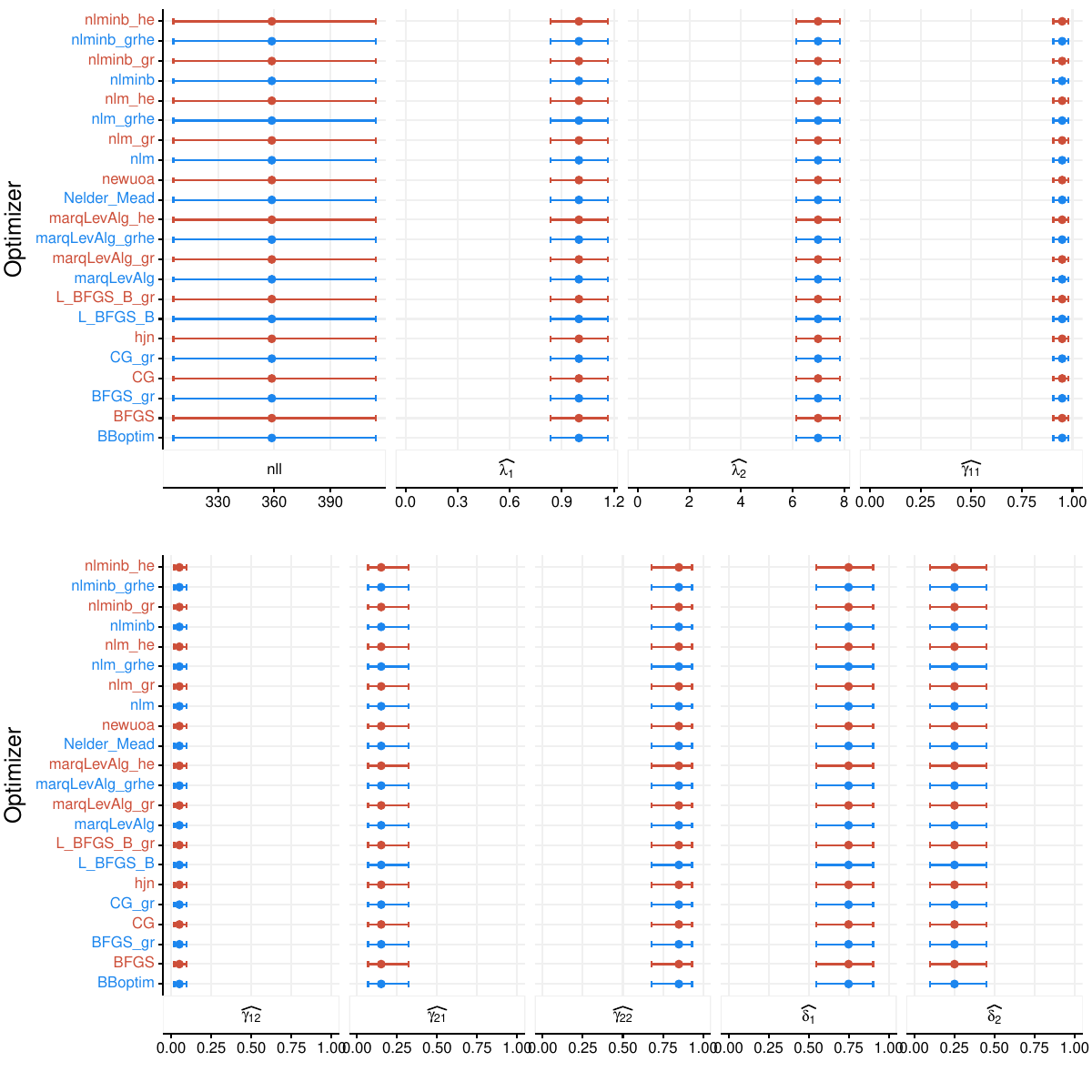} 

}

\caption[Plots of estimates and NLL when estimating a two-state Poisson HMM in the second study design, with (200 observations), over 1000 realizations]{Plots of estimates and NLL when estimating a two-state Poisson HMM in the second study design, with (200 observations), over 1000 realizations. The columns display in order the NLL, Poisson rates, TPM elements, and the stationary distribution. The dots represent the medians, and the lines display the 95\% percentile CIs.}\label{fig:bootstrap-graph-simu1}
\end{figure}

\end{knitrout}

\begin{knitrout}
\definecolor{shadecolor}{rgb}{0.969, 0.969, 0.969}\color{fgcolor}\begin{figure}[htb]

{\centering \includegraphics[width=\maxwidth]{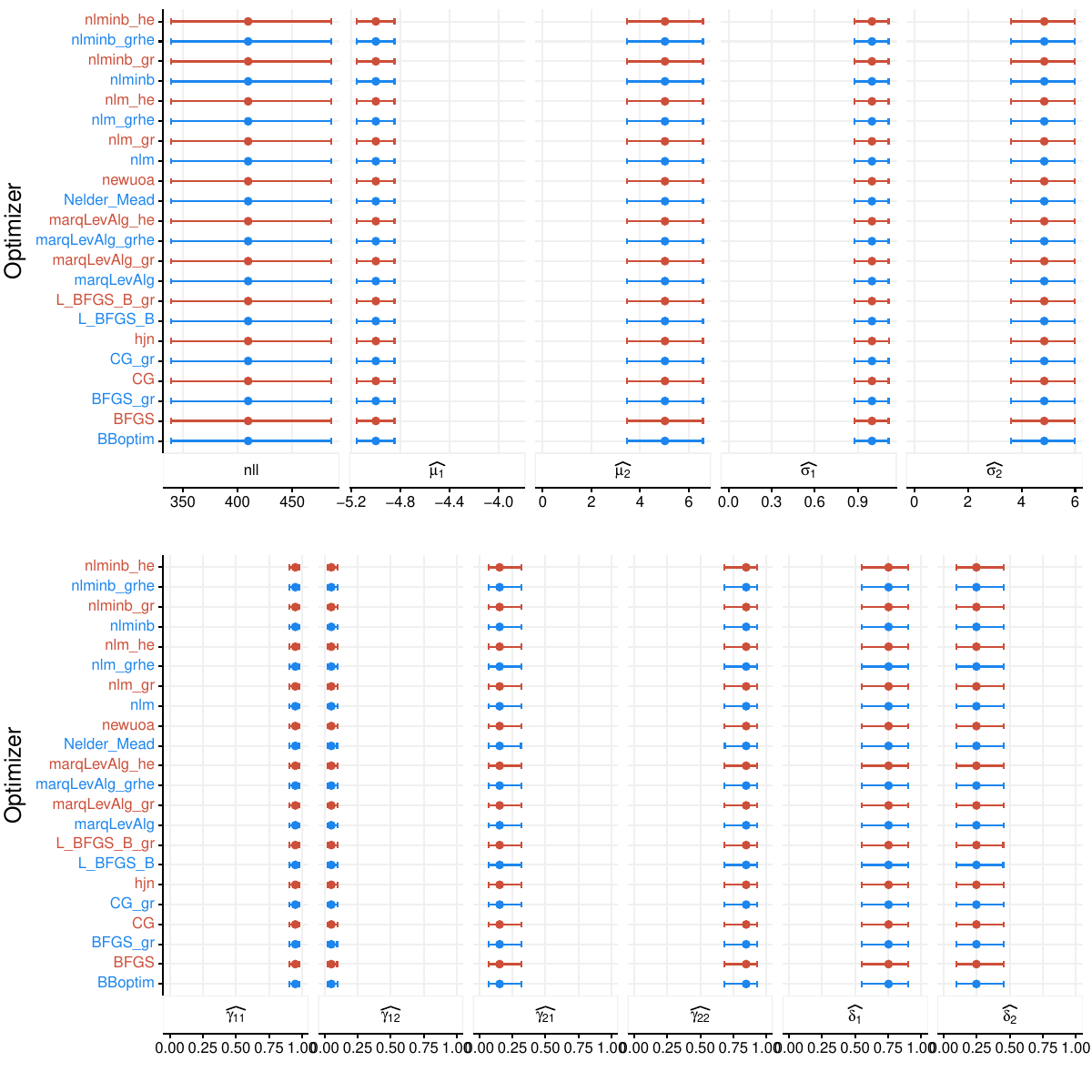} 

}

\caption[Plots of estimates and NLL when estimating a two-state Poisson HMM in the third study design, with (200 observations), over 1000 realizations]{Plots of estimates and NLL when estimating a two-state Poisson HMM in the third study design, with (200 observations), over 1000 realizations. The columns display in order the NLL, Poisson rates, TPM elements, and the stationary distribution. The dots represent the medians, and the lines display the 95\% percentile CIs.}\label{fig:bootstrap-graph-simu3}
\end{figure}

\end{knitrout}

\restoregeometry

\clearpage

%%%%%%%%%%%%%%%%%%%%%%%%%%%%%%%%%%%%%%%%%%%%%%%%%%%%%%%%%%%%%%%%%
\end{document}